\documentclass[12pt]{iopart}

\usepackage[utf8]{inputenc}
\usepackage{xcolor}
\usepackage{url}
\usepackage{subcaption}

\usepackage[sort&compress,numbers]{natbib}

\usepackage{latexsym}

\expandafter\let\csname equation*\endcsname\relax
\expandafter\let\csname endequation*\endcsname\relax

\usepackage{amsmath}
\usepackage{amssymb}
\usepackage{amsfonts}

\usepackage{dsfont}
\usepackage{slashed}
\usepackage{bm}
\usepackage{urwchancal}
\DeclareMathAlphabet{\mathpzc}{OT1}{pzc}{m}{it} % this line is necessary for the \mathpzc command (CA)

\usepackage[vcentermath]{youngtab}
\usepackage{supertabular}
\usepackage{placeins}
\usepackage{epsfig}
\usepackage{graphicx}

\usepackage{ctable}
\usepackage{multirow}
\usepackage{makecell}
\usepackage{rotating}
\usepackage{tabu}

\usepackage{enumitem}
\usepackage[textwidth = 155mm]{geometry}

\definecolor{purple}{rgb}{0.5,0,0.5}
\definecolor{blue}{rgb}{0.0,0,0.9}
\definecolor{prdblue}{rgb}{0.133,0.118,0.498}
\usepackage[colorlinks=true, pdfstartview=FitV, linkcolor=prdblue, citecolor= prdblue, urlcolor=prdblue]{hyperref}
\usepackage{academicons}
\newcommand{\orcid}[1]{\href{https://orcid.org/#1}{\textcolor[HTML]{A6CE39}{\aiOrcid}}}
\newcommand\newblock{\hskip .11em\@plus.33em\@minus.07em}
\begin{document}

\title[Revealing the structure of light pseudoscalar mesons at the EIC]{Revealing the structure of light pseudoscalar mesons at the Electron-Ion Collider}

% SJDKK - Changed author list to match IOP style guidelines, they explicitly state that initials should NOT be followed by a full stop.
%%  Also ... must use "equation" & "figure" & "table" & "section" and citations should not have "Ref." or any such thing before them.
%%
\author{J~Arrington$^{1}$, C~Ayerbe~Gayoso$^2$, PC~Barry$^{6,21}$,
V~Berdnikov$^3$, 
D~Binosi$^4$, 
L~Chang$^5$, 
M~Diefenthaler$^6$, 
M~Ding$^4$, 
R~Ent$^6$, 
T~Frederico$^7$,
Y~Furletova$^6$,
TJ~Hobbs$^{6,8,20}$, T~Horn$^{3,6,*}$, 
GM~Huber$^{9}$, 
SJD~Kay$^{9}$, 
C~Keppel$^6$, 
H-W~Lin$^{10}$,
C~Mezrag$^{11}$,
R~Montgomery$^{12}$,
IL~Pegg$^3$, 
K~Raya$^{5,13}$, 
P~Reimer$^{14}$, %
DG~Richards$^6$,
CD~Roberts$^{15,16}$,
J~Rodr\'{\i}guez-Quintero$^{17}$%\footnote{\url{https://orcid.org/0000-0002-1651-5717} - J Rodr\'{\i}guez-Quintero},
D~Romanov$^6$,
G~Salm{\`e}$^{18}$,
N~Sato$^6$,
J~Segovia$^{19}$,
P~Stepanov$^3$, AS~Tadepalli$^6$ and RL~Trotta$^3$ % and R~Yoshida$^{14}$
}

\address{$^1$ Lawrence Berkeley National Laboratory, Berkeley, CA 94720, USA}
\address{$^2$ Mississippi State University, Starkville, MS, USA}
\address{$^3$ Catholic University of America, Washington, DC, USA}
\address{$^4$ European Centre for Theoretical Studies in Nuclear Physics and Related Areas (ECT$^\ast$) and Fondazione Bruno Kessler Villa Tambosi, Strada delle Tabarelle 286, I-38123 Villazzano (TN) Italy}
\address{$^5$ School of Physics, Nankai University, Tianjin 300071, China}
\address{$^6$ Thomas Jefferson National Accelerator Facility, Newport News, Virginia 23606, USA}
\address{$^7$ Instituto Tecnol\'ogico de Aeron\'autica, 12.228-900 S\~ao Jos\'e dos Campos, Brazil}
\address{$^8$ Southern Methodist University, Dallas, TX 75275-0175, USA}
\address{$^9$ University of Regina, Regina, SK S4S~0A2, Canada}
\address{$^{10}$ Michigan State University, East Lansing, MI 48824, USA}
\address{$^{11}$ IRFU, CEA, Universit\'e Paris-Saclay, F-91191 Gif-sur-Yvette, France}
\address{$^{12}$ SUPA School of Physics and Astronomy, University of Glasgow, Glasgow \mbox{G12 8QQ}, United Kingdom}
\address{$^{13}$ Instituto de Ciencias Nucleares, Universidad Nacional Aut{\'o}noma de M{\'e}xico, Apartado Postal 70-543, C.P. 04510, CDMX, M{\'e}xico }
\address{$^{14}$ Argonne National Laboratory, Lemont, IL 60439, USA}
\address{$^{15}$ School of Physics, Nanjing University, Nanjing, Jiangsu 210093, China}
\address{$^{16}$ Institute for Nonperturbative Physics, Nanjing University, Nanjing, Jiangsu 210093, China}
\address{$^{17}$ Department of Integrated Sciences and Center for Advanced Studies in Physics, Mathematics and Computation, University of Huelva, E-21071 Huelva, Spain}
\address{$^{18}$ Istituto Nazionale di Fisica Nucleare, Sezione di Roma, P.le A. Moro 2, I-00185 Rome, Italy}
\address{$^{19}$ Departamento de Sistemas F\'isicos, Qu\'imicos y Naturales, Universidad Pablo de Olavide, E-41013 Sevilla, Spain}
\address{$^{20}$ Department of Physics, Illinois Institute of Technology, Chicago, IL 60616, USA}
\address{$^{21}$ North Carolina State University, Raleigh, NC 27607, USA}
\ead{$^{*}$hornt@cua.edu}

\begin{indented}
%% 2021 01 30
%% 2021 01 27
\item[]2021 January \\[2ex]%\footnotetext{\hspace*{-\parindent}
\begin{footnotesize}
\url{https://orcid.org/0000-0002-0702-1328} - J. Arrington\\
\url{https://orcid.org/0000-0001-8640-5380} - C. Ayerbe Gayoso\\
\url{https://orcid.org/0000-0001-6933-9166} - PC Barry\\
\url{https://orcid.org/0000-0003-4916-6194} - VV Berdnikov\\
\url{https://orcid.org/0000-0003-1742-4689} - D Binosi\\
\url{https://orcid.org/0000-0002-4339-2943} - L Chang\\
\url{https://orcid.org/0000-0002-4717-4484} - M Diefenthaler\\
\url{https://orcid.org/0000-0002-3690-1690} - M Ding\\
\url{https://orcid.org/0000-0001-7015-2534} - R Ent\\
\url{https://orcid.org/0000-0002-5497-5490} - T Frederico\\
\url{https://orcid.org/0000-0001-9032-1999} - Y Furletova\\
\url{https://orcid.org/0000-0002-2729-0015} - TJ Hobbs\\
\url{https://orcid.org/0000-0003-1925-9541} - T Horn\\
\url{https://orcid.org/0000-0002-5658-1065} - GM Huber\\
\url{https://orcid.org/0000-0002-8855-3034} - SJD Kay\\
\url{https://orcid.org/0000-0002-7516-8292} - C Keppel\\
\url{https://orcid.org/0000-0002-0899-3866} - HW Lin\\
\url{https://orcid.org/0000-0001-8678-4085} - C Mezrag\\
\url{https://orcid.org/0000-0002-2007-6833} - R Montgomery\\
\url{https://orcid.org/0000-0002-1195-3013} - IL Pegg\\
\url{https://orcid.org/0000-0001-8225-5821} - K Raya\\
\url{https://orcid.org/0000-0002-0301-2176} - P Reimer\\
\url{https://orcid.org/0000-0002-2937-1361} - CD Roberts\\
\url{https://orcid.org/0000-0002-1651-5717} - J Rodr\'{\i}guez-Quintero\\
\url{https://orcid.org/0000-0001-6715-3448} - D Romanov\\
\url{https://orcid.org/0000-0002-9209-3464} - G Salm\`e\\
\url{https://orcid.org/0000-0002-1535-6208} - N Sato\\
\url{https://orcid.org/0000-0001-5838-7103} - J Segovia\\
\url{https://orcid.org/0000-0002-0806-1743} - P Stepanov\\
\url{https://orcid.org/0000-0002-5312-8943} - AS Tadepalli\\
\url{https://orcid.org/0000-0002-8193-6139} - RL Trotta\\

\end{footnotesize}
\end{indented}

\begin{abstract}
%Rolf, Paul, John, CDR

How the bulk of the Universe's visible mass emerges and how it is manifest in the existence and properties of hadrons are profound questions that probe into the heart of strongly interacting matter. Paradoxically, the lightest pseudoscalar mesons appear to be the key to the further understanding of the emergent mass and structure mechanisms.  These mesons, namely the pion and kaon, are the Nambu-Goldstone boson modes of QCD.  Unravelling their partonic structure and the interplay between emergent and Higgs-boson mass mechanisms is a common goal of three interdependent approaches -- continuum QCD phenomenology, lattice-regularised QCD, and the global analysis of parton distributions -- linked to experimental measurements of hadron structure. Experimentally, the foreseen electron-ion collider will enable a revolution in our ability to study pion and kaon structure, accessed by scattering from the ``meson cloud'' of the proton through the Sullivan process. With the goal of enabling a suite of measurements that can address these questions, we examine key reactions to identify the critical detector system requirements needed to map tagged pion and kaon cross sections over a wide range of kinematics. The excellent prospects for extracting pion structure function and form factor data are shown, and similar prospects for kaon structure are discussed in the context of a worldwide programme. Successful completion of the programme outlined herein will deliver deep, far-reaching insights into the emergence of pions and kaons, their properties, and their role as QCD's Goldstone boson modes.
\end{abstract}

% SJDK - Tweak/add keywords as needed

% CA - From http://li.mit.edu/Archive/Activities/PubFormat/MSMSE/IOPLaTeXGuidelines.pdf
% choose one style

%\noindent{\it Keywords:\/} Meson structure, EIC, Hadrons


%\keywords{Meson structure, EIC, Hadrons}
\submitto{\jpg}

%\setcounter{tocdepth}{2}
%\tableofcontents

\maketitle

%%%-%%%
\section{Introduction}

%%%---%%%
\subsection{Mass budgets}
The Standard Model has two mechanisms for mass generation.  One is connected with the Higgs boson~\cite{Higgs:1964pj}, discovered at the large hadron collider in 2012~\cite{Aad:2012tfa, Chatrchyan:2012xdj}.  In the context of strong interactions, the Higgs produces the Lagrangian current-mass for each of the quarks. Yet, regarding the kernels of all known nuclei, these current masses account for less than 2\% of the mass of a neutron or proton.  More than 98\% of visible mass emerges as a consequence of strong interactions within QCD~\cite{Roberts:2020udq, Roberts:2020hiw, Roberts:2021xnz}: this is emergent hadronic mass (EHM).

Consider, therefore, the Lagrangian of QCD in the absence of Higgs couplings to the quarks.  Classically, it defines a scale invariant theory; and scale invariant theories do not support dynamics. Therefore, bound states are impossible and the Universe cannot exist.

The process of renormalisation in the quantisation of chromodynamics introduces a mass which breaks the scale invariance of the classical theory.  Hence, in the absence of quark couplings to the Higgs boson, i.e. in the chiral limit, the QCD stress-energy tensor, $T_{\mu\nu}$, exhibits a trace anomaly~\cite{tarrach}:
\begin{equation}
\label{SIQCD}
T_{\mu\mu} = \beta(\alpha(\zeta))  \tfrac{1}{4} G^{a}_{\mu\nu}G^{a}_{\mu\nu} =: \Theta_0 \,,
\end{equation}
where $\beta(\alpha(\zeta))$ is QCD's $\beta$-function, $\alpha(\zeta)$ is the associated running-coupling, $G^{a}_{\mu\nu}$ is the gluon field strength tensor, and $\zeta$ is the renormalisation scale.  The consequences of equation~\eqref{SIQCD} are wide-ranging and heavy in impact.

\begin{figure}[!t]
\begin{center}
\hspace*{-1ex}\begin{tabular}{l}
{\sf A} \\[-0.9ex]
\includegraphics[clip, width=0.49\textwidth]{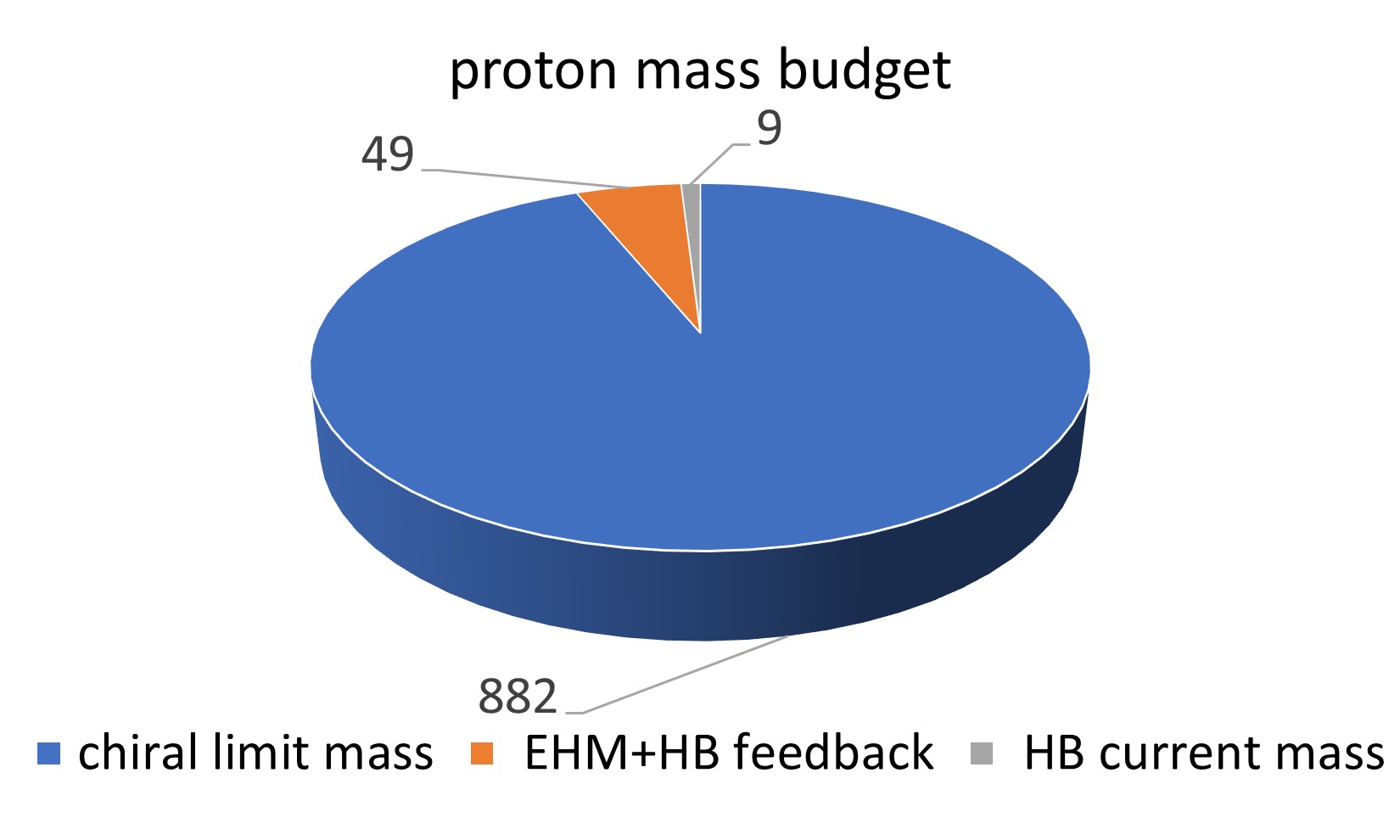}
\end{tabular}\\
\hspace*{-1ex}\begin{tabular}{ll}
{\sf B} & {\sf C} \\[-0.9ex]
\includegraphics[clip, width=0.49\textwidth]{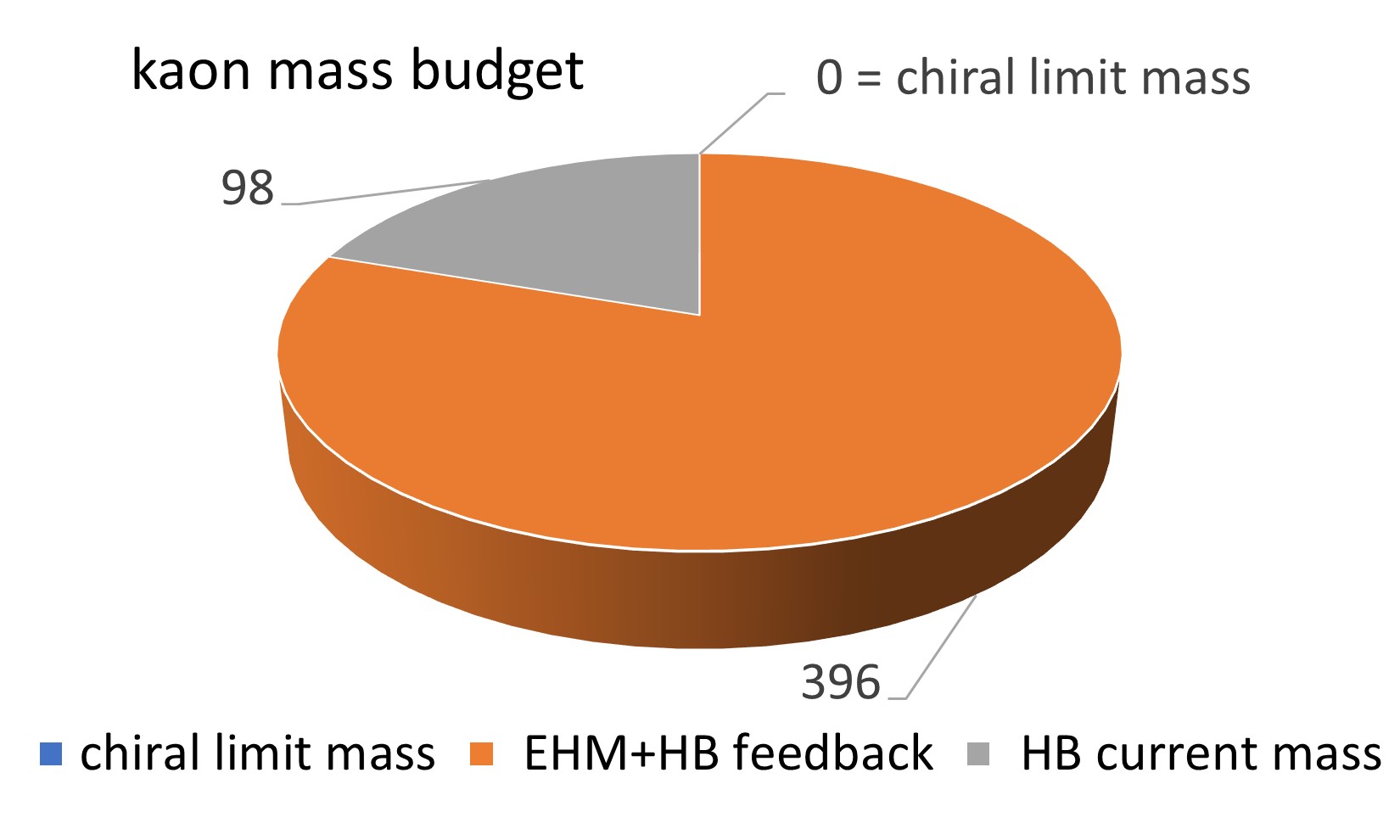} &
\includegraphics[clip, width=0.49\textwidth]{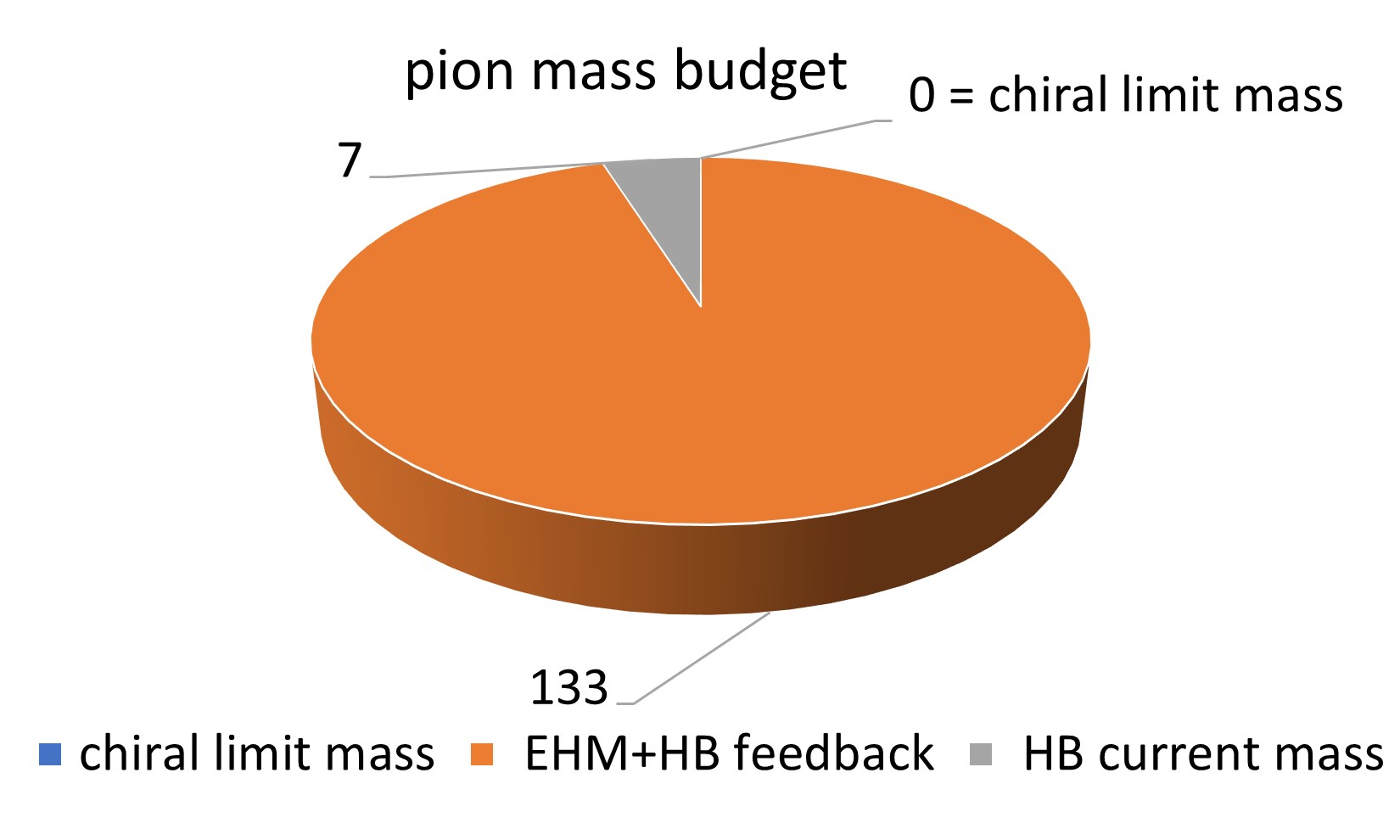}
\end{tabular}
\end{center}
%
%%\hspace*{-1ex}\begin{tabular}{lll}
%%{\sf A} & {\sf B} & {\sf C} \\[-1.1ex]
%%\includegraphics[clip, width=0.34\textwidth]{F1ACDR.jpg} &
%%\includegraphics[clip, width=0.32\textwidth]{F1BCDR.jpg} &
%%\includegraphics[clip, width=0.33\textwidth]{F1CCDR.jpg}
%%\end{tabular}
%
\caption{\label{F1CDR}
Mass budgets for the proton -- {\sf A}, kaon -- {\sf B} and pion -- {\sf C}.  The differences are stark.  Owing to EHM, the proton's mass is large in the chiral limit.  Conversely and yet still owing to EHM via its corollary dynamical chiral symmetry breaking, the kaon and pion are massless in the absence of quark couplings to the Higgs boson.  (Units MeV, Poincar\'e-invariant separation at $\zeta=2\,$GeV, breakdowns produced using information from~\cite{RuizdeElvira:2017stg, Zyla:2020}.)
}
\end{figure}

A first question to ask is whether the magnitude of $\Theta_0$ can be measured and understood. Measurement is straightforward. Consider the following in-proton expectation value:
\begin{equation}
\label{EPTproton}
\langle p(P) | T_{\mu\nu} | p(P) \rangle = - P_\mu P_\nu\,,
\end{equation}
where the equations-of-motion for a one-particle proton state produce  the right-hand-side.  In the chiral limit
\begin{equation}
\label{anomalyproton}
\langle p(P) | T_{\mu\mu} | p(P) \rangle  = - P^2  = m_p^2 = \langle p(P) |  \Theta_0 | p(P) \rangle\,.
\end{equation}
As highlighted by the blue domain in figure~\ref{F1CDR}A, this expectation value is 94\% of the proton's measured mass.  Since $\Theta_0$ is expressed solely in terms of gluons when a (large-$\zeta$) parton basis is used, then one might conclude that the chiral-limit value of $m_p$ is generated completely by glue.

However complex that might seem, Nature is even more subtle.  This may be seen by returning to equation~\eqref{EPTproton} and replacing the proton by the pion
\begin{equation}
\label{EPTpion}
\langle \pi(q) | T_{\mu\nu} | \pi(q) \rangle = - q_\mu q_\nu \quad  \Rightarrow \quad
 \langle \pi(q) |  \Theta_0 | \pi (q) \rangle = m_\pi^2 \;\; \stackrel{\mbox{\rm chiral\;limit}}{=} \;\; 0
\end{equation}
because the chiral-limit pion is a massless Nambu--Goldstone (NG) mode~\cite{Nambu:1960tm, Goldstone:1961eq}.  This feature is highlighted by the complete absence of a blue domain for the pion in figure~\ref{F1CDR}C.
Conceivably, this could mean that the scale anomaly is trivially zero in the pion; to wit, strong gluon-gluon interactions have no effect in the pion because each term required to express $\Theta_0$ vanishes separately~\cite{Ji:1995sv}.  However, such an explanation would sit uncomfortably with known QCD dynamics, which expresses both attraction and repulsion, often remarkable fine tuning, but never inactivity.  (Additional discussion of these points may be found, e.g.\ in \cite[section\,V]{Roberts:2020udq}, \cite[sections\,4.4,\,4.5]{Roberts:2020hiw}).

Switching on the Higgs boson couplings to light quarks, then one encounters the other two wedges in figure~\ref{F1CDR}A: grey shows the sum of Higgs-generated valence-quark current-masses in the proton, which amounts to just $0.01 \times m_p$; and orange indicates the contribution generated by constructive interference between EHM and Higgs-boson (HB) effects, 5\%.  Again, the picture for the pion is completely different, with EHM+HB interference being responsible for 95\% of the pion's mass.  The kaon lies somewhere between these extremes.  It is a would-be Nambu-Goldstone mode, so there is no blue-domain in figure~\ref{F1CDR}\,B; but the sum of valence-quark and valence-antiquark current-masses in the kaon amounts to 20\% of its measured mass -- four times more than in the pion, with EHM+HB interference producing 80\%.

Equations~\eqref{anomalyproton},~\eqref{EPTpion}, and the mass budgets drawn in figure~\ref{F1CDR} demand interpretation.  They stress that any explanation of the proton's mass is incomplete unless it simultaneously clarifies equation~\eqref{EPTpion}.  Moreover, both phenomena are coupled with confinement, which is fundamental to the proton's stability.  These remarks highlight the ubiquitous influence of EHM.  They emphasise that in order to finally complete the Standard Model, it is crucial to understand the emergence of mass within the strong interaction and the modulating effects of Higgs boson mass generation, both of which are fundamental to understanding the evolution of our Universe.

In facing these questions, unique insights can be drawn by focusing on the properties of QCD's (pseudo-)\,Nambu-Goldstone modes, i.e.\ pions and kaons; and diverse phenomenological and theoretical approaches are now being deployed in order to develop a coherent image of these bound states.  Complete understanding demands that tight links be drawn between dynamics in QCD's gauge sector and pion and kaon light-front wave functions, and from there to observables, such as pion and kaon elastic form factors and distribution amplitudes and functions.
Herein, we propose an array of measurements and associated analyses designed to deliver significant progress toward these goals~\cite{Aguilar:2019teb, Brodsky:2020vco}.

It is worth remarking here that \emph{measurements} of form factors, distribution amplitudes and functions, spectra, charge radii, etc., are all on the same footing.  Theory supplies predictions for such quantities.  Experiments measure precise cross sections; and cross-sections are expressed, via truncations that optimally have the quality of approximations, in terms of the desired quantity.  At question is the reliability of the truncation/approximation employed in relating the measured cross section to this quantity.  The phenomenology challenge is to ensure that every contribution known to have a material effect is included in building the bridge.  The quality of the phenomenology can alter neither that of the experiment nor the theory.  However,
inadequate phenomenology can deliver results that mislead interpretation.   The reverse is also true.  Thus, progress requires the building of a positive synergy between all subbranches of the programme.

%%%-%%%

\subsection{EIC context}
%% -- Science goal relation to NAS (0.5 page)
The electron-ion collider (EIC) \cite{PhysicsToday73} will be capable of addressing an array of profound questions that probe into the heart and reach out to the frontiers of strong interactions within the Standard Model.  Looming large in this array are the emergence of the bulk of visible mass and its manifestations in the existence and properties of hadrons and nuclei.  The research described herein aims to build a path toward answers by focusing on the properties of pions and kaons, the Standard Model's would-be Nambu-Goldstone modes.  It combines experiment, phenomenology and theory in a synergistic effort to reveal: how the roughly 1\,GeV mass-scale that characterises atomic nuclei appears; why it has the observed value; why ground-state pseudoscalar mesons are unnaturally light in comparison; and the role of the Higgs boson in forming hadron properties.

The focus on pions and kaons acknowledges that these states are unique expressions of Standard Model dynamics, exhibiting a peculiar dichotomy.  Namely, they are hadron bound states defined, like all others, by their valence quark and/or antiquark content, making calculation of their properties no different, in principle, from proton computations; but the mechanism(s) which give all other hadrons their roughly $1\,$GeV mass-scale are obscured in these systems.  This elevates studies of pion and kaon structure to the highest levels of importance.  Yet, although discovered more than seventy years ago \cite{Lattes:1947mw, Rochester:1947mi}, remarkably little is known about their structure.  The EIC, with its high-luminosity and wide kinematic range, offers an extraordinary new opportunity to eliminate that ignorance.  There is much to be learnt: pions and kaons are not pointlike; their internal structure is more complex than is usually imagined; and the properties of these nearly-massless strong-interaction composites provide the clearest windows onto EHM and its modulation by Higgs-boson interactions.

This report identifies a raft of measurements and associated phenomenology and theory that will exploit the distinctive strengths of the EIC in driving toward answers to some of the most basic questions in Nature.  Successful completion of the programme will deliver deep, far-reaching insights into
the distributions and apportionment of mass and spin within the pion and kaon;
the similarities and differences between such distributions in these (almost) Nambu-Goldstone modes and the benchmark proton;
the symbiotic relationship between EHM and confinement;
and the character and consequences of constructive interference between the Standard Model's two mass-generating mechanisms.

\section{Meson structure as a QCD laboratory - status and prospects}
% SJDK - Need a suitable subheading title here
% Names on section for reference
% (Tobias, {\underline{Jorge}})
%\subsection{Suitable Subheading Title Here}

\subsection{Pion and kaon structure - theory status}
Emergent hadronic mass (EHM) is an elemental feature of the Standard Model.  As reviewed elsewhere \cite{Roberts:2020udq, Roberts:2020hiw, Roberts:2021xnz}: it is the origin of a running gluon mass; the source of dynamical chiral symmetry breaking (DCSB); and very probably crucial to any explanation of confinement. DCSB is basic to understanding the notion of constituent quarks and the successes of related models; and it provides the foundation for the existence of nearly-massless pseudo-Goldstone modes.
%% (pions and kaons in the approximated $SU(3)_{\text{flavour}}$ symmetry).
%% CDR approximate SU(3)-flavour symmetry is not necessary for these states to be pseudo NG modes.
Confinement is related to the empirical fact that all attempts to remove a single quark or gluon from within a hadron and isolate it in a detector have failed. The mechanisms responsible for EHM must be expressed, besides in hadron masses, in their wave functions with the associated Fock-space representation in terms of quarks and gluons; especially, in the corresponding light-hadron structure observables. In the following, examples of such measurable quantities will be presented, focused largely on pion and kaon observables accessible at EIC. The kaon is very interesting because therein a competition between emergent and Higgs-driven mass generation is taking place.  All differences between the pion and kaon are driven by Higgs-induced modulation of EHM.

%%%%%%%%%%%%%%%%%%%%%%%%%%%%%%%%%%%%%%%%%%%%%%%%%%%%%%%%%%%%%%%%%%%%%%%%%%%%%%%%%%%%%%%%%%%%%%%%%%%%%%

\subsubsection{Pion and kaon distribution amplitudes.}

The cross sections for many hard exclusive hadronic reactions can be expressed in terms of the parton distribution amplitudes (DAs) of the hadrons involved. For instance, in the case of the electromagnetic form factor of light pseudoscalar mesons~\cite{Lepage:1979zb, Efremov:1979qk, Farrar:1979aw, Lepage:1980fj}:
\begin{equation}
\label{eq:pionFFUV}
\exists \, Q_0 > \Lambda_{\rm QCD} \; |\;   Q^2 F_P(Q^2) \stackrel{Q^2 > Q_0^2}{\approx} 16 \pi \alpha_s(Q^2)  f_P^2 w_\varphi^2 \,, \text{ with } w_\varphi = \frac{1}{3} \int_0^1 dx\, \frac{1}{x}\, \varphi_P(x) \,,
\end{equation}
where $\alpha_s(Q^2)$ is the strong running coupling, $f_P$ is the pseudoscalar meson's leptonic decay constant and $\varphi_P(x)$ is the DA of the pseudoscalar meson. However, the value of $Q_0$ is not predicted by QCD and the DAs are not determined by the analysis framework; perturbative QCD (pQCD) only states that $\varphi_P(x) \approx \varphi_{\text{as}}(x) = 6x(1 - x)$ for $Q^2\gg Q_0^2$.

\begin{figure}[!t]
\includegraphics[width=0.47\textwidth]{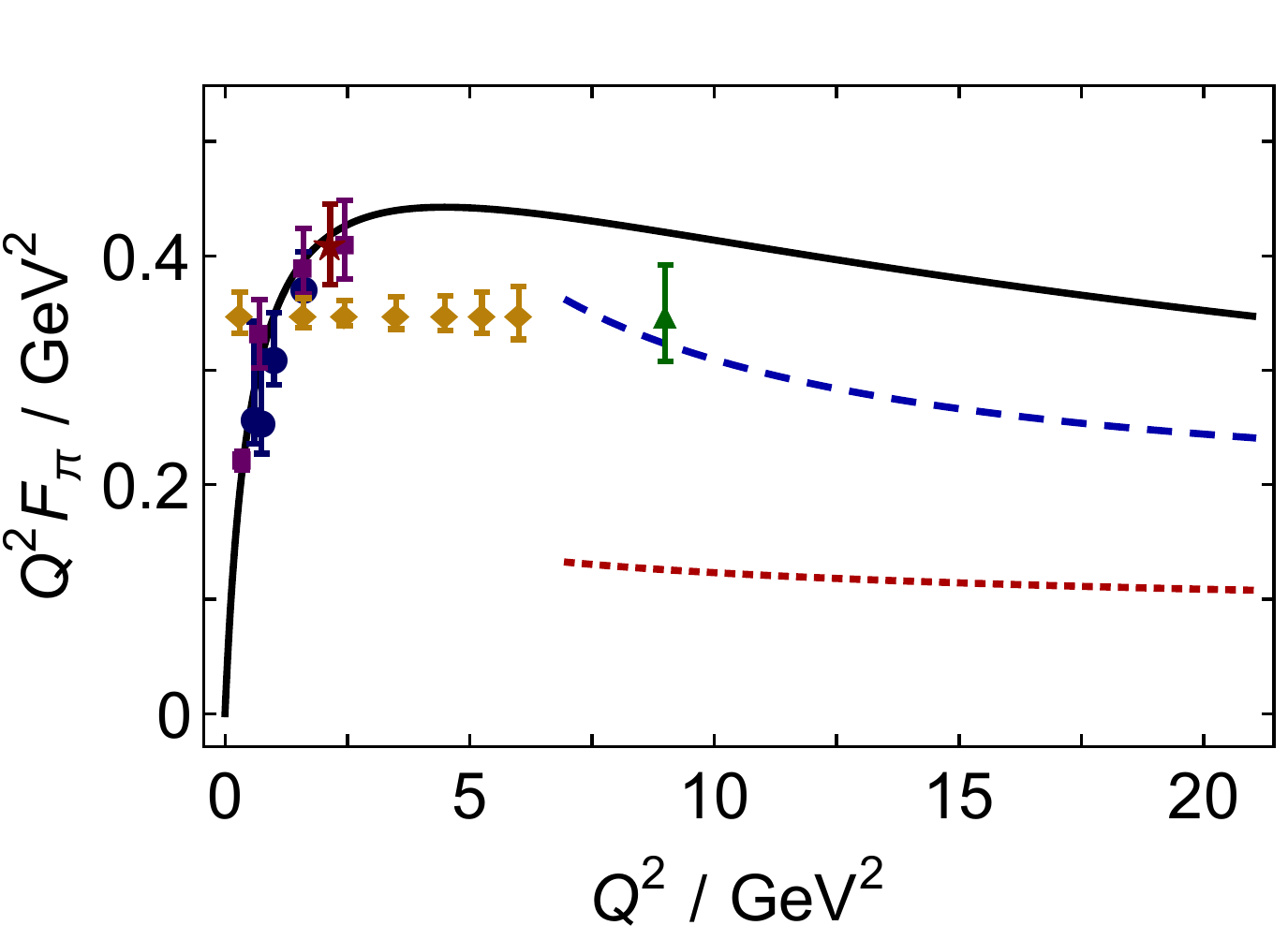}
\hspace*{0.20cm}
\includegraphics[width=0.47\textwidth]{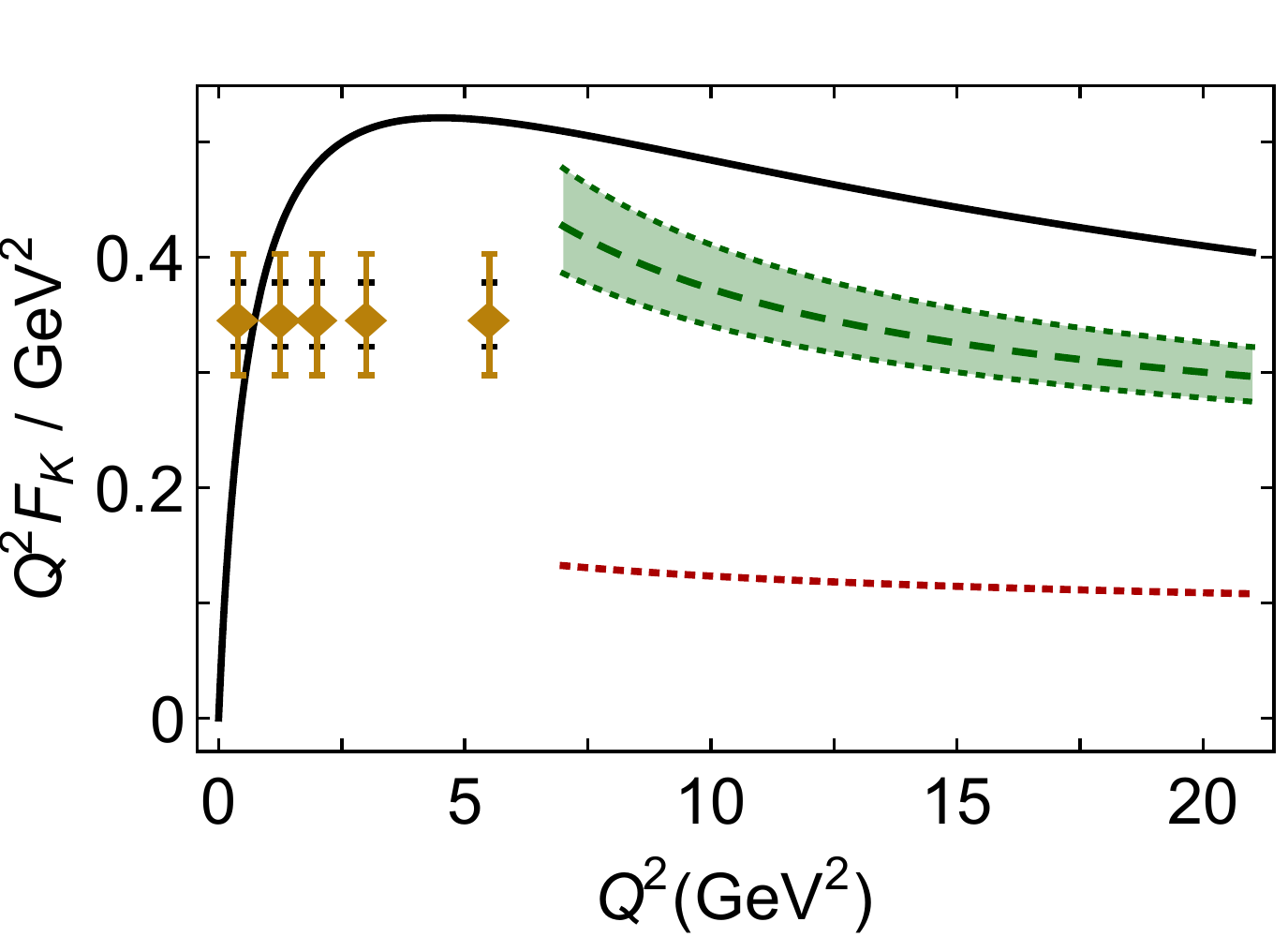}
\caption{\label{fig:F1JorSeg}
The left (right) panel show calculations, measurements, and projected precision of future measurements for $Q^2 F_\pi(Q^2)$  ($Q^2 F_K(Q^2)$).
Solid curve -- prediction from~\cite{Chang:2013nia, Chen:2018rwz};
dotted curve -- result produced by the hard scattering formula, equation~\eqref{eq:pionFFUV}, using the asymptotic DA;
dashed curve -- result produced by the hard scattering formula using the DA calculated in the CSM framework at a scale relevant to the experiment.
%\emph{Left panel}.
%Solid curve -- prediction for $Q^2 F_\pi(Q^2)$~\cite{Chang:2013nia, Chen:2018rwz};
%dotted curve -- result produced by the hard scattering formula, equation~\eqref{eq:pionFFUV}, using the asymptotic PDA;
%dashed curve -- result produced by the hard scattering formula using the PDA calculated in the CSM framework at a scale relevant to the experiment.
%
%\emph{Right panel}. Same as left panel but for the kaon form factor.
%-- prediction for $Q^2 F_K(Q^2)$~\cite{Gao:2017mmp};
%dotted curve (red) -- result produced by the hard scattering formula, %equation~\eqref{eq:pionFFUV}, using the asymptotic PDA;
%and dashed curve and band (green) -- result produced by the hard scattering formula, equation~\eqref{eq:pionFFUV}, but using the PDA calculated in the CSM framework, again at a scale relevant to the experiment.
%
Stars~\cite{Horn:2007ug}, circles and squares~\cite{Huber:2008id} show existing data; diamonds and triangle show the anticipated reach and accuracy of forthcoming experiments~\cite{E12-06-101, E12-07-105, E12-09-011}.
%
%In both panels, the dotted curve (red) is equation~\eqref{eq:pionFFUV} computed with the asymptotic PDA, $\varphi_{\text{as}}(x)$.
}
\end{figure}

One may alternatively use continuum Schwinger function methods (CSMs) for QCD to describe exclusive reactions in terms of Poincar\'e-covariant hadron bound-state amplitudes (BSAs). This approach has been formulated in both Euclidean~\cite{Roberts:2015lja, Horn:2016rip, Eichmann:2016yit, Burkert:2017djo, Qin:2020rad} and Minkowski~\cite{Sales:1999ec, Frederico:2009fk, Carbonell:2010zw, Salme:2014sha, Salme:2014cfa, dePaula:2016oct} space. Moreover, recent progress within CSMs has established that the hadron DA, which is essentially nonperturbative, can be obtained as a light-front projection of the hadron's BSA~\cite{Chang:2013pq}, an approach first employed in \cite{tHooft:1974pnl}. Using this connection, the solid curves of figure~\ref{fig:F1JorSeg} are the CSM's predictions for the pion~\cite{Chang:2013nia,Chen:2018rwz} and kaon~\cite{Gao:2017mmp} elastic electromagnetic form factors to arbitrarily large-$Q^2$. Also depicted (dashed curves) are the results obtained using equation~\eqref{eq:pionFFUV} and the DAs calculated in the CSM framework at a scale relevant to the experiment. These DAs are very different from $\varphi_{\text{as}}(x)$, being markedly broader owing to EHM. The dotted curve in both panels is the hard scattering formula, equation~\eqref{eq:pionFFUV}, computed with the asymptotic profile, $\varphi_{\text{as}}(x)$.

EIC is capable of providing precise pion form factor data that will probe deep into the region where $F_{\pi}(Q^2)$ exhibits strong sensitivity to EHM and the evolution of this effect with scale. In particular, as more results from Euclidean and Minkowski space approaches within the CSM framework become available, an estimate of a lower bound for $Q_0$ can be found by comparing the full covariant form factor with the valence one, which should dominate at large momenta, with the difference quantifying the contribution from the higher Fock-components of the light-front wave function to the pion charge distribution. The extraction of the pion form factor via a Sullivan process involves the extrapolation to an off-mass-shell pion that can be quantified and validated~\cite{Qin:2017lcd, Choi:2019nvk}. Moreover, the EIC will be the first facility to measure the size and range of nonperturbative EHM--Higgs-boson interference effects in hard exclusive processes if high-$Q^2$ kaon form factor measurements can be feasible at EIC.

%%%%%%%%%%%%%%%%%%%%%%%%%%%%%%%%%%%%%%%%%%%%%%%%%%%%%%%%%%%%%%%%%%%%%%%%%%%%%%%%%%%%%%%%%%%%%%%%%%%%%%%

\subsubsection{Pion and kaon distribution functions.}

The pion valence quark distribution function (DF), $q^{\pi}(x,\zeta)$, expresses the probability density that a valence $q$-quark in the pion carries a light-front fraction $x$ of the system's total momentum at a resolving scale $\zeta$~\cite{Ellis:1991qj}.
In this connection, capitalising on the known behaviour of hadron wave functions at large valence-quark relative momenta~\cite{Lane:1974he, Politzer:1976tv, Lepage:1980fj, Maris:1997hd}, numerous analyses within a diverse array of frameworks predict the following large-$x$ behaviour (see e.g.~\cite{Ezawa:1974wm, Farrar:1975yb, Berger:1979du, Brodsky:1994kg, Yuan:2003fs}):
\begin{equation}
\label{pionPDFlargex}
q_{\pi}(x;\zeta=\zeta_H) \stackrel{x\simeq 1}{\sim} (1-x)^{\beta} \,, \text{ with } \beta=2 \,,
\end{equation}
where $\zeta_H$ is the hadronic scale, which is not accessible in experiment because certain kinematic conditions must be met in order for the data to be interpreted in terms of $q_{\pi}(x,\zeta)$~\cite{Ellis:1991qj}. Hence, any result for a distribution function at $\zeta_H$ must be evolved to $\zeta_E\,(>\zeta_H)$ for comparison with experiment~\cite{Dokshitzer:1977sg, Gribov:1971zn, Lipatov:1974qm, Altarelli:1977zs}. Under such evolution, the exponent grows, viz.\ $\beta=2+\delta$, where $\delta$ is an anomalous dimension that increases logarithmically with $\zeta$. Significantly, within DF fitting uncertainties, the analogous behaviour for the proton's valence-quark distribution function has been confirmed~\cite{Ball:2016spl}.
%% Eq. (6) is consistent with the updated Drell-Yan-West relation.  See, e.g. [Landshoff:1973pw, Brodsky:1994kg]

It is worth noting here that what has come to be known as the Drell-Yan-West relation provides a link between the large-$x$ behaviour of DFs and the large-$Q^2$ dependence of hadron elastic form factors~\cite{Drell:1969km, West:1970av}.  In its original form, the relation was discussed for the $J=1/2$ proton.  It has long been known that this original form is invalid when, e.g.\ the target is a $(J=0)$ pseudoscalar meson and the valence-parton scatterers are $J=1/2$ objects~\cite{Ezawa:1973qx, Landshoff:1973pw}.
The generalisation to spin-$J$ targets constituted from $J=1/2$ quarks may be found in \cite{Brodsky:1994kg}: for a hadron $H$ defined by $n+1$ valence $J=1/2$ partons, so that its leading elastic electromagnetic form factor scales as $(1/Q^2)^n$:
\begin{equation}
q_{H}(x;\zeta_H) \stackrel{x\simeq 1}{\sim} (1-x)^{p} \,, p= 2 n - 1 + 2 \Delta S_z\,,
\end{equation}
where $\Delta S_z = |S_z^q - S_z^H|$.  For a pseudoscalar meson, $n=2$, $S_z^{H}=0$, so $p=2$.  One thereby recovers Eq.\,\eqref{pionPDFlargex}.

Experiments interpretable in terms of $q_{\pi}(x,\zeta)$ were completed more than thirty years ago~\cite{Badier:1983mj, Conway:1989fs}.
Notably, phenomenological analyses of that data which ignore soft-gluon (threshold) resummation effects return a DF that roughly resembles a profile with $\beta \approx 1$~\cite{Barry:2018ort, Novikov:2020snp, Han:2020vjp}, in conflict with equation~\eqref{pionPDFlargex}.
% SJDK - Probably need to tweak how the citation is incorporated here -style guide recommends against "reference [X] or ref [X] so maybe something like "work by X~\cite[}" or something
On the other hand, \cite{Aicher:2010cb}, which included such next-to-leading-logarithm resummation using the ``cosine method'', yields $\beta > 2$, in accord with equation~\eqref{pionPDFlargex}.
The dependence of the inferred large-$x$ behaviour of DFs on the resummation prescription is being explored
\cite{Barry-resummation-20}: preliminary findings suggest that, depending on the method adopted (``double Mellin'', ``expansion'' or ``cosine''), the apparent $\beta$ exponent can range between $\sim 1$ and $\sim 2.5$ at the input scale.
Importantly, however, all methods yield softened large-$x$ behaviour, with both the \emph{expansion} and \emph{cosine} approaches producing $\beta > 2$.
Additional remarks on these issues are presented in section\,\ref{sec:QCDfit}.

\begin{figure}[!t]
\includegraphics[width=0.47\textwidth]{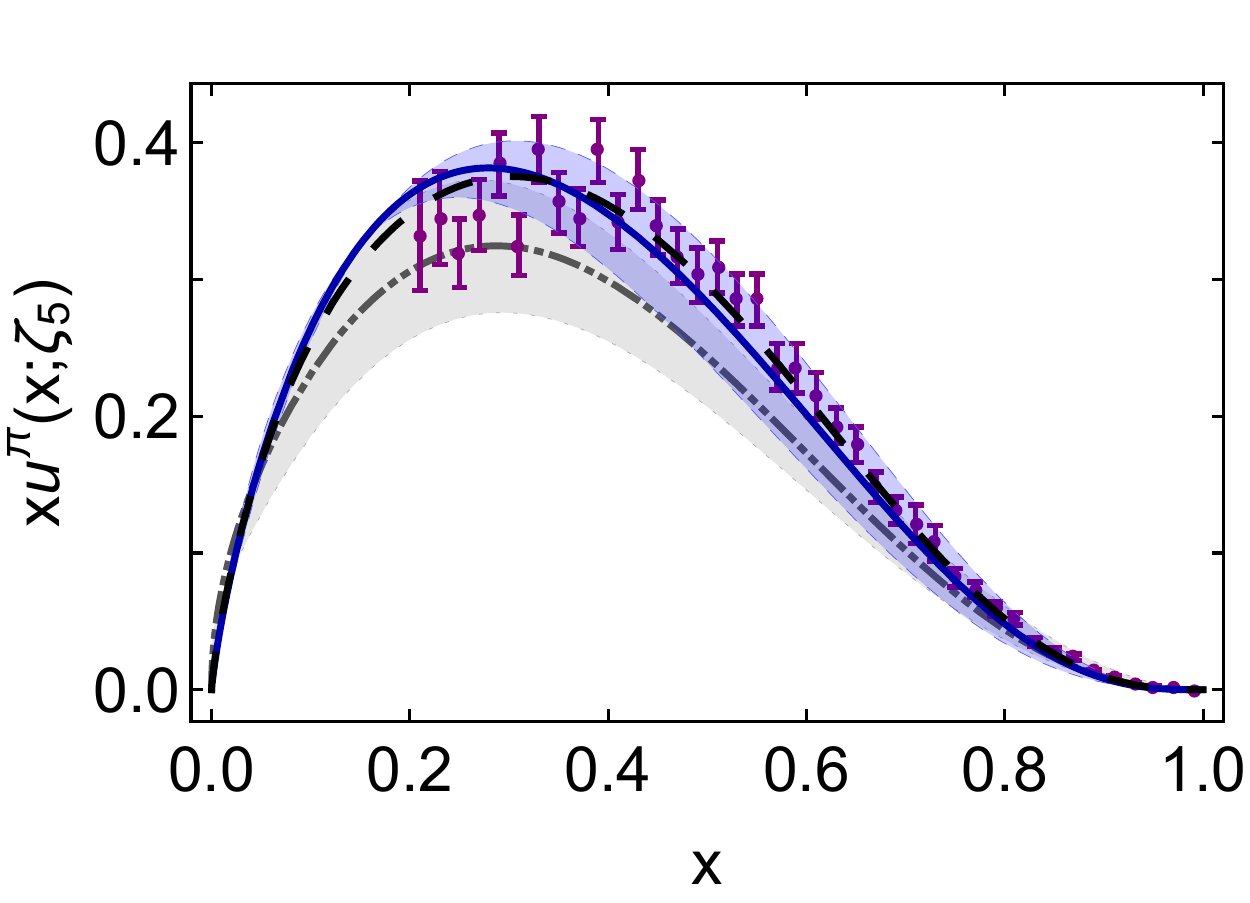}
\hspace*{0.20cm}
\includegraphics[width=0.47\textwidth]{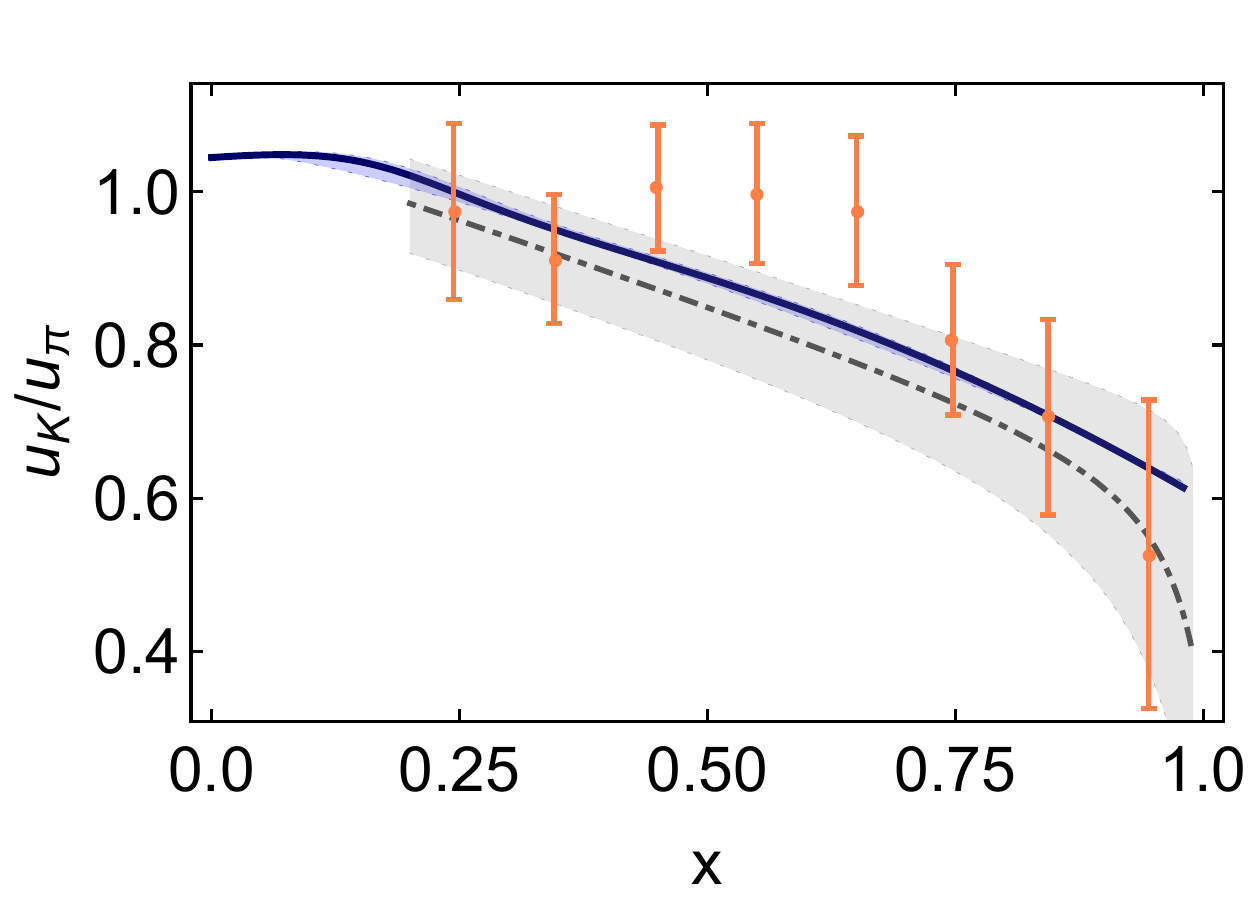}
\caption{\label{fig:F2JorSeg}
\emph{Left panel}. Pion valence-quark momentum distribution function, $xq_\pi(x;\zeta_5=5.2\,\text{GeV})$: solid blue curve -- modern continuum calculation~\cite{Cui:2020dlm, Cui:2020tdf}; long-dashed black curve -- early continuum analysis~\cite{Hecht:2000xa}; and dot-dot-dashed grey curve -- lattice QCD result~\cite{Sufian:2019bol}. Data (purple) from~\cite{Conway:1989fs}, rescaled according to the analysis in~\cite{Aicher:2010cb}.
\emph{Right panel}. $u_K(x;\zeta_5)/u_\pi(x;\zeta_5)$. Solid blue curve -- prediction from~\cite{Cui:2020dlm, Cui:2020tdf}. Dot-dashed grey curve within grey band -- lattice QCD result~\cite{Lin:2020ssv}. Data (orange) from~\cite{Badier:1980jq}.
}
\end{figure}

The CSM prediction~\cite{Cui:2020dlm, Cui:2020tdf} for $u_\pi(x;\zeta_5)$ is depicted in the left panel of figure~\ref{fig:F2JorSeg}.  Its large-$x$ behaviour agrees with equation~\eqref{pionPDFlargex} and the pointwise form matches that determined in \cite{Aicher:2010cb}. Regarding glue and sea DFs, \cite{Cui:2020dlm, Cui:2020tdf} provide parameter-free predictions for all pion DFs.

Notably, as described in section~\ref{lQCDpiK}, lattice-regularised QCD is now beginning to yield results for the pointwise behaviour of the pion's valence-quark distribution~\cite{Sufian:2019bol, Chen:2018fwa}, with that delivered by the approach in \cite{Sufian:2019bol} being in fair agreement with both equation~\eqref{pionPDFlargex} and the CSM prediction.  This is highlighted by the comparison between the  blue CSM result and the dot-dot-dashed (grey) curve in the left panel of figure~\ref{fig:F2JorSeg}.

Parameter-free predictions for all kaon DFs are also provided in \cite{Cui:2020dlm, Cui:2020tdf}.  Concerning valence-quarks, there are qualitative similarities between $u_K(x)$, $\bar s_K(x)$ and $u_\pi(x)$, e.g.\ all three DFs are consistent with equation~\eqref{pionPDFlargex} so that $\bar s_K(x)$ is much softer than the lattice-QCD result drawn in figure~\ref{fig:xqvpi}\,--\,left.  There are also quantitative differences between the valence distributions, as highlighted by the prediction for $u_K(x)/u_\pi(x)$ drawn in figure~\ref{fig:F2JorSeg}\,--\,right and compared with the result determined from a measurement of the $K^-/\pi^-$ structure function ratio~\cite{Badier:1980jq}. 

The first lattice-QCD results for $u_K(x)/u_\pi(x)$ are also drawn in figure~\ref{fig:F2JorSeg}\,--\,right.  The relative difference between the central lattice QCD result and the continuum prediction~\cite{Cui:2020dlm, Cui:2020tdf} is $\approx 5$\%, despite the fact that the individual DFs from these two sources are qualitatively and quantitatively different. This feature highlights a long known characteristic, i.e.\ $u_K(x)/u_\pi(x)$ is quite forgiving of even large differences between the individual DFs used to produce the ratio. Evidently, more precise data is crucial if this ratio is to be used effectively to inform and test the modern understanding of pion and kaon structure; and results for $u_\pi(x;\zeta_5)$, $u_K(x;\zeta_5)$ separately have greater discriminating power. These remarks are amplified by the fact that the lone $K^-/\pi^-$ structure function experiment was performed forty years ago.  Hence, new precision data and extractions must be a high priority.

Significantly, \cite{Cui:2020dlm, Cui:2020tdf} also provides the first parameter-free predictions for the ratios of glue and sea DFs in the pion and kaon.  The kaon's glue and sea distributions are similar to those in the pion; but the inclusion of mass-dependent splitting functions, expressing Higgs-induced current-quark mass splittings, introduces differences on the valence-quark domain.  Today, no empirical information is available that would enable these predictions to be tested.  Hence, experiments sensitive to glue and sea distributions in the kaon and pion would be of enormous value.

Euclidean-space based CSMs obtain the pion's DFs considering it as a bound-state of a dressed-quark and dressed-antiquark at the hadronic scale, with the sea and glue distributions being zero at $\zeta_H$ and generated by evolution on $\zeta > \zeta_H$~\cite{Ding:2019qlr, Ding:2019lwe}.  This is also the case for the kaon. In contrast, a Minkowski space analysis of the Bethe-Salpeter equation finds that the valence-quark probability in the pion state is about 70\%~\cite{dePaula:2020a} with the remaining normalisation distributed among higher Fock-space components, carrying gluons at the hadronic scale. A resolution of this puzzle will likely be found in the mapping between the different quasi-particle degrees-of-freedom that serve in each calculation.

Related analyses of pseudoscalar meson generalised transverse momentum dependent DFs (GTMDs) are also becoming available~\cite{Zhang:2020ecj}.  They indicate that GTMD size and shape are also prescribed by the scale of EHM.  Proceeding from GTMDs to generalised parton distributions (GPDs)~\cite{Brommel:2007xd, Nam:2010pt, Dorokhov:2011ew, Fanelli:2016aqc, Zhang:2020ecj, Raya:2021:letter}, it is found that the pion's mass distribution form factor is harder than its electromagnetic form factor, which is harder than the gravitational pressure distribution form factor; the pressure in the neighbourhood of the pion's core is similar to that at the centre of a neutron star; the shear pressure is maximal when confinement forces become dominant within the pion; and the spatial distribution of transversely polarised quarks within the pion is asymmetric.

Regarding transverse momentum dependent distribution functions (TMDs), these studies indicate that their magnitude and domain of support decrease with increasing twist~\cite{Zhang:2020ecj}. Consistent with intuition~\cite{Zhang:2020ecj}, at $\zeta_H$, the simplest Wigner distribution associated with the pion's twist-two dressed-quark GTMD is sharply peaked on the kinematic domain associated with valence-quark dominance; has a domain of negative support; and broadens as the transverse position variable increases in size.

More sophisticated studies are beginning to appear.  For instance, \cite{Raya:2021:letter} computes and compares pion and kaon GPDs built using the overlap representation from light-front wave functions constrained by the one-dimensional valence distributions described above.  It finds, \emph{inter alia}, that $K$ pressure profiles are spatially more compact than $\pi$ profiles and near-core pressures in both NG modes are of similar magnitude to that found in neutron stars.
Plainly, now is the right time to plan on exploiting the capacities of EIC to probe these higher-dimensional aspects of pion and kaon structure.

%%%%%%%%%%%%%%%%%%%%%%%%%%%%%%%%%%%%%%%%%%%%%%%%%%%%%%%%%%%%%%%%%%%%%%%%%%%%%%%%%%%%%%%%%%%%%%%%%%%%%%%
%%%%%%%%%%%%%%%%%%%%%%%%%%%%%%%%%%%%%%%%%%%%%%%%%%%%%%%%%%%%%%%%%%%%%%%%%%%%%%%%%%%%%%%%%%%%%%%%%%%%%%%

%\newpage
% SJDK - Need a suitable subheading title here
% Names on section for completeness
% (Huey-Wen, {\underline{David}})
%\subsection{Suitable Subheading Title Here}
\subsection{Pion and kaon Structure - lattice QCD status}
\label{lQCDpiK}

Quantising QCD on a finite-volume discrete lattice in Euclidean space-time enables the numerical calculation of correlation functions defined by the functional integral \cite{Wilson:1974sk}.  Accessing hadron structure information using lattice QCD has been a very challenging task, since  distribution functions are light-cone quantities and cannot be calculated directly on a Euclidean lattice.  Over the years, a range of methods have been proposed to overcome this obstacle, such as studies based on the hadronic tensor~\cite{Liu:1993cv, Liu:1998um, Liu:1999ak}, auxiliary quark field approaches~\cite{Detmold:2005gg, Braun:2007wv}, large-momentum effective theory (LaMET)~\cite{Ji:2013dva, Ji:2014gla, Ji:2020ect} (quasi-PDFs), pseudo-PDFs~\cite{Radyushkin:2016hsy}, an operator-product-expansion based method~\cite{Chambers:2017dov}, and the good lattice cross sections (LCS) approach~\cite{Ma:2014jla,Ma:2014jga,Ma:2017pxb}.  These methods have some common ground, but also differences.  Interested readers may consult~\cite{Lin:2017snn, Cichy:2018mum, Ji:2020ect} for more details.  This subsection describes a few examples of recent progress in meson structure studies and indicates calculations that will be important for the success of EIC science.

\subsubsection{Meson distribution amplitudes}
\begin{figure}[t]
	\centering
	\includegraphics[width=0.45\linewidth]{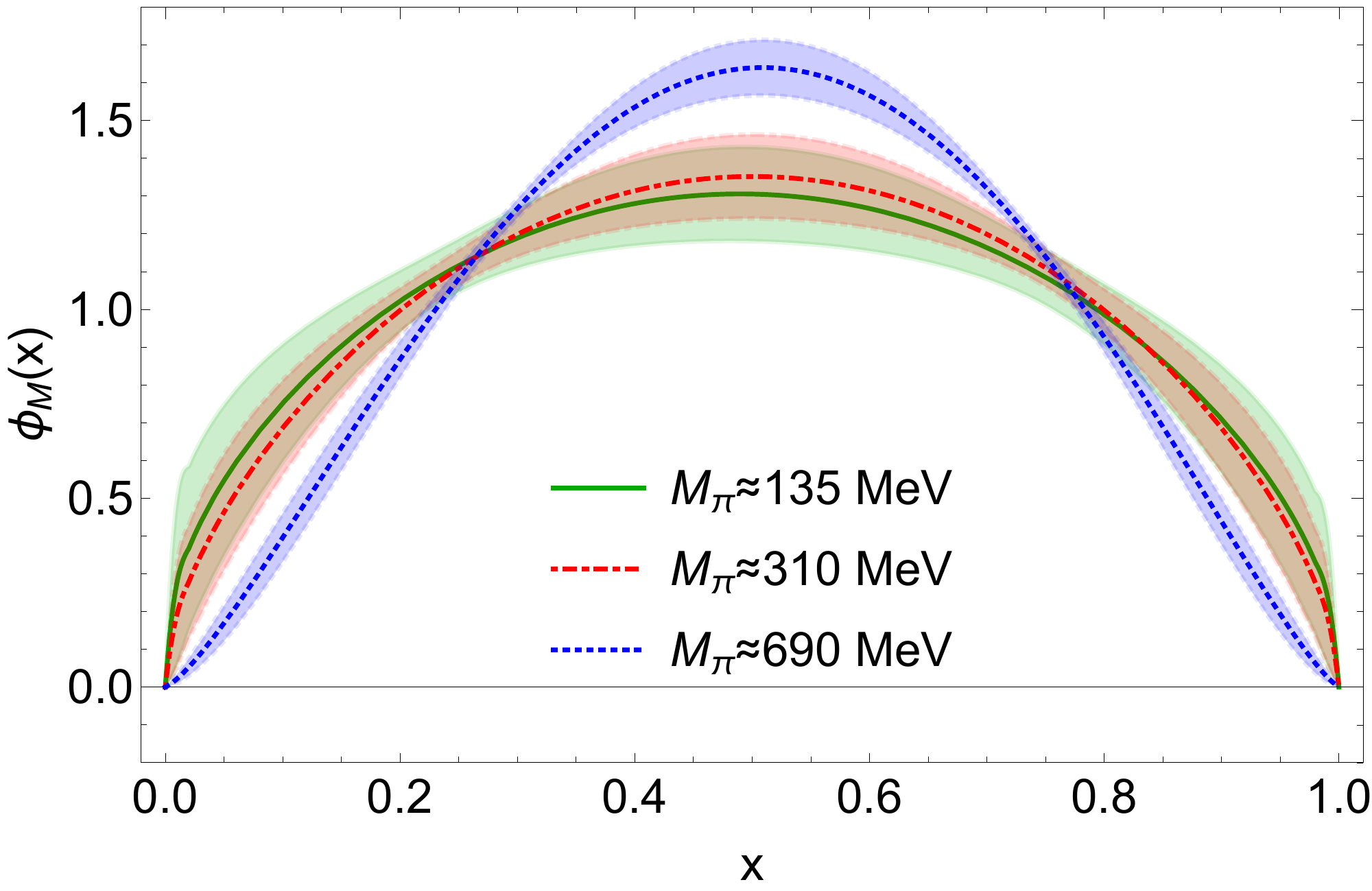}
	\includegraphics[width=0.45\linewidth]{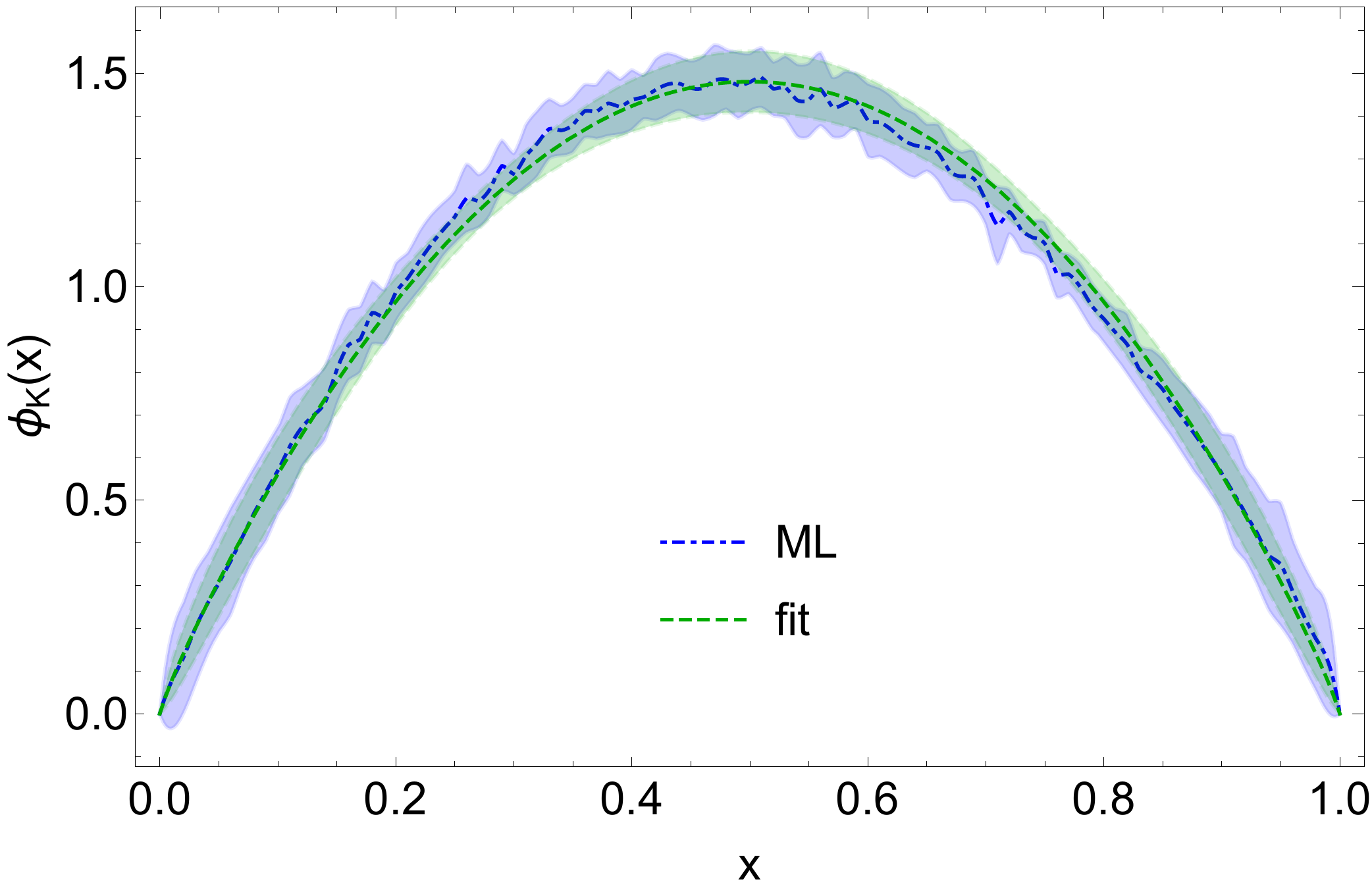}
	\caption{
	\emph{Left panel}. $x$-dependent pion DA at two different pion masses, along with an extrapolation to  the physical pion mass.
	%mass as a function of Bjorken-$x$ from MSULat Group.
	As the pion mass decreases, the distribution amplitude become broader.  The calculations use meson boosted momentum of $P_z=1.73\,{\rm GeV}$ and are renormalised at 2 GeV in $\overline{MS}$ scheme.
\emph{Right panel}.  $x$-dependent kaon DA obtained using a fit to lattice results obtained through a machine-learning approach~\cite{Zhang:2020gaj}.
	\label{fig:LQCD-DA-Mpi}}
\end{figure}

The $x$-dependent quark distribution amplitudes (DAs) of the pseudoscalar mesons have been calculated both in the LaMET approach~\cite{Chen:2018fwa, Zhang:2020gaj} and using a current-current approach analogous to LCS~\cite{Bali:2017gfr, Bali:2018spj}.  \cite{Zhang:2020gaj} studied the pion-mass dependence of the pion distribution amplitude on the lattice in the continuum limit as determined from three lattice spacings: $0.06$, $0.09$, $0.12\,$fm.  Figure~\ref{fig:LQCD-DA-Mpi} shows pion DA results at pion masses of 690\,MeV and 310\,MeV, together with their extrapolation to 135~MeV.  Note that the chiral extrapolation of \cite{Zhang:2020gaj} is dominated by the 310-MeV calculations.  The lattice kaon DA is shown on the right-hand-side of figure~\ref{fig:LQCD-DA-Mpi}.  The kaon DA is narrower than that of the pion, as suggested in \cite{Roberts:2019ngp, Roberts:2021xnz}.  The variation of the DA shapes with quark mass helps to understand the origin of mass~\cite{Aguilar:2019teb}.

\subsubsection{Meson parton distribution functions.}

The advances in computing hadron structure from calculations on a Euclidean lattice are transforming our ability to study the DFs of mesons within lattice QCD.  Many of the challenges in capitalising on these advances mirror those encountered in the global fitting community, most notably in obtaining a faithful description of DFs from incomplete data; in the case of lattice calculations, the advent of exascale computing, the application of novel methods -- such as Bayesian approaches, and machine learning promise to enable us to address and overcome these challenges.

\subsubsection{Valence quark distribution.}

\begin{figure}[tb]
\includegraphics[width=0.49\textwidth]{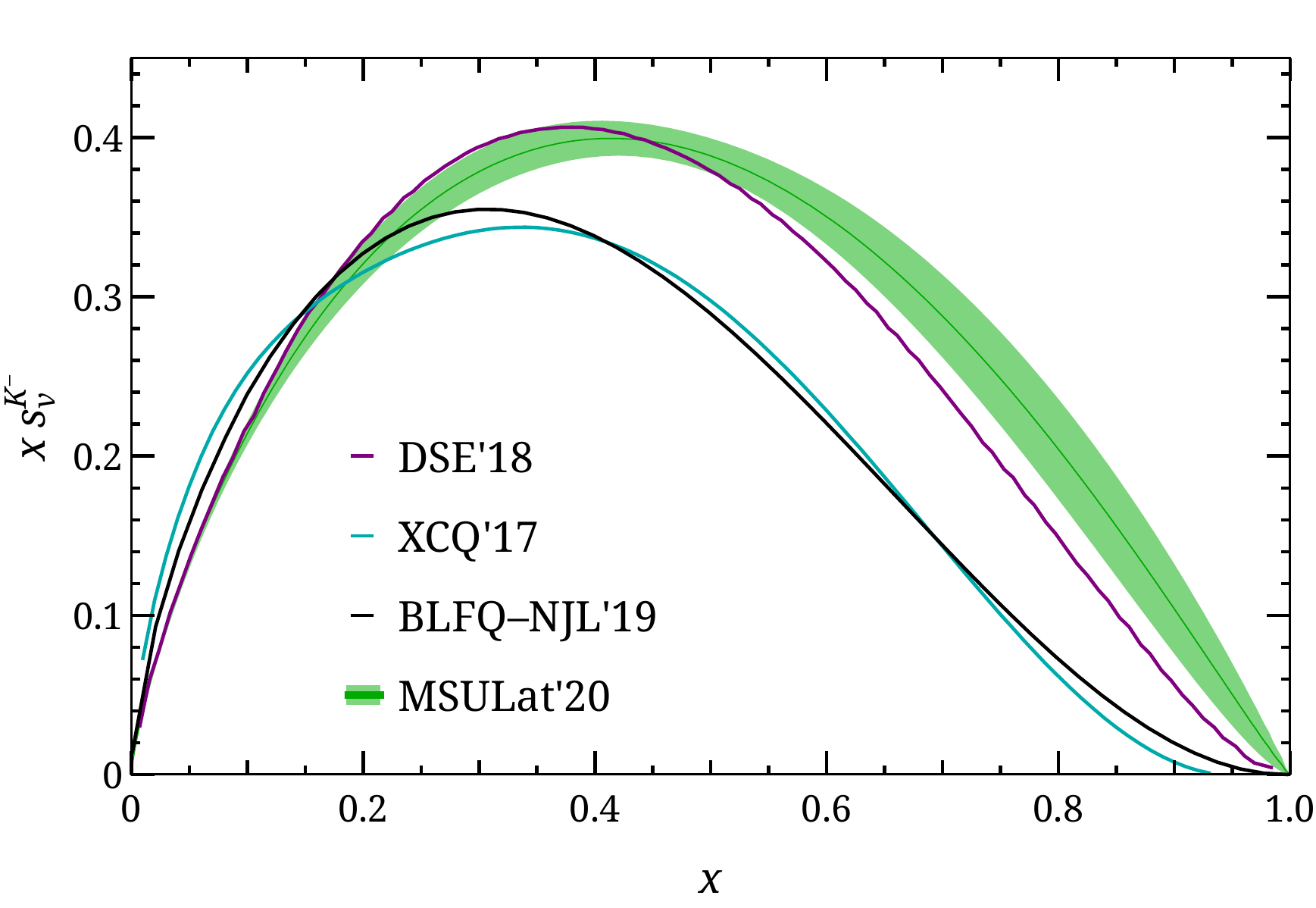}
\includegraphics[height=12.2em, width=0.49\textwidth]{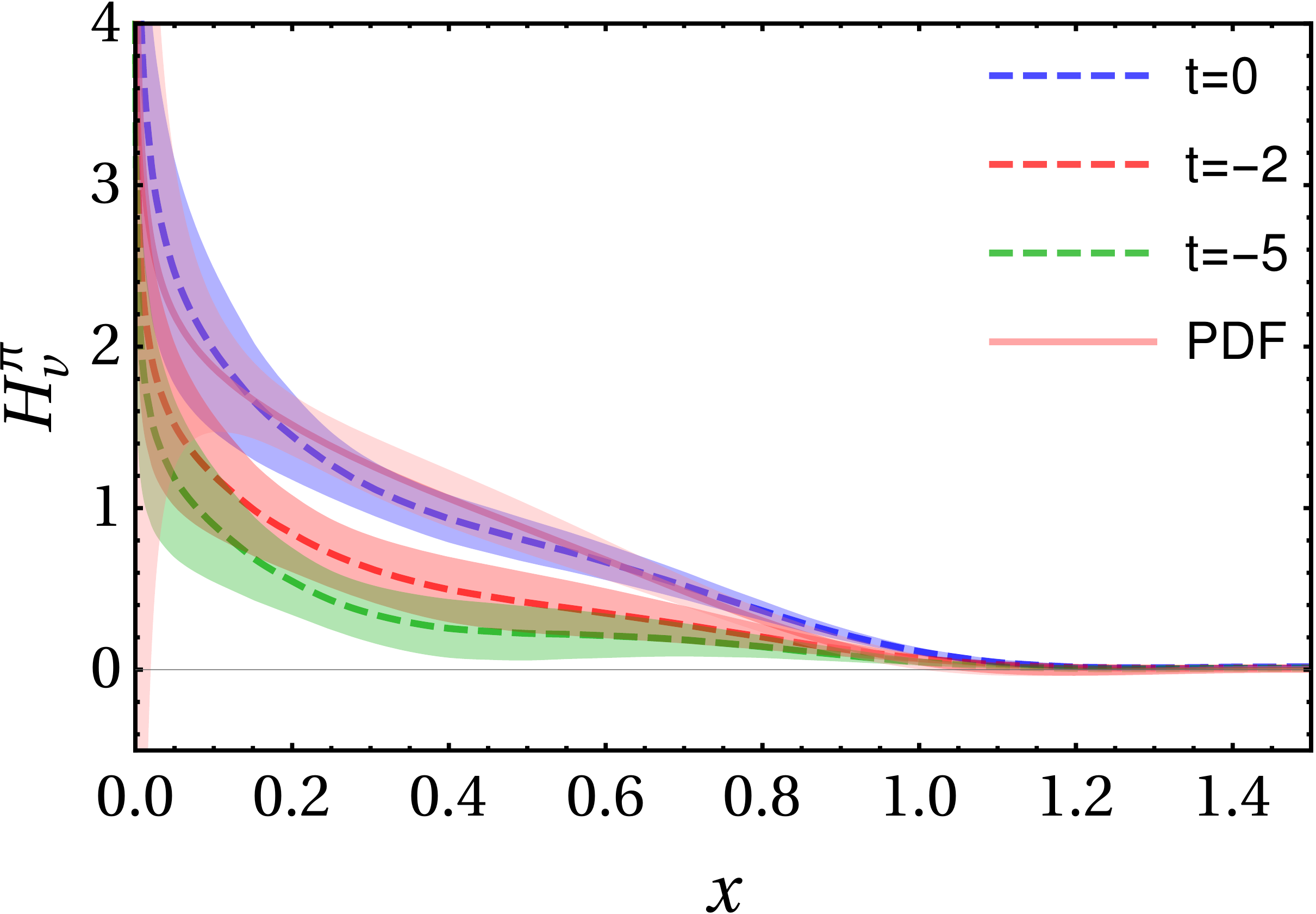}
\caption{  \label{fig:xqvpi}
\emph{Left panel}. Lattice results for $x \overline{s}_v^K(x)$ as a function of $x$ (labelled as ``MSULat'20'')~\cite{Lin:2020ssv}.  Some model studies are shown for comparison.  Additional discussion may be found elsewhere \cite[section\,7.2]{Cui:2020tdf}.
\emph{Right panel}.  Lattice results for the zero-skewness pion valence quark GPD $H^{\pi^{+}}_{v}(x,\xi=0,t,\zeta=4\,\text{GeV})$ for $t=\{0, -2, -5\}(2\pi/L)^2$ after one-loop matching and meson-mass corrections \cite{Chen:2019lcm}. ``PDF" denotes the pion DF result in \cite{Chen:2018fwa}.}
\end{figure}

The valence quark distributions are the most widely studied distributions within lattice QCD and where these new approaches have shown the most immediate impact.  Notably, calculations of the $x$-dependent DFs of the pion have been performed at close-to-physical light-quark masses, with increasing control over the systematic uncertainties arising from the finite-volume and discretisation systematic uncertainties.  Calculations have been performed within the LaMET~\cite{Chen:2018fwa, Izubuchi:2019lyk, Lin:2020ssv}, pseudo-PDF~\cite{Joo:2019bzr} and LCS frameworks~\cite{Sufian:2019bol, Sufian:2020vzb}.  Most recently, these methods have been applied to the valence quark distributions of the kaon~\cite{Lin:2020ssv}, as illustrated in Fig.~\ref{fig:xqvpi}, and the $u_K/u_\pi$ ratio, as discussed in connection with Fig.~\ref{fig:F2JorSeg}.  These calculations are becoming comparable with extant data over a large range of $x$; and arrival of the exascale era will enable them to be refined, especially with a view to determination of the large-$x$ behaviour of the PDFs, whose significance is discussed in connection with Eq.~\eqref{pionPDFlargex}.

\subsubsection{Gluon distribution.}
Within lattice QCD, gluonic and flavour-singlet quantities are much noisier than valence-quark distributions.  Thus, a far larger statistical sample is required to reveal a signal.
The first exploratory gluon DF study applied the quasi-PDF approach to the gluon DFs~\cite{Fan:2018dxu}, using ensembles with unphysically heavy quark masses corresponding to pion masses 340 and 678~MeV.
%calculations were done using overlap fermions on gauge ensembles with 2+1 flavours of domain-wall fermion and spacetime volume $24^3\times 64$, $a=0.1105(3)$~fm, and $M_\pi^\text{sea}=330$~MeV. The gluon operators were calculated for all volumes and high statistics: 207,872 measurements were taken of the two-point functions with valence quarks at the light sea and strange masses (corresponding to pion masses 340 and 678~MeV, respectively).
Unfortunately, the noise-to-signal ratio grows rapidly with the dimensionless parameter $zP_z$ and only coordinate-space gluon quasi-PDF matrix element ratios results are presented.  %Fig.~\ref{fig:gluonPDF}, and compared to the corresponding Fourier transform of the gluon PDF based on two global fits at NLO: the {\sc PDF4LHC15} combination~\cite{Butterworth:2015oua} and the {\sc CT14} set~\cite{Dulat:2015mca}. Up to perturbative matching and power corrections $O(1/P_z^2)$, the lattice results are compatible with global fits within the statistical uncertainty at large $z$. The results at the lighter pion mass (at the unitary point) of 340~MeV are also shown in Fig.~\ref{fig:gluonPDF}. These are consistent with those from the strange point but have larger uncertainties. The gluon quasi-PDFs in the pion were also studied for the first time in Ref.~\cite{Fan:2018dxu} and features similar to those observed for the proton were revealed. Finally,
Since then, there have also been developments in improving the operators for the gluon DF lattice calculations~\cite{Balitsky:2019krf, Wang:2019tgg, Zhang:2018diq}, which should enable evaluation of the continuum limit in future lattice calculations of gluon DFs. The pseudo-PDF approach developed in \cite{Balitsky:2019krf}, along with improved methods of calculation for reaching higher boosted momentum, have recently been used to provide the first result for the nucleon gluon DF~\cite{fan2020nucleon}.
%TF:\cite{Fan:2020}.
The prospects for applications to the cases of the pion and kaon appear promising, so that one may anticipate the appearance of increasingly precise calculations over the next few years.

\subsubsection{Pion GPDs.}
In \cite{Chen:2019lcm}, the pion valence quark GPD at zero skewness was calculated using clover valence fermions on an ensemble of gauge configurations with $2+1+1$ flavours (degenerate up/down, strange and charm) of highly-improved staggered quarks with lattice spacing $a \approx 0.12$~fm, box size $L \approx 3$~fm and pion mass $m_\pi \approx 310$~MeV.  The result is shown in figure~\ref{fig:xqvpi}\,--\,right.  It turns out that, with current uncertainties, the result does not show a clear preference amongst different model assumptions about the kinematic dependence of the
GPD.  To distinguish between different models, further studies with higher-statistics will be crucial.

\begin{figure}
\includegraphics[width=0.95\textwidth]{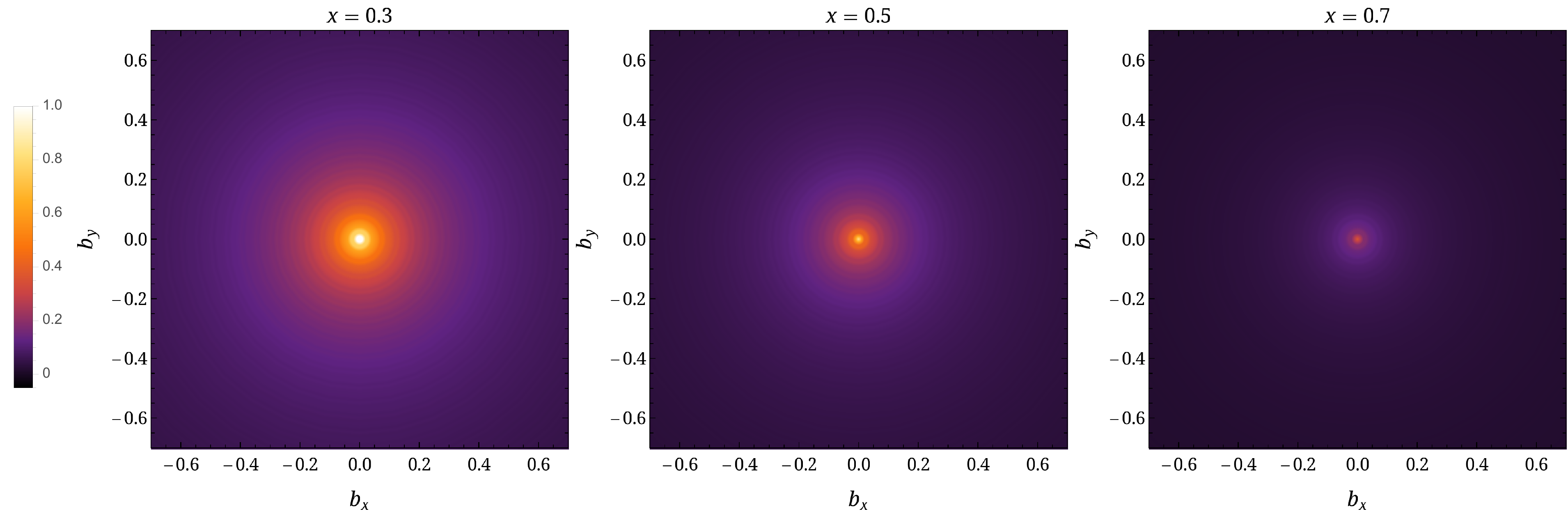}
\hspace*{0.20cm}
\caption{\label{fig:impact-distribution}
The two-dimensional impact-parameter--dependent distribution, $\mathsf{q}(x,b)$, for $x=0.3$, 0.5 and 0.7 determined from a lattice-QCD calculated pion GPD at physical pion mass.
}
\end{figure}

One may therefore anticipate that there will be lattice calculations of the pion's valence quark GPD, $H_\pi(x,\xi=0,Q^2)$, within the next few years.  The Fourier transform of this GPD gives the impact-parameter--dependent distribution, $\mathsf{q}(x,b)$~\cite{Burkardt:2002hr, Diehl:2003ny}:
\begin{equation}\label{eq:impact-dist}
\mathsf{q}(x,b) = \int \frac{ d \mathbf{q}}{(2\pi)^2} H_\pi(x,\xi=0,t=-\mathbf{q}^2) e^{i\mathbf{q}\,\cdot \, \mathbf{b} },
\end{equation}
where $b$ is the light-front distance from the transverse center of momentum (CoTM).
Figure~\ref{fig:impact-distribution} shows the two-dimensional distributions at $x=0.3, 0.5,0.7$.  The impact-parameter--dependent distribution describes the probability density for a parton with momentum fraction $x$ to be found in the transverse plane at distance $b$ from the CoTM.  It provides a snapshot of the pion in the transverse plane and indicates what might be expected from nucleon tomography.

% SJDK - Removed names from title, commented below as a reminder, delete once no longer needed
% {(\underline{Tim}}, Wally)
\subsection{Global QCD analysis}
\label{sec:QCDfit}

Extracting information about DFs (or any partonic content of hadrons) from experimental data is a challenging problem in % nuclear and particle
hadronic physics.
Since the Lagrangian partons can never be isolated as free particles, details about their properties must be inferred indirectly by exploiting theoretical tools, such as QCD factorisation theorems~\cite{Collins:1989gx}.
The latter allow experimental observables in certain kinematic regions, {\it viz}.~$M_N^2/[(1-x) Q^2] \ll 1$, to be written as convolutions of perturbatively-calculable hard scattering cross sections and nonperturbative DFs parametrising long-distance quark-gluon physics.
The most robust method to extract information about DFs from experiment is through global QCD analyses of various QCD-factorisable hadronic processes that are sensitive to different combinations of DFs~\cite{Owens:1984zj, Aurenche:1989sx, Sutton:1991ay, Gluck:1991ey, Gluck:1999xe, Wijesooriya:2005ir, Barry:2018ort, Novikov:2020snp, Barry-resummation-20}.

Historically, the main experimental observables that have been used to constrain pion DFs have come from pion-nucleus collisions with inclusive production of lepton pairs or prompt photons~\cite{Badier:1983mj, Betev:1985pf, dy1}.
More recently, leading neutron electroproduction data~\cite{Chekanov:2002pf, Aaron:2010ab} have been used to constrain pion DFs at small $x$, assuming the validity of pion exchange at small values of the transverse momentum and large longitudinal momentum of the produced neutron~\cite{Chekanov:2002pf, Aaron:2010ab, McKenney:2015xis}.

Important questions remain, however, concerning the fraction of the pion's momentum carried by gluons relative to the valence and sea quarks, and the behaviour of DFs at small and large values of $x$.
For the latter, many calculations have been completed, with the exponent on $(1-x)^\beta$ ranging from $\beta \sim 0$ to $\beta \sim 2$.  This was highlighted in \cite{Holt:2010vj} and in a raft of calculations since then, e.g.
\cite{Noguera:2015iia, Hutauruk:2016sug, Hobbs:2017xtq, Xu:2018eii, Ding:2019qlr, Ding:2019lwe,  PhysRevLett.122.172001, Chang:2020kjj, Kock:2020frx, Cui:2020dlm, Cui:2020tdf, Zhang:2020ecj}.
In model calculations \cite{Noguera:2015iia, Hutauruk:2016sug, Hobbs:2017xtq, Xu:2018eii, PhysRevLett.122.172001, Chang:2020kjj, Kock:2020frx, Zhang:2020ecj}, the energy scale and the value of $x$ at which the asymptotic behaviour should be evident is {\it a priori} unknown.  On the other hand, as noted in connection with Eq.\,\eqref{pionPDFlargex}, all calculations which enable a connection to be drawn between the underlying meson-binding dynamics and the valence-quark DF show that $\beta$ is determined by the behaviour of the quark-antiquark interaction.
%
%%% CDR ... Bednar:2018mtf violates a raft of symmetries; hence, cannot be judged reliable.  If it is mentioned, then its defects must also be highlighted so that readers are properly informed. ...
%%% Moreover, it was observed~\cite{Bednar:2018mtf} that a linear $\beta=1$ dependence over a finite range in $x$ could nevertheless be consistent with a formal $\beta=2$ behaviour in the $x \to 1$ limit, so that it is important to understand the extent to which these can be tested or constrained by data.
Concerning data analyses, as already noted, inclusion and/or treatment of soft-gluon resummation can affect the inferred large-$x$ PDF behaviour~\cite{Aicher:2010cb, Aicher:2011ai, Bonvini:2015ira, Westmark:2017uig, werner_vogelsang_2020_4019432, Barry-resummation-20}, and the interplay of resummation and fixed-order calculations needs to be better understood.

% Other attempts have been made to extract information about pion PDFs from inclusive $J/\psi$ production within models such as the colour evaporation model~\cite{Chang:2020rdy}; however, it is not clear in this case how to uniquely determine the PDFs independent of the framework and parameters of the model.

The detailed $x$ dependence of pion DFs is clearly a topic of considerable theoretical and phenomenological interest, and more data over a large range of kinematics would be helpful to unravel this structure.
A programme of leading baryon production in inclusive deep inelastic scattering (DIS) from the deuteron with proton tagging (``TDIS'') at Jefferson Lab~\cite{E12-15-006} aims to explore the structure of pions emitted from the bound neutron~\cite{Hobbs:2014fya},
% in a mirror image of the HERA leading neutron production~\cite{Hobbs:2014fya},
with generalisation to the hyperon case~\cite{C12-15-006A} aimed at investigating corresponding kaon structure observables.
Complementing HERA and JLab measurements, EIC data on leading neutron and hyperon production can provide information on the role of the nucleon's peripheral structure in a unique region interpolating between these kinematics.  Especially in this interpolation region, EIC's combination of high precision and wide kinematical coverage from $Q^2\!\sim$\! [few GeV$^2$] to $Q^2\!\sim\! {\rm O}(100\,\mathrm{GeV}^2)$ suggests a significant potential to constrain scaling violations in the pion structure function.  This, in turn, may afford a higher level of discriminating power in unravelling the pion's gluon content from the corresponding valence-/sea-quark contributions.

\begin{figure}[t]
\includegraphics[width=0.99\textwidth, trim={20mm 0mm 50mm 0mm}, clip]{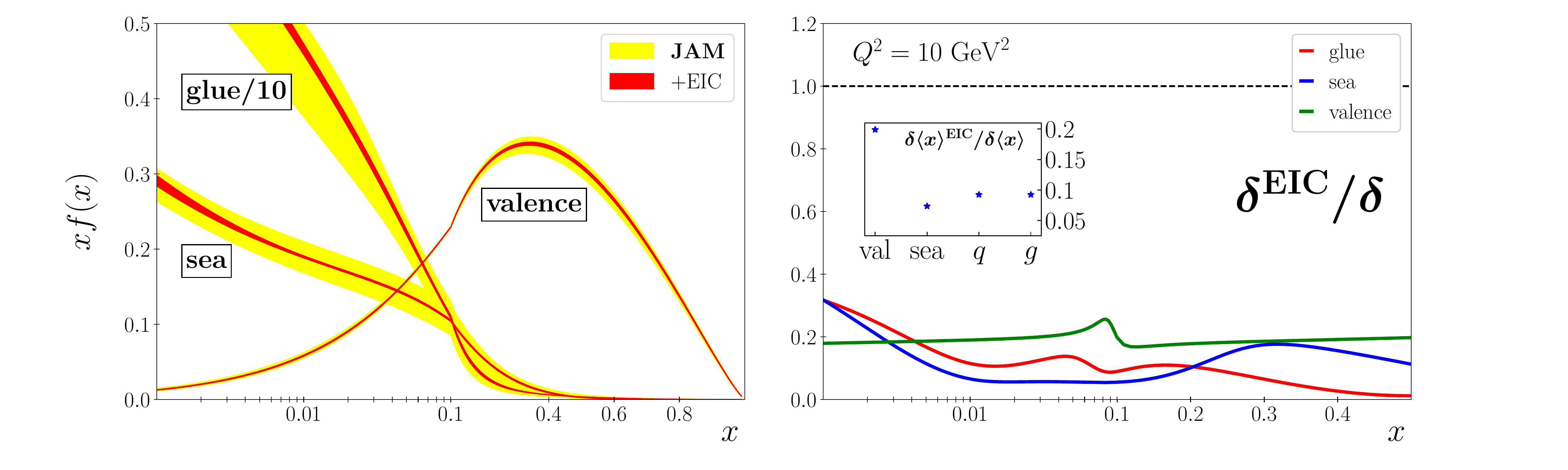}
\caption{
\label{fig:pion-pdf-impact}
    \emph{Left panel}.  Comparison of uncertainties on the pion's valence, sea quark and gluon PDFs before (yellow bands) and after (red bands) inclusion of EIC data.
    \emph{Right panel}. Ratio of uncertainties with EIC data to without, $\delta^{\rm EIC}/\delta$, for the valence (green line), sea quark (blue) and gluon (red) PDFs,
    %RE 01/24/21 changed based on discussions last week: assuming 1.2\% systematic uncertainty,
    assuming 1.2\% experimental systematic uncertainty but no model systematic uncertainty,
    and (inset) the corresponding ratios of the momentum fraction uncertainties, $\delta\langle x \rangle^{\rm EIC}/\delta\langle x \rangle$, for valence, sea, total quark and gluon PDFs~\cite{Barry-resummation-20}, at a scale $Q^2=10$~GeV$^2$.}
\end{figure}

The potential impact of EIC neutron production data is illustrated in figure~\ref{fig:pion-pdf-impact}, which shows the valence, sea quark and gluon PDFs in the pion from the JAM global QCD analysis at the evolved scale $Q^2=10$~GeV$^2$~\cite{Barry:2018ort}, comparing current uncertainties with those expected following the addition of EIC data~\cite{Barry-resummation-20}.
The analysis of the existing data includes pion-nucleus Drell-Yan cross sections, both $p_T$-differential and $p_T$-integrated, and the leading neutron structure functions from HERA~\cite{Cao2020}.
The analysis assumes a centre-of-mass (CM) energy $\sqrt{s}=73.5$~GeV for an integrated luminosity $\mathcal{L} = 100~\mathrm{fm}^{-1}$ and a 1.2\% systematic uncertainty across all kinematics.
% systematic uncertainties similar to those from the HERA leading neutron data~\cite{Aaron:2010ab, Chekanov:2002pf}.
For both the sea quark and gluon distributions, the PDF uncertainties reduce by a factor $\sim 5-10$ for most of the $x$ range, with a similar factor $\sim 5$ reduction in the valence sector.
For a decomposition of the pion mass written in terms of QCD stress-energy tensor matrix elements \cite{Yang:2014xsa}, the first moments, $\langle x \rangle_{q,g}$, are relevant.  However, as discussed in connection with Fig.~\ref{F1CDR}, the meaning of such a frame and scale dependent decomposition is uncertain \cite[section\,V]{Roberts:2020udq}, \cite[sections\,4.4,\,4.5]{Roberts:2020hiw}.
Notwithstanding that, such moments are interesting in themselves, so it is worth noting that the reduction in associated uncertainties is a factor $\approx 10$ for both the total quark and gluon contributions, as can be seen in the inset of figure~\ref{fig:pion-pdf-impact}\,--\,right.
Note, however, that the errors do not include uncertainties associated with the model dependence of the ``pion flux,'' which may be of the order $10\%-20\%$~\cite{Aaron:2010ab, Chekanov:2002pf}, and might reduce the impact of the projected data on the pion PDF uncertainties.
A similar analysis may be performed for the PDFs in the kaon, which can be obtained from leading hyperon production in the forward region.
In this case, the near-absence of empirical information on the parton structure of kaons will mean an even more striking impact of new EIC data.

\begin{figure}
\begin{center}
\includegraphics[width=0.95\textwidth]{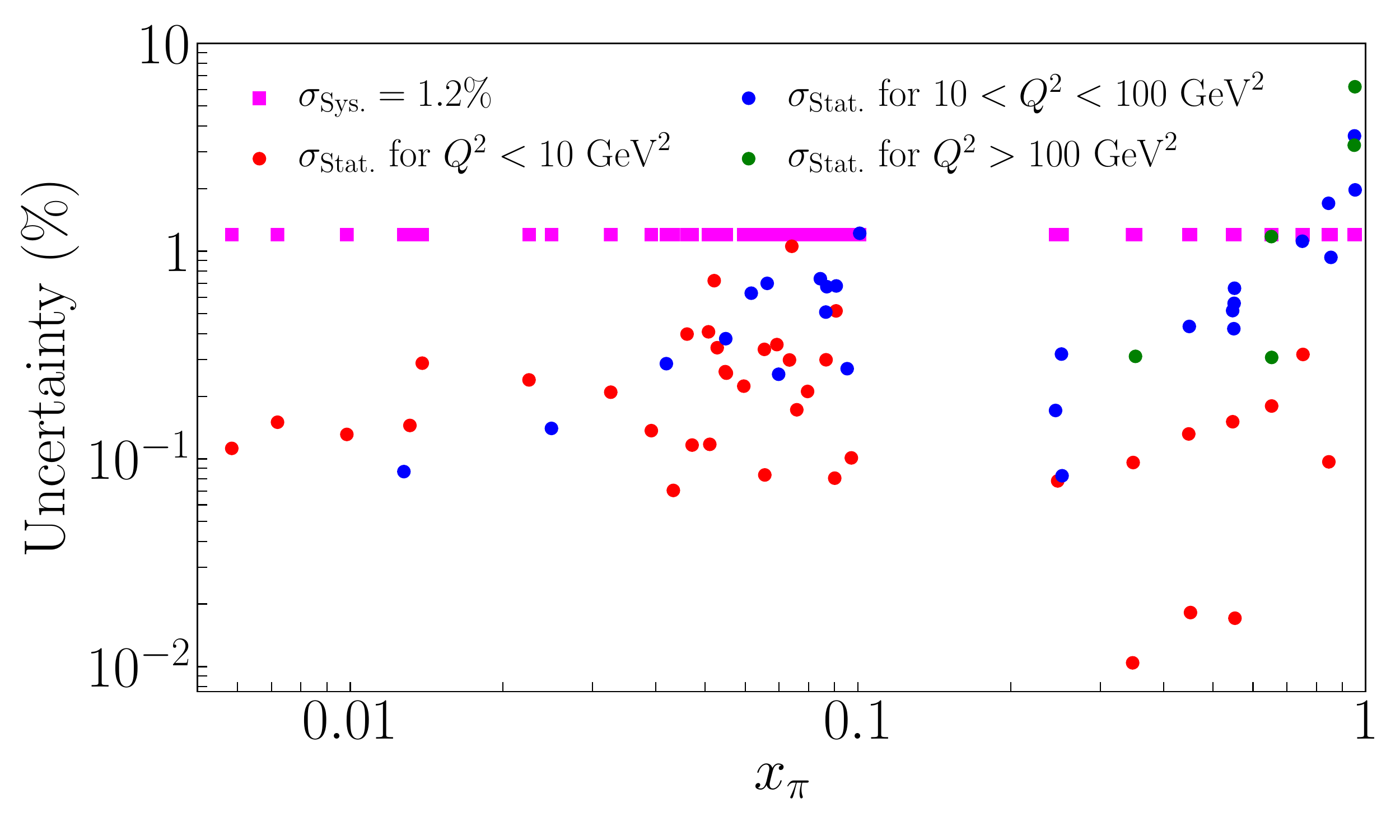}
\caption{
\label{fig:cross_section_uncertainties}
%Percent uncertainties of cross section data used for the pion PDF impact analysis in figure~\ref{fig:pion-pdf-impact} as a function of $x_\pi$.
Fractional uncertainties of the cross section projected data as used for the pion PDF impact analysis in figure ~\ref{fig:pion-pdf-impact}.
    The statistical uncertainties $\sigma_{\rm Stat.}$ are shown as circles,
    separated in color by the ranges in $Q^2$.
    The assumed systematic uncertainty $\sigma_{\rm Sys.}$
    of 1.2\% are shown in the magenta squares.}
\end{center}
\end{figure}

%\textcolor{red}{In the impact study for the pion PDFs, we use the uncertainties on the differential cross section shown in %figure~\ref{fig:cross_section_uncertainties} as a function of $x_\pi$.
%The systematic uncertainties of 1.2\% are shown in the magenta squares,
%while the statistical uncertainties are shown in the filled circles.
%The data points with small $x_\pi$ and $Q^2$ tend to give the smallest
%statistical uncertainties
%as shown by the red circles, while the blue and green circles indicate
%that as $Q^2$ and $x_\pi$ grow,
%generally the statistical uncertainties grow as well.
%Clearly for much of the range $x_\pi \lesssim 0.8$, the 
%systematic uncertainties dominate the statistical uncertainties.
%This was not the case with the same observable in the HERA experiments~\cite{Aaron:2010ab, Chekanov:2002pf}.
%Because the integrated luminosity for the EIC 
%is projected to be about three
%orders of magnitude greater than that for HERA,
%the total uncertainty quantification will be largely driven by the
%systematic uncertainties.}

For the impact study of the pion PDFs the uncertainties on the differential cross section were used. These are shown in figure~\ref{fig:cross_section_uncertainties} as a function of $x_\pi$ and further discussed in section~\ref{sec:meson-structure-functions}. The systematic uncertainty is 1.2\% (magenta squares); the statistical uncertainties are on average less than 0.5\% and vary as a function of $x_\pi$ and $Q^2$ (filled circles). The statistical uncertainty is smallest at small $x_\pi$ and $Q^2$ and increases with increasing values of $x_\pi$ and $Q^2$. Clearly for much of the range $x_\pi \lesssim 0.8$, the 
systematic uncertainties dominate the statistical uncertainties.
This was not the case with the same observable in the HERA experiments~\cite{Aaron:2010ab, Chekanov:2002pf}.
Because the integrated luminosity for the EIC 
is projected to be about three
orders of magnitude greater than that for HERA,
the total uncertainty quantification will be largely driven by the
systematic uncertainties.

%\newpage
% SJDK - Strip the names and clean up title once finished editing/writing the section
%\subsection{QCD phenomenology, lattice and global analysis summary - Tobias, David and Tim}
%\subsection{Synergy between QCD phenomenology, lattice and global data analysis}
\subsection{Synergy between theory calculations and data analysis}
\begin{figure}[htb]
  \centering
 \includegraphics[width=1.2\textwidth]{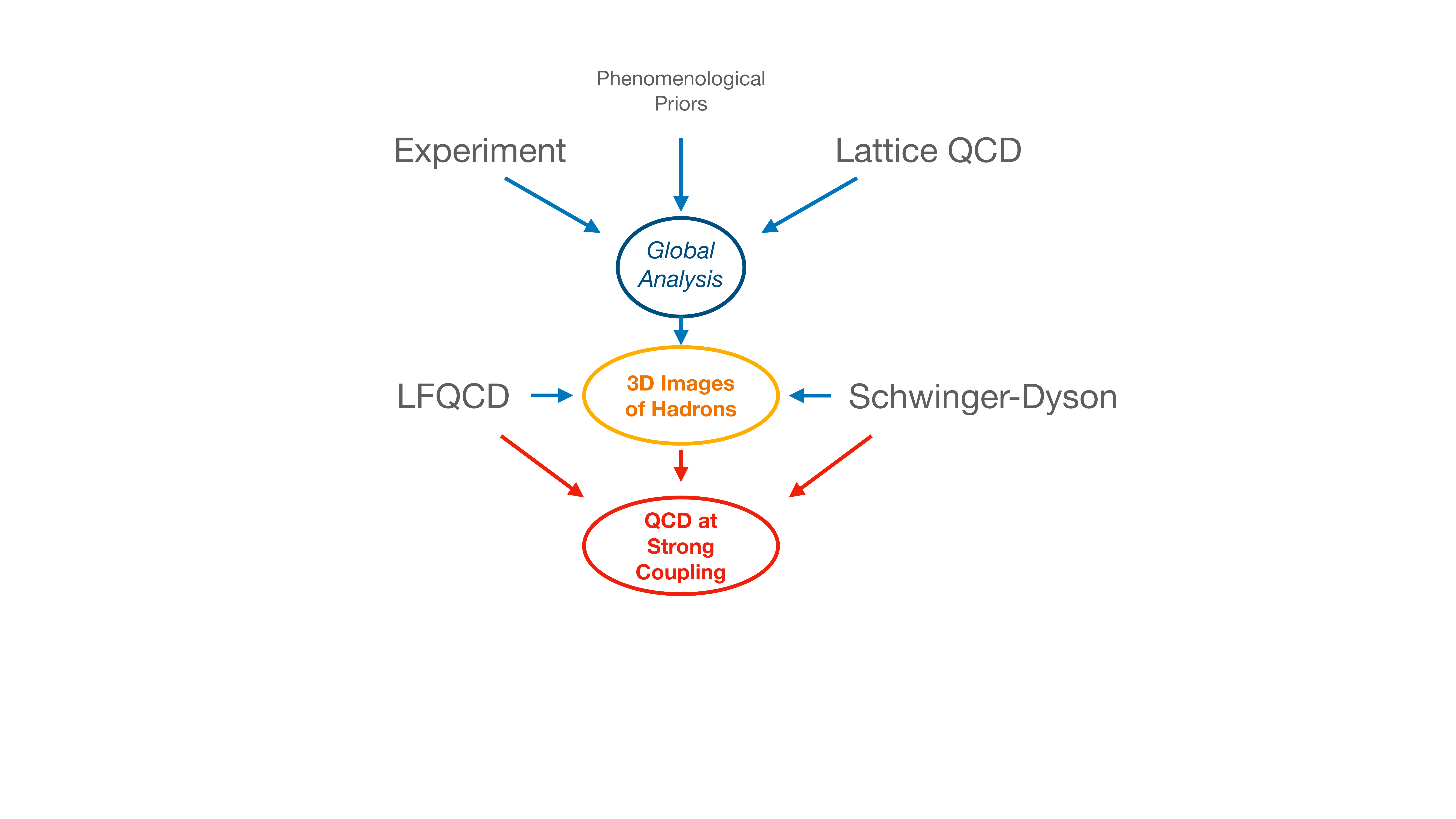}
  \vspace{-1.3in}
  \caption{Potential interplay between QCD phenomenology, continuum approaches to QCD, lattice, global analysis and experiment.}
  \label{fig:QCD}
\end{figure}

In comparison with the nucleon, experimental probes of the pion and kaon have been relatively sparse, with
substantial ambiguities remaining regarding their partonic, quark-gluon substructure. Partly for
this reason, one may expect future knowledge of the pion to derive from an interplay among
several methods (see Fig.~\ref{fig:QCD}): QCD phenomenology, including QCD-inspired models and continuum methods in Euclidean and Minkowski spaces;
recent developments from lattice QCD; and QCD global analyses of both contemporary and future data.
For the latter, EIC can be expected to furnish a significant amount of valuable data; and therefore to be a crucial driver of global theory efforts.
Naturally, in the context of this discussion, the focus here is on the structure of NG modes as quantified via collinear parton distribution functions (PDFs), or, in the case of three-dimensional structure, GPDs and TMDs.

To explore possible synergies among the approaches sketched above, a number of theoretical issues require further development and understanding.  The following are especially prominent.
\begin{itemize}

\item
%On the QCD lattice, a firmer grasp is needed of the relation(s) between lattice-calculable quantities %({\it e.g.}, Ioffe-time distributions or
%quasi-distributions --- the analogous object within LaMET) and corresponding extractions from %experiments; {DGR TO DO}
%
{\bf Direct lattice calculation of distributions}. A major advance in the ability to study the internal structure of hadrons from lattice QCD computations was the realisation that PDFs, and their three-dimensional extensions, described as matrix elements of operators separated along the light cone, could be related to quantities calculable in Euclidean space~\cite{Ji:2013dva, Ji:2014gla, Radyushkin:2017cyf, Ma:2017pxb}.
%
%{\bf Need to cite Xiangdong, Anatoly, Jianwei, etc}.
%
Such calculations yield not pointwise evaluation of the PDFs at each Bjorken-$x$, but rather convolutions of those PDFs with some calculable kernel, together with modifications in the manner of higher-twist and mass corrections.  The resulting convolutions in practical lattice calculations yield functions % of Ioffe time %{\bf Need to think a little about terminology - sensitive topic}
%%% It's not always and not only Ioffe time.
%
that are incomplete and limited by the finite volume and discretisation of the lattice.  The extraction of the $x$-dependent PDFs from such calculations therefore requires that an inverse problem be addressed, whose solution requires additional information.

The situation somewhat mirrors that of global fits to experimental data, when the desired PDFs generally arise, as in Drell-Yan processes, from a convolution with a perturbatively calculated kernel.  For the case of one-dimensional distributions, such as PDFs, the additional information can be provided through an assumed PDF parametrisation so that extraction of the PDF becomes a parameter fitting exercise.  That method has been central to the strategy of the global fitting community and has likewise been adopted by several lattice collaborations.  More recently, there has been an effort to incorporate different schemes, such as machine learning and Bayesian reconstruction~\cite{Karpie:2019eiq, DelDebbio:2020rgv}.

 \item
%Lattice computations as inputs into global fits; (DGR/TIM)
%
{\bf Inclusion of lattice results as data in QCD global analyses}.  An evolving opportunity is to introduce results from lattice computations into the global fits~\cite{Lin:2020rut}, either in the same manner as experimental data, in the case of the nucleon PDFs~\cite{Bringewatt:2020ixn}, or as a Bayesian prior in the fit to experimental data, as was accomplished for the nucleon tensor charge~\cite{Lin:2017stx}.  The aim, at least for the nucleon, is not to test QCD, but rather to use both experiment and lattice computations to provide more information about key measures of hadron structure than either can alone.
% {\emr Following is completely speculative!}
For pions and kaons, for which there are no free meson targets and structure is probed indirectly through the Sullivan process at EIC, lattice computations of the PDFs of a lone, isolated pion or kaon may assist in validating the experimental analysis, potentially providing benchmarks to quantify the effects of off-shellness or kinematical extrapolations in $t$;

\vspace{0.1cm}

%\item
%Complementarity between Lattice QCD and experiments; (DGR/TIM)
Lattice computation and experiment may also provide complementary information on hadron structure.  Thus an important effort is the computation of gluon and flavour-singlet contributions to hadron structure, and those computations may predict the outcomes of experiment.  Similarly, recent development have enabled the calculation of the $x$-dependent GPDs, both for the nucleon~\cite{Alexandrou:2020zbe,Lin:2020rxa} and the pion~\cite{Chen:2019lcm} -- see Fig.\,\ref{fig:xqvpi}\,--\,right, and the frameworks allow these distributions to be extracted at definite non-zero skewness~\cite{Radyushkin:2019owq, Alexandrou:2020zbe}.

\item
{\bf Benchmarking calculations with QCD fits and phenomenology}.
PDF (or GPD/TMD) phenomenology can offer benchmarks for use in developing lattice and/or continuum QCD computations; % (TIM);
in particular, the NNLO precision of contemporary {\it nucleon} PDF analyses in the unpolarised sector is such that these extractions can play an important role in testing analogous calculations
of lattice quantities, such as PDF Mellin moments~\cite{Hobbs:2019gob}.  Similar arguments apply to recent QCD global analyses of meson structure noted above.

\item
{\bf Relating Euclidean lattice QCD and continuum methods}.
A number of formal developments related to treatments of the pion using CSMs, including Dyson-Schwinger equations and light-front quantisation,
%based on the Dyson-Schwinger and Bethe-Salpeter Equations and light-front quantization
would also be helpful.  Lattice studies long ago confirmed the continuum predictions of nonperturbative infrared dressing of gluons and quarks,
%%% I don't seen anything relevant here ... (see e.g. the lattice QCD review  in \cite{Zyla:2020zbs}),
which are the effective degrees-of-freedom in the pion/kaon exploited by continuum methods.
% in Euclidean space (an analytic extension to Minkowski  has to be developed).
%% Actually exploited in any mathematically accessible space.
%
The light-front projection of a hadron's Bethe-Salpeter amplitude~\cite{Sales:1999ec,Chang:2013pq} is linked to a Fock component in a basis whose character is specified by the resolving scale \cite[section\,2]{Cui:2020tdf}.  This projection is a gauge invariant probability density~\cite{RevModPhys.92.045003}.  The hadron image on the null-plane~\cite{RevModPhys.21.392}, expressed by Ioffe time and transverse coordinates of dressed constituents, may be accessible along this path~\cite{PhysRevC.102.022201, PhysRevD.103.014002};

\item
{\bf Mapping the pion light-front wave function.}
Understanding  QCD on the null-plane will place the concept of the light-front wave function on firmer ground, allowing access to pion PDFs, GPDs, TMDs, and more~\cite{doi:10.1142/S0218301320300064} within a unified, invariant representation of the meson.  A clear identification of the unitary transformation from the free light-front Fock-space basis to the one that entails dressed and confined constituents is necessary.  To this end, investigations on the non-triviality of the vacuum and the role zero-modes in the light-front quantisation are on-going (see e.g.\ \cite{de_Melo_1998, collins2018nontriviality, PhysRevD.102.116010, Martinovic:2018apr, mannheim2020comparing}).

\end{itemize}
It is worth reiterating that data supplied by EIC will provide a firm basis for new insights into pion and kaon structure. By significantly expanding the world's pool of data sensitive to light-meson structure functions,
EIC will provide a setting to explore and refine the synergies enumerated above and sketched in figure~\ref{fig:QCD}.
These refinements will occur along several tracks, providing new constraints to QCD fits of meson parton distributions, which will then be used to inform and validate continuum analyses in QCD and phenomenological calculations while also benchmarking rapidly-developing lattice efforts.  These in turn can be expected to serve reciprocally as guidance for and constraints on QCD fits of data, including the highly anticipated EIC measurements envisioned in this work.

\section{Key EIC measurements}

It is here necessary to summarise the experimental requirements for critical EIC measurements that tackle some outstanding questions in the study of pion and kaon mass and structure.  This will lead subsequently to an explanation of how these meson structure measurements, which serve as a laboratory in which fundamental aspects of QCD can be elucidated, complement and strengthen ongoing and foreseen programmes worldwide.

To facilitate this discussion, it is useful to translate current theory understanding of light meson structure and emergent hadron mass (and structure) mechanisms into a set of critical science questions.  Currently, not all these science questions are rigorously defined theoretically; but they do reflect the current state of understanding.  These questions come from community discussions at a series of dedicated pion/kaon structure workshops (2017, 2018, 2019, and 2020), and at meetings related to the ongoing EIC Yellow Report activities. They represent outstanding mysteries that require further experimental (and theoretical) examination, and illustrate the impact of a coherent study of pion and kaon structure yielding results similar to present studies of proton structure.

Table~\ref{tab:EIC_science_questions_table} on page~\pageref{tab:EIC_science_questions_table} lists the key science questions along with specific measurements required to advance community understanding.  It also presents the high-level experimental needs, providing the minimum experimental requirements as well as improvements that could further expand these studies.
Later sections will examine other important considerations aimed at demonstrating that one can extract pion and kaon structure information independent of the phenomenology ansatz, independent of physics background contributions, and independent of Mandelstam-$t$.
Some interesting science questions that may be more challenging to address are listed at the bottom of the table; they are considered more speculative because validating the reaction mechanism will be more challenging than the other cases, owing to considerations such as competing reaction and background mechanisms.

For all observables, a luminosity well above 10$^{33}$ is required to compensate for the (few times) 10$^{-3}$ fraction of the proton wave function related to the pion (kaon) Sullivan process.  Also, a large range in $x_L$ (the longitudinal energy fraction carried by the produced particle) is required, up to $x_L \sim 1$ for $ep$ reactions and $x_L$ at least $\sim 0.5$ for $ed$ reactions.  Data on negatively-charged pions (e.g.\ $e + d \rightarrow e^\prime + p + p + X$) and on neutral-pion channels (e.g.\  $e + p \rightarrow e^\prime + p + X$) are crucial to constrain reaction mechanisms and theory backgrounds in extracting the physical pion (kaon) target information.

% SJDK - Removed names from title, included below for completeness. Delete once no longer needed
%(Rachel, Tanja)
\subsection{Sullivan process}\label{sec:sullivan}

In specific kinematic regions, the observation of recoil nucleons (N) or hyperons (Y) in the semi-inclusive reaction $e p \to e^\prime (N\,{\rm or}\,Y) X$ can reveal features associated with correlated quark-antiquark pairs in the nucleon, referred to as the ``meson cloud'' of the nucleon.
At low values of $|t|$, the four-momentum transfer from the initial proton to the final nucleon or hyperon, the cross section displays behaviour characteristic of meson pole dominance. The reaction in which the electron scatters off the meson cloud of a nucleon target is called the Sullivan process~\cite{Sullivan:1971kd}.
For elastic scattering ($X = \pi^+$ or $K^+$), this process carries information on the pion or kaon form factor, and could be tagged by detection of a recoil nucleon or hyperon, respectively. For DIS, the typical interpretation is that the nucleon parton distributions contain a mesonic parton content. To access pion or kaon partonic content via such a structure function measurement requires scattering from a meson target, which again could be facilitated in the Sullivan process by detection of a recoil nucleon or hyperon.

The Sullivan process can provide reliable access to a meson target in the space-like $t$ region, if the pole associated with the ground-state meson remains the dominant feature of the process and the structure of the related correlation evolves slowly and smoothly with virtuality. To check whether these conditions are satisfied empirically, one can take data covering a range in $t$, particularly low $|t|$, and compare with phenomenological and theoretical expectations. A recent calculation~\cite{Qin:2017lcd} explored the circumstances under which these conditions should be satisfied. For the pion (kaon) Sullivan process, low $-t$ equates to $-t <$ 0.6 (0.9) GeV$^2$ to be able to cleanly extract pion (kaon) structure, and data over a range of $-t$ down to the lowest accessible are needed to verify pion (kaon) structure extraction.

\subsection{Theoretical backgrounds in extracting the data}\label{sec:theory_background}

Extraction of the mesonic structure of the nucleon from the tagged DIS cross section is inherently model dependent.  It will, therefore, be necessary to examine all reasonable models that are available (such as Regge models of baryon production and Dyson-Schwinger equation inspired models), or that may be available in the future, to evaluate the theoretical uncertainty associated with extracting meson structure functions from the tagged deep inelastic data. To clarify this model dependence, one can formally write, e.g.\ the measured semi-inclusive structure function of the leading proton, $F_{2}^{LP(4)}$, related to the measured cross-section as:
\begin{equation}
\frac{d^4\sigma (ep \rightarrow e^{'}X p^{'}) }{dxdQ^2dydt} = \frac{4\pi \alpha^2}{xQ^4}\left[ 1-y+\frac{y^2}{2\left(1+R\right)}\right] F^{LP(4)}_{2}(x,Q^2,y,t),
\label{eq:Xsect}
\end{equation}
$y=P\cdot q/P\cdot l$, where
$P(P')$ are the initial (scattered) proton four-vectors,
$l~ (l')$ are the initial (scattered) lepton vectors,
and $R$ is the ratio of the cross section for longitudinally and transversely polarised virtual photons. 
The measured cross section can be integrated over the proton momentum (which is effectively an integration over $t$~\cite{dis2}) to obtain the leading proton structure function $F_{2}^{LP(3)}$. The pion structure function $F_{2}^{\pi}$ can then be extracted from $F_{2}^{LP(3)}$ {\it using models}, such as the Regge
model of baryon production.  In the Regge model, the contribution of a specific exchange $i$ is defined by the product of its flux $f_{i}(y,t)$ and its structure function $F_{2}^{i}$ evaluated at $(x_{i},Q^2)$.  Thus,
\begin{equation}
F_{2}^{LP(3)}= \sum_{i}\left[\int_{t_0}^{t_{\rm min}}f_{i}(z,t)dt\right]F_{2}^{i}(x_{i},Q^2) ,
\label{eq:TSF}
\end{equation}
where $i$ is the pion, $\rho$-meson etc, and the $t$ corresponds to the range of $p_{T}$ analyzed. 

Neglecting uncertainties in the evaluation of $R = \sigma_L/\sigma_T$, which should be a small quantity, the extraction of the pion structure function will have to be corrected for a number of complications to the simple Sullivan picture. These include non-pion pole contributions, $\Delta$ and other $N^*$ resonances, absorptive effects, and uncertainties in the pion flux. For example, the cross section for leading charged pion production from the neutron is about twice reduced by absorptive corrections from other mesons. While these corrections can be large and one cannot extract the pion structure function without their inclusion, detailed calculations do exist~\cite{Boris}. Moreover, these corrections are minimised by measuring at the lowest $-t$ or tagged nucleon momentum possible from the reaction. This minimises the absorptive correction since, at lower momenta, the pion cloud is further from the bare nucleon. In addition, the low momentum ensures that the higher meson mass exchanges are suppressed by the energy denominator. Also, the charged pion exchange process has the advantage of less background from Pomeron and Reggeon processes~\cite{nikolaev}, and the charged pion cloud is expected to be roughly double the neutral pion cloud in the proton.

Having data from {\it both} protons and deuterons will provide essential cross-checks for the models used in the extraction of the pion structure function. In the Regge model it is assumed that the neutral pion, the Pomeron and the $f_2$ will be the leading contributions to the cross section from the proton while the charged pion, $\rho$ and $a_2$ are the leading contributions from the neutron~\cite{GolecBiernat:1997vy,Kazarinov:1975kw}.  However, Regge phenomenology also predicts that the flux of Reggeons with isospin one ($\rho$ and $a_2$) account for only $\approx 3$\% of the flux of Reggeons with isospin zero ($\omega$ and $f_2$)~\cite{GolecBiernat:1997vy}.  It also predicts that, for the neutron, the contributions from charged pion exchange are an order of magnitude larger than the contributions from the $\rho$ and $a_2$ \cite{Boris}. Pomeron exchange does not give a significant contribution since diffractive dissociation is believed to be $\approx 6$\% of the pion exchange contribution~\cite{Boris}.

The measured tagged cross sections and extracted tagged structure functions can be analyzed within a Regge framework where, assuming the dominance of a single Regge exchange, the differential cross section for recoil baryon production as a function of $z$ at fixed $t$ should be proportional to $z - n$, where $n = 2\alpha(t) - 1$, and $\alpha(t)$ specifies the Regge trajectory of the dominant exchange. For pion exchange, the $n$ averaged over the $t$ dependence is expected to be $n\approx -1$, while other Reggeons are expected to have $n > -1$. Thus, by comparing the $z$ dependence of the cross sections from proton and neutron (deuteron) scattering, it should be possible to determine the dominant exchange mechanism(s). Further, if the predictions for pion exchange are found to describe the data, the pion flux from the Regge model fits to hadron-hadron data may be safely used to extract the pion structure function.

The largest uncertainty in extracting the pion structure function, however, will likely arise from the (lack of) knowledge of the pion flux in the framework of the pion cloud model. One of the main issues is whether to use the $\pi NN$ form factor or the Reggeised form factor. The difference between these two methods can be as much as $20\%$~\cite{diff}. From the $N-N$ data the $\pi NN$ coupling constant is known to $5\%$~\cite{Stoks:1999bz}. If we assume that all corrections  can be performed with a 50\% uncertainty, and we assume a 20\% uncertainty in the pion flux factor, the overall theoretical, systematic uncertainty could approach $25\%$. The superior approach is to have a direct measurement of the pion flux factor by comparing with pionic Drell-Yan data. For example the pion structure function at $x=0.5$ has been measured using pionic Drell-Yan data to an accuracy of $5\%$ (see, e.g.\ \cite{dy1,dy2}).  New data from COMPASS should enable this possibility to be leveraged further and likely reduce projected uncertainties even more.

\begin{sidewaystable}[t!]
\scriptsize
\vspace{15cm}
\hspace{-2.5cm}
{\tabulinesep=0.3mm
\begin{tabu}{p{7cm} p{5.5cm} p{7.3cm}}
\hline
\hline
\textbf{Science Question}  & \textbf{Key Measurement{[}1{]}}    & \hspace{2ex}\textbf{Key Requirements{[}2{]}}  \\
\hline
\multirow{4}{\linewidth}{What are the quark and gluon energy contributions to the pion mass?}                                                   & \multirow{4}{0.75\linewidth}{Pion structure function data over a range of $x$ and $Q^2$.}                      & \multirow{4}{\linewidth}{\begin{tabular}[c]{p{7.3cm}}$\bullet$ Need to uniquely determine\\ \hspace{2ex}e + p $\rightarrow$ e’ + X + n (low -t)\\$\bullet$ CM energy range $\sim$10-100 GeV\\ $\bullet$ Charged- and neutral currents desirable\end{tabular}} \\
 &  & \\
 &  & \\
 &  & \vspace{1ex}\\
\hline
\multirow{2}{0.7\linewidth}{Is the pion full or empty of gluons as viewed at large $Q^2$?}                                                            & \multirow{2}{0.7\linewidth}{Pion structure function data at large $Q^2$.}                                       & \multirow{2}{\linewidth}{\begin{tabular}[c]{p{7cm}}$\bullet$ CM energy $\sim$100 GeV\\$\bullet$ Inclusive and open-charm detection\end{tabular}}  \\
&  & \vspace{1ex}\\
\hline
\multirow{3}{\linewidth}{What are the quark and gluon energy contributions to the kaon mass?}                                                   & \multirow{3}{0.75\linewidth}{Kaon structure function data over a range of $x$ and $Q^2$.}                      & \multirow{3}{\linewidth}{\begin{tabular}[c]{p{7cm}}$\bullet$ Need to uniquely determine $\Lambda$, $\Sigma^0$: \\\hspace{2ex}e + p  $\rightarrow$ e’ + X + $\Lambda /\Sigma^0$ (low -t)\\$\bullet$ CM energy range $\sim$10-100 GeV\end{tabular}} \\
 &  & \\
 &  & \vspace{1ex}\\
 \hline
\multirow{2}{\linewidth}{Are there more or less gluons in kaons than in pions as viewed at large Q$^2$?}                                           & \multirow{2}{0.7\linewidth}{Kaon structure function data at large $Q^2$.}                                        & \multirow{2}{\linewidth}{\begin{tabular}[c]{p{5cm}}$\bullet$ CM energy $\sim$100 GeV\\$\bullet$ Inclusive and open-charm detection\end{tabular}} \\
 &  & \vspace{1ex}\\
 \hline
\multirow{4}{\linewidth}{Can we get quantitative guidance on the emergent pion mass mechanism?}                                                    & \multirow{4}{0.6\linewidth}{Pion form factor data for $Q^2$ = 10-40 (GeV/c)$^2$.}                & \multirow{4}{\linewidth}{\begin{tabular}[c]{p{7.3cm}}$\bullet$ Need to uniquely determine exclusive process \\ \hspace{2ex}e + p  $\rightarrow$ e’ + $\pi^+$ + n (low -t)\\$\bullet$ e-p and e-d at similar energies\\$\bullet$ CM energy $\sim$10-75 GeV\end{tabular}}                                          \\
 &  & \\
 &  & \\
 &  & \vspace{1ex}\\
 \hline
\multirow{4}{\linewidth}{What is the size and range of interference between emergent-mass and the Higgs-mass mechanism?}                        & \multirow{4}{0.65\linewidth}{Kaon form factor data for $Q^2$ = 10-20 (GeV/c)$^2$.}               & \multirow{4}{\linewidth}{\begin{tabular}[c]{p{6cm}}$\bullet$ Need to uniquely determine exclusive process\\ \hspace{2ex}e + p  $\rightarrow$ e’ + K$^+$ + $\Lambda$ (low -t)\\$\bullet$  L/T separation at CM energy $\sim$10-20 GeV\\$\bullet$  e-p  $\Lambda /\Sigma^0$ ratios at CM energy $\sim$10-50 GeV\end{tabular}}            \\
 &  & \\
 &  & \\
 &  & \vspace{1ex}\\
 \hline
\multirow{3}{\linewidth}{What is the difference between the impacts of emergent- and Higgs-mass mechanisms on light-quark behaviour?}            & \multirow{3}{0.75\linewidth}{Behaviour of (valence) up quarks in pion and kaon at large $x$}                     & \multirow{3}{\linewidth}{\begin{tabular}[c]{p{7cm}}$\bullet$  CM energy $\sim$20 GeV (lowest CM energy to access\\ \hspace{2ex}large-x region)\\$\bullet$  Higher CM energy for range in $Q^2$ desirable\end{tabular}} \\
 &  & \\
 &  & \vspace{1ex}\\
 \hline
\multirow{3}{\linewidth}{What is the relationship between dynamically chiral symmetry breaking and confinement?}                     & \multirow{3}{0.8\linewidth}{Transverse-momentum dependent Fragmentation Functions of quarks into pions and kaons} & \multirow{3}{\linewidth}{\begin{tabular}[c]{p{7cm}}$\bullet$ Collider kinematics desirable (as compared to\\ \hspace{2ex}fixed-target kinematics)\\$\bullet$  CM energy range $\sim$20-140 GeV\end{tabular}}\\
& & \\
& & \vspace{1ex}\\
\hline
\specialrule{.1em}{.05em}{.05em}
\multicolumn{3}{l}{\textbf{More speculative observables}}                                                      \\
\hline
\specialrule{.1em}{.05em}{.05em}
\multirow{4}{0.75\linewidth}{What is the trace anomaly contribution to the pion mass?}                                                              & \multirow{4}{0.6\linewidth}{Elastic J/$\psi$ production at low W off the pion.}                                  & \multirow{4}{\linewidth}{\begin{tabular}[c]{p{6cm}}$\bullet$ Need to uniquely determine exclusive process \\ \hspace{2ex}e + p  $\rightarrow$ e’ + $\pi^+$ + J/$\Psi$ + n (low -t)\\$\bullet$  High luminosity (10$^{34+}$)\\$\bullet$  CM energy $\sim$70 GeV\end{tabular}}                                               \\
& & \\
& & \\
& & \vspace{1ex}\\
\hline
\multirow{4}{\linewidth}{Can we obtain tomographic snapshots of the pion in the transverse plane? What is the pressure distribution in a pion?} & \multirow{4}{\linewidth}{Measurement of DVCS off pion target as defined with Sullivan process}                 & \multirow{4}{\linewidth}{\begin{tabular}[c]{p{6cm}}$\bullet$ Need to uniquely determine exclusive process \\ \hspace{2ex}e + p  $\rightarrow$ e’ + $\pi^+$ + $\gamma$ + n (low -t)\\$\bullet$  High luminosity (10$^{34+}$)\\$\bullet$  CM energy $\sim$10-100 GeV\end{tabular}}                                             \\
& & \\
& & \\
& & \vspace{1ex}\\
\hline
\multirow{5}{0.75\linewidth}{Are transverse momentum distributions universal in pions and protons?}                                                 & \multirow{5}{\linewidth}{Hadron multiplicities in SIDIS off a pion target as defined with Sullivan process}    & \multirow{5}{\linewidth}{\begin{tabular}[c]{p{6.3cm}}$\bullet$ Need to uniquely determine scattered off pion: \\ \hspace{2ex}e + p $\rightarrow$ e + h + X + n (low -t)\\$\bullet$  High luminosity (10$^{34+}$)\\$\bullet$ e-p and e-d at similar energies desirable\\$\bullet$  CM energy $\sim$10-100 GeV\end{tabular}} \\
& & \\
& & \\
& & \\
& & \vspace{2ex}\\
\hline
\hline
\end{tabu}
}
\captionsetup{margin=1cm}
\caption{Science questions related to pion and kaon structure and understanding of the emergent-hadron mass mechanism possible accessible at an EIC, with the key measurement and some key requirements. Further requirements are addressed in the text.
\label{tab:EIC_science_questions_table}}
\end{sidewaystable}
\clearpage

\subsection{Kinematics of interest to address specific theory questions}

The science questions of interest summarised in  table~\ref{tab:EIC_science_questions_table} require a range of physics processes, spanning from (tagged) inclusive structure function measurements to (tagged) exclusive measurements such as required for a form factor determination or meson femtography. In general, a large range of CM energies is required to access a wide range in $x$ and $Q^2$, as relevant for pion (kaon) structure function measurements or hadron multiplicity measurements for a TMD programme. This has to be balanced against the requirement to uniquely determine the remnant nucleon (or $\Lambda$ or $\Sigma^0$) to ensure the scattering process occurs off a pion (kaon). The latter favours not-too-high CM energies to be able to uniquely determine the remnant $\Lambda$ (or $\Sigma^\circ$), both for missing-mass determination and to ensure their decays occur before detection.  In addition, there is need for both $ep$ and $ed$ measurements at similar CM energies to validate the reaction mechanism and understanding. This drives the ``typical'' CM energy range for pion and kaon structure function measurements to $\sim 10-100\,$GeV. Higher CM energies will increase the range in $Q^2$. On the other hand, lower CM energies are preferable for accessing the large-$x$ region to determine the behaviour of the valence quarks in pions (or kaons). In this case, the figure of merit, folding in all kinematic effects, is optimised at the lowest CM energy that provides a sufficiently large $Q^2$ for a clean interpretation of the data.

For pion (kaon) fragmentation processes, the collider kinematics greatly facilitate transverse-momentum dependent measurements at low scales ($p_T < 1$GeV), and the largest range in CM energy is required. For some processes the exact CM energy is not that important, so long as one obtains sufficient phase space for particle electroproduction to boost the experimental cross section. For instance, this is true for the (deep) exclusive $J/\Psi$ measurements to possibly constrain the QCD trace anomaly, and also for access to charged-current cross sections.

For pion (kaon) form factor determination, the situation is different. The standard method relies on Rosenbluth L/T-separated cross sections as the longitudinal (L) cross section enhances pion (kaon) pole sensitivity. Such measurements are best done at a relatively low CM energy range ($\sim 10-20\,$GeV). An alternate method to extract the pion form factor makes use of direct comparison of charged-pion cross sections for $ep$ and $ed$. This method may be applicable up to higher CM energies (and higher $Q^2$ values). Similarly, for the kaon form factor it may be possible to increase the $Q^2$ range (as compared to that from L/T-separated cross sections) from the $\Lambda/\Sigma^0$ cross section ratios. The latter requires further study, but is only possible at CM energies, $\sim 10-50$\,GeV, where the $\Lambda$ and $\Sigma^0$ may be cleanly isolated.

\subsection{Complementarity with other facilities}

The broad science programme to understand pion and kaon structure and the QCD mechanism behind the emergent hadron masses requires a strong interplay between experiment and theory, matching experimental prospects by new theoretical insights, rapid computational advances, and high-level QCD phenomenology. The EIC will play a key role in the experimental programme to chart in-pion and in-kaon distributions of, $\sl inter alia$, mass, charge, magnetisation and angular momentum. Nonetheless, to provide experimental measurements guiding theoretical understanding requires a coherent, worldwide effort.

The unique role of EIC is its access to pion and kaon structure over a wide range of large CM energies: $\sim 20-140\,$GeV. Jefferson Lab will provide, at its CM energy $\sim 5\,$GeV, tantalising data for the pion (kaon) form factor up to $Q^2 \sim 10 (5)\,$GeV$^2$, and measurements of the pion (kaon) structure functions at large-$x$ ($> 0.5$) through the Sullivan process.

AMBER will play a crucial role as they can uniquely provide pion (kaon) Drell-Yan measurements in the CM energy region $\sim 10-20\,$GeV \cite{Adams:2676885}. Some older pion and kaon Drell-Yan measurements exist, but for the kaon this is limited to less than 10 data points worldwide, so these measurements are essential for a global effort aimed at pion structure function measurements (also providing a handle on determination of the so-called ``pion flux'' for EIC Sullivan process measurements) and a \emph{sine qua non} for any kaon structure function data map. The AMBER data in themselves will already give new fundamental insights into the emergent hadron mass mechanism.

An Electron-Ion Collider in China (EicC) is under consideration with a similar CM energy range as AMBER ($\sim 10-20 \,$GeV) and bridging the energy range from Jefferson Lab to EIC \cite{Chen:2020ijn}. EicC on its own, and even more in combination with AMBER, can provide good access to the region $x \gtrsim $ 0.01 for pion, and especially kaon, structure function determination and the impact on emergent hadron mass mechanisms on valence quark and gluon structure. In addition, EicC can extend the Rosenbluth L/T-separated cross section technique beyond Jefferson Lab and access pion and kaon form factors to higher $Q^2$ values, perhaps by a factor of $2-4$.

The EIC, with its larger CM energy range, will clearly have the final word on the contributions of gluons in pions and kaons as compared to protons.  It will finally settle questions relating to the gluon content of Nature's NG modes when they are viewed with very high resolution, and vastly extend the $(x,Q^2)$ range of pion and kaon charts and meson structure knowledge.

% I've tried to make the beam energies from here onward consistent in being of the format X$\times$Y. I also altered the units to use non line breaking spaces - SK 05/10/20
%%%%%%%%%%%%%%%%%%%%%%%%%%%%%%%%%%%%%%%%%%%%%%%%%%%%%%%%%%%%
%Kinematics section was merged with Detector Requirements
%%%%%%%%%%%%%%%%%%%%%%%%%%%%%%%%%%%%%%%%%%%%%%%%%%%%%%%%%%%%

%============= D E T E C T O R   R E Q U I R M E N T S ============
% As with the previous section, tried to make the X$\times$Y format of the beam on beam descriptions consistent - here that meant dropping the units. I've also added non line breaking spaces between value and unit - SK 05/10/20

% SJDK - Removed names from title, commented below as a reminder, delete once no longer needed
%  ({\underline{Yulia}}, Dmitry, Arun, Carlos, Richard)
\section{Kinematic coverage and detector requirements}

\subsection{Far-forward area setup.}
\begin{figure}[!ht]
  \centering
% The original picture is located here
% \includegraphics[width=0.8\textwidth]{reco-Meson-FF-setup-old.png}

% This one is 01/18/2020 edit. The source is located here:
% https://gitlab.com/eic/escalate/plugins/meson_structure
  \includegraphics[width=0.8\textwidth]{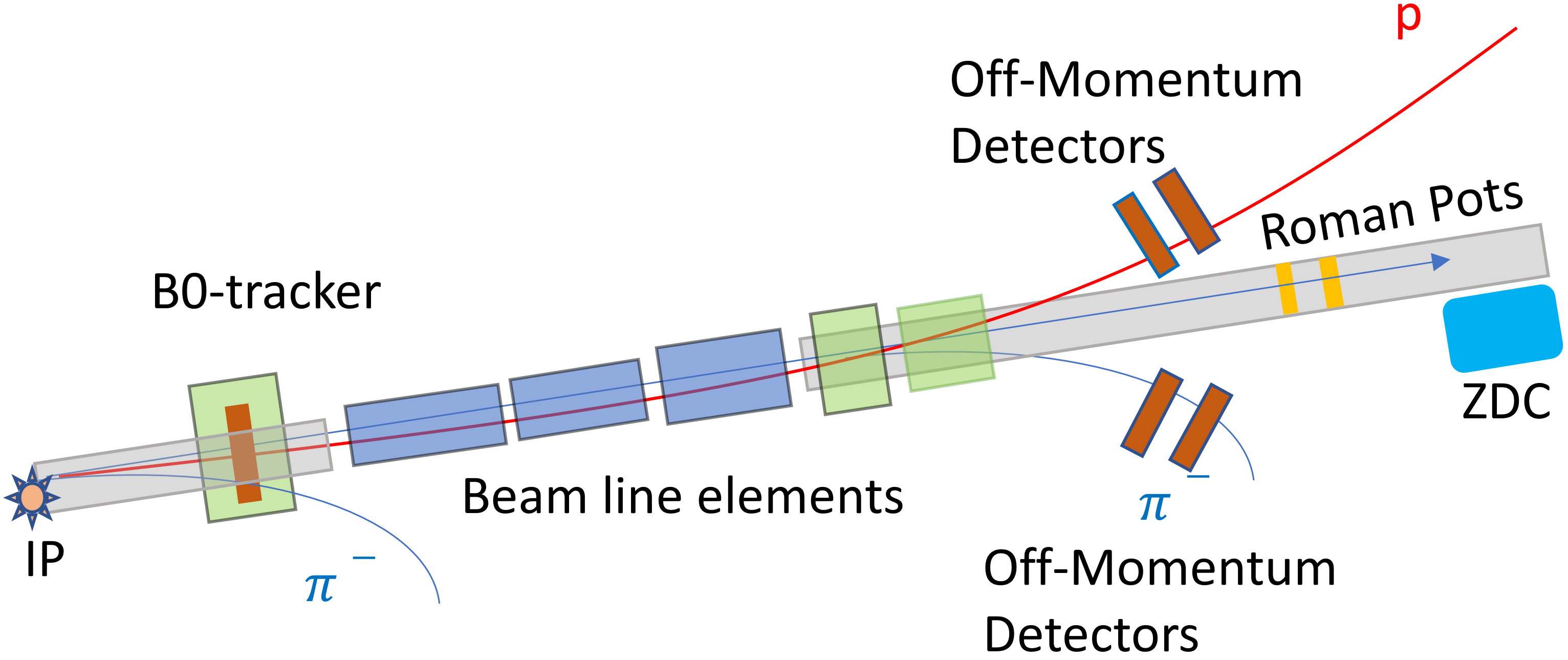}

  \caption{A sketch of the integrated beam line and detector setup in the Far-Forward area, along the direction of the proton/ion beam. The sketch is not to scale. The initial B0-tracker is integrated in the warm area of a combined electron-proton/ion beam magnet. Then a set of beam line magnetic elements follows that is integrated in one cryostat. This is followed by off-momentum detectors that capture the charged-particle decay products, roman pots that capture far-forward going protons with nearly the energy of the proton/ion beams, and the Zero-Degree Calorimeter to capture far-forward-going neutral particles.
  \label{fig:far-forw}}
\end{figure}

The far-forward EIC detector is described in detail in the EIC Yellow Report \cite{EIC-Yellow-Report}.
%  restore if/when have clear citation details ~\cite{EIC-Yellow-Report}.
Figure~\ref{fig:far-forw} shows the main elements of this far-forward region.  For the detection of particles of relevance to meson structure studies, all sub-components of the far-forward area play an important role: detection in the B0 area, detection of decay products with off-momentum detectors, and detection of forward-going protons and neutrons with the Roman Pots and the Zero-Degree Calorimeter (ZDC).

\subsection{e p $\rightarrow $ e$'$ + X + n.}

\begin{figure}[!ht]
\begin{center}
\includegraphics[width=4.5in,trim={25mm 0mm 0mm 0mm}]{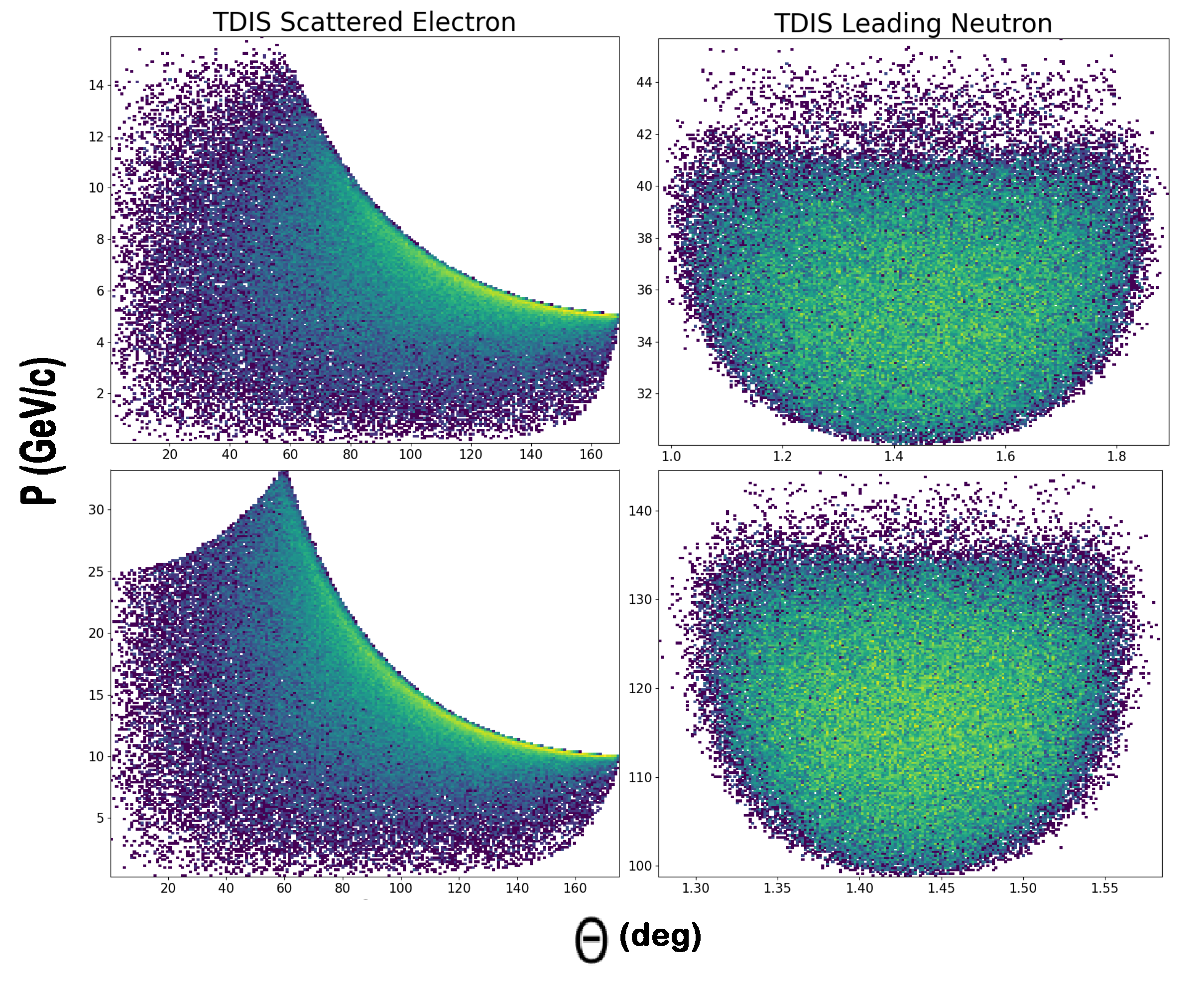}
\caption{\label{fig:TDIS_kin_scat_plot}
A comparison of the scattered electron (left) and leading neutron (right) kinematics for two energy settings - 10$\times$135 (bottom) and 5$\times$41 (top). The momentum, P, and angle, $\theta$, are defined in the lab frame. In both cases, the scattered electrons are within the acceptance of the central detector and the leading neutrons are at small forward angles and carry most of the proton beam energy after the scattering process.
}
\end{center}
\end{figure}

%%%%%%%%%%%%%%%%%%%%%%%%%%%%%%%
%Merged Section from Kinematics
%%%%%%%%%%%%%%%%%%%%%%%%%%%%%%%
The initial pion structure studies were conducted at the highest energy of 18$\times$275 (corresponding to the electron and proton beam energy, respectively, both in GeV) to maximise the kinematics coverage. However, to improve access to the high $x_{\pi}$ region, alternate lower beam energies 10$\times$135 and 5$\times$41 were also selected. These lower beam energies allow access to this high $x_{\pi}$ regime over a wider range of $Q^2$. For a comparison, the 18$\times$275 energies allow access to high $x_{\pi}$ data over a $Q^2$ range of $\sim 200-1000\,$GeV$^{2}$, while with the 10$\times$135 energies that range was increased to $\sim 30-1000\,$GeV$^{2}$, and with the 5$\times$41 energies to $\sim 5-1000\,$GeV$^{2}$. The lower-energy combination of 5$\times$41 is even more beneficial for tagging kaon structure by allowing detection of the leading $\Lambda$ events.
%, as will be outlined later.

The kinematics for the more advantageous lower energy settings, 10$\times$135 and 5$\times$41, are shown in  figure~\ref{fig:TDIS_kin_scat_plot}. While the scattered electrons are within the acceptance of the central detector, the leading neutrons for these two energy settings are at a very small forward angle while carrying nearly all of the proton beam momentum. These leading neutrons will be detected by the ZDC.

\begin{figure}[!h]
  \centering

\includegraphics[width=0.325\textwidth,trim={5mm 3mm 0mm 10mm}, clip]{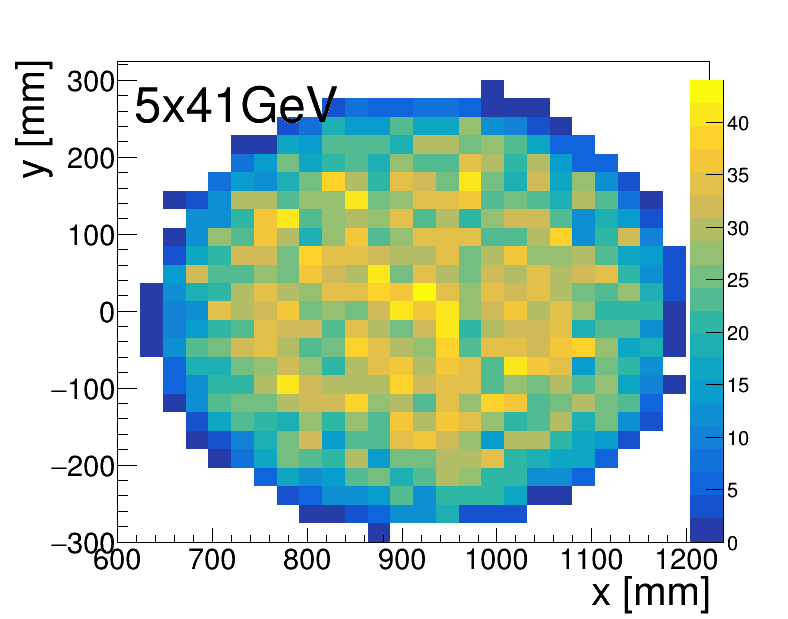}
\includegraphics[width=0.325\textwidth,trim={5mm 3mm 0mm 10mm}, clip]{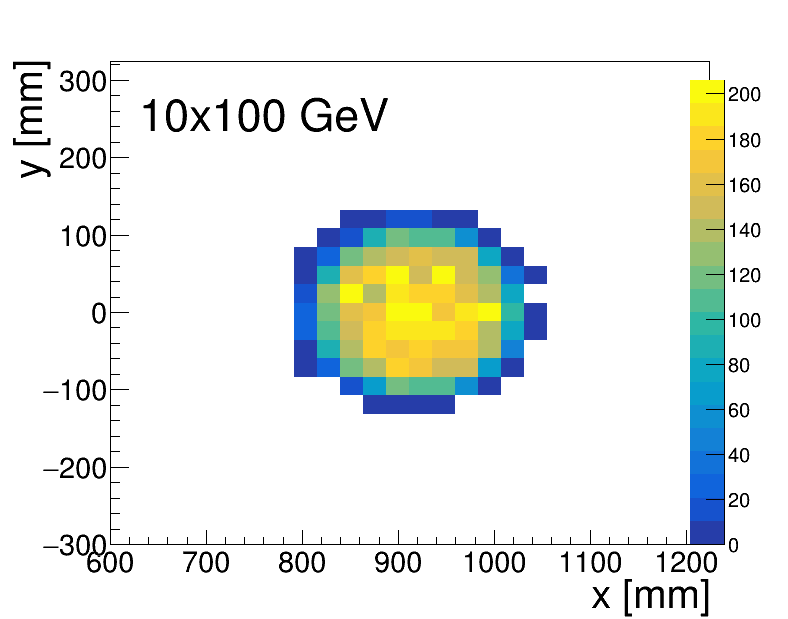}
\includegraphics[width=0.325\textwidth,trim={5mm 3mm 0mm 5mm}, clip]{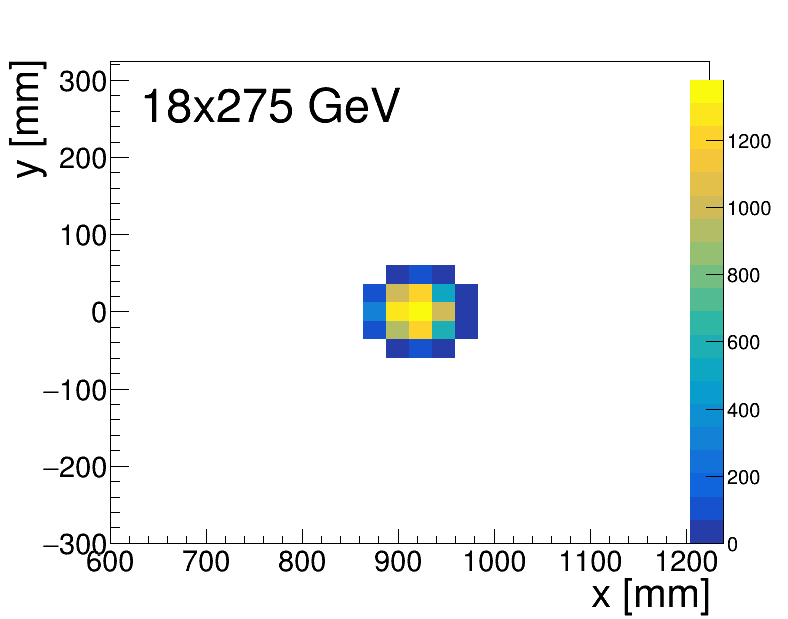}
%  \includegraphics[width=0.3\textwidth]{pict/reco-n-5x41-lowres.png}
%\includegraphics[width=0.3\textwidth]{pict/reco-n-10x100-lowres.png}
%\includegraphics[width=0.3\textwidth]{pict/reco-n-18x275-lowres.png}
%  \caption{Acceptance plot for neutrons in 60$\times$260~cm$^2$  ZDC, with a spatial resolution of 3~cm, for different energy settings. }
%  \label{fig:n-ZDC-low-res}
%\end{figure}

%\begin{figure}[!ht]
%  \centering
  \includegraphics[width=0.325\textwidth,trim={5mm 3mm 0mm 10mm}, clip]{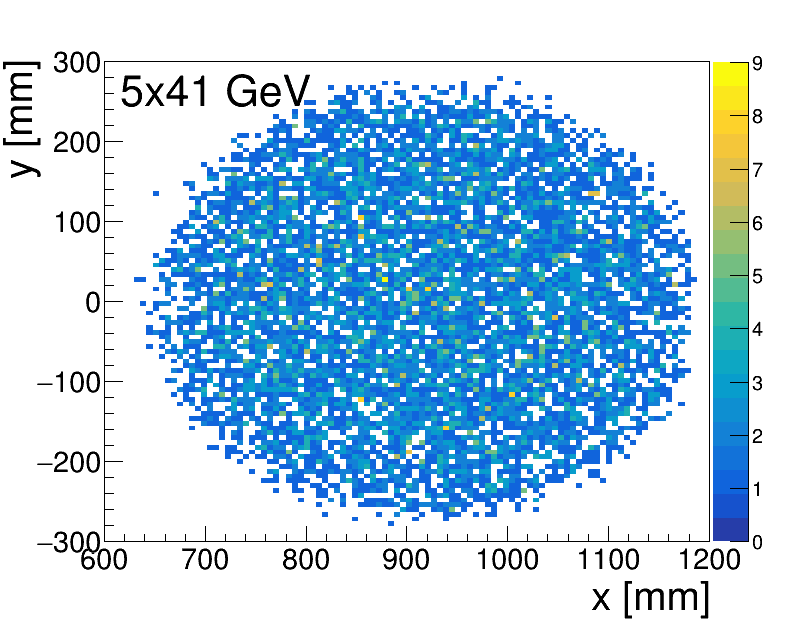}
  \includegraphics[width=0.325\textwidth,trim={5mm 3mm 0mm 10mm}, clip]{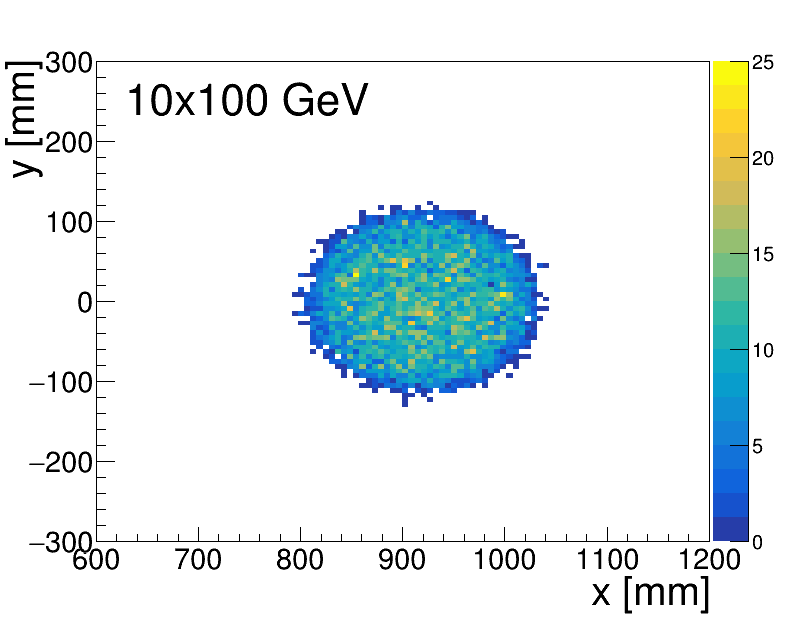}
  \includegraphics[width=0.325\textwidth,trim={5mm 3mm 0mm 5mm}, clip]{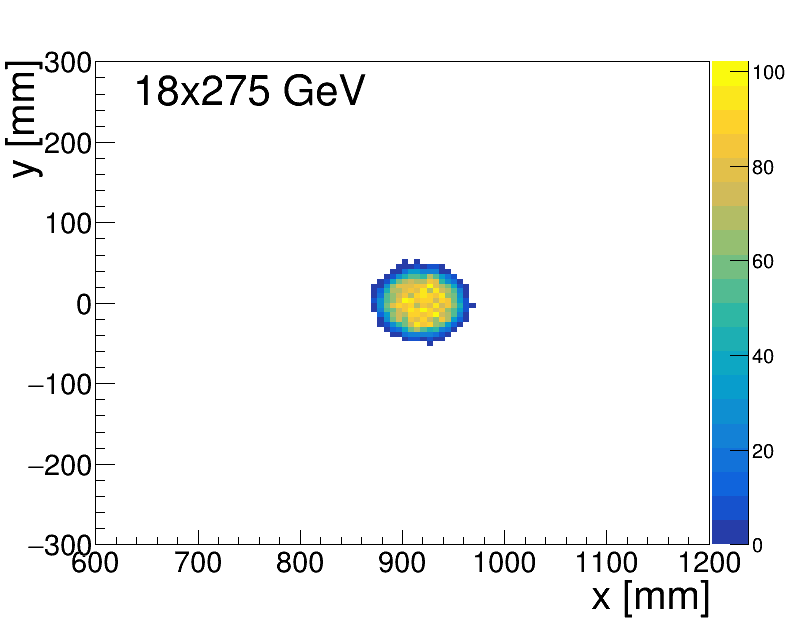}
  \caption{Acceptance plot for neutrons in the 60$\times$60~cm$^2$ ZDC, with a low spatial resolution of 3~cm (upper panels) and with a high spatial resolution of 0.6~cm (lower panels), for different energy settings, from left to right, of 5$\times$41, 10$\times$100, and 18$\times$275. The acceptance plot for 5$\times$100 would be similar as shown for 10$\times$100. The lower proton (ion) energies set the requirement for the size of the ZDC, whereas the higher proton (ion) energies drive the spatial resolution requirement.
  \label{fig:n-ZDC-res}}
\end{figure}

%Figs.~\ref{fig:n-ZDC-low-res} and ~\ref{fig:n-ZDC-high-res}
Figure~\ref{fig:n-ZDC-res} shows the acceptance plots for neutrons in the ZDC for all three energy settings. As one can see, the spatial resolution of the ZDC plays an important role for the highest energy setting, since it is directly related to the measurements of $p_T$ or $t$. For the lowest energy setting, the total acceptance coverage of the ZDC is important. This sets a requirement for the total size of ZDC to be a minimum of 60$\times$60~cm$^2$. Such a configuration of the ZDC provides nearly 100\% neutron detection efficiency for this channel.

%===========================================================

%\clearpage

%===========================================================
\subsection{$\Lambda$ tagging}

%%%%%%%%%%%%%%%%%%%%%%%%%%%%%%%
%Merged Section from Kinematics
%%%%%%%%%%%%%%%%%%%%%%%%%%%%%%%
For the case of a leading $\Lambda$ event, to elastic or DIS scattering from a kaon, both $\Lambda$ decay products must be detected at small forward angles owing to the nature of two-body decay kinematics. The detection of these decay products requires high-resolution and granularity because of the small angle of separation of decay products.

Detection of the decay channel $\Lambda\rightarrow n+ \pi^0$ is feasible, but will require electromagnetic calorimetry before the ZDC to distinguish the neutron and the two photons coming from $\pi^0$ decay.
Detection of the other decay channel, $\Lambda\rightarrow p+\pi^-$, poses a more challenging measurement owing to its requirement of additional charged-particle trackers or a veto trigger on the path to ZDC.

The reconstruction of the $\Lambda$ event in the far-forward detection area is one of the most challenging tasks. This comes mainly from the fact that these leading $\Lambda$s have energy close to the initial beam energy, and thus their decay lengths can be tens-of-metres along the Z-axis (or beam line). This complicates detection of the decay products; hence, the final $\Lambda$ mass reconstruction.

\begin{figure*}[!htb]
    \centering
    \includegraphics[width=0.7\textwidth]{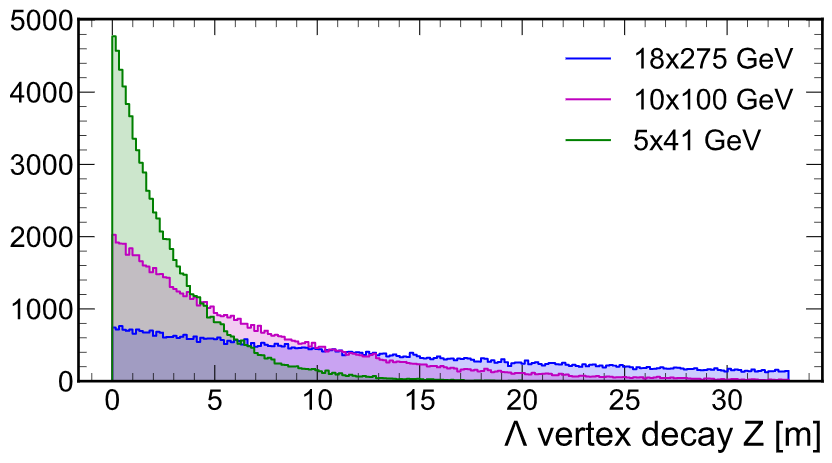}
    \caption[]{
        The $\Lambda$-decay spectrum along the beam line for different beam energies.
        \label{fig:lam_z_decay}
    }
\end{figure*}

Figure~\ref{fig:lam_z_decay} illustrates this further, showing the Z-coordinate of where the $\Lambda$-decay occurs for different beam energies. For the lower beam energy settings (5$\times$41) most $\Lambda$ decays are within the central detector region, but at the higher proton (ion) beam energies the $\Lambda$ decays happen more in the forward-detection area, with tails of the decay process reaching to near the ZDC location. Table~\ref{tab:lambda_decay} shows the percentage of decayed $\Lambda$ for different energies and different Z ranges: $Z_{vtx} < 5$\,m, 5\,m $< Z_{vtx} < 30$\,m and $Z_{vtx} > 30$\,m.

\begin{table}[hbt]
    %    \begin{tabular}  {  |l| c| c | c| c| c| c |   } % original
    \begin{center}
    \begin{tabular}  {  l c c c c c c    }
   \hline
   \hline
    $E_\text{beams}$ & & $Z_{vtx}<5$\,m & & 5\,m\,$< Z_{vtx}<30$\,m  & &$Z_{vtx}>30$\,m\\
    \hline
    5$\times$41   & &83.0\% & & 16.6\% & & 0.4\% \\
    10$\times$100  & &52.1\% & & 46.7\% & & 1.2\% \\
    18$\times$275 & &23.3\% & & 56.2\% & & 20.5\% \\
    \hline
    \hline
    \end{tabular}
    \end{center}
    \caption{ Percentage of decayed $\Lambda$'s in different detection ranges. }
    \label{tab:lambda_decay}
\end{table}

To study the possibility of $\Lambda$ mass reconstruction further, both main decay modes have been examined: $\Lambda \rightarrow p + \pi ^{-}$, with a branching ratio of $63.9\%$, and $\Lambda \rightarrow n + \pi ^{0}$, with a branching ratio of $35.8\%$. Both channels can be cleanly separated by the different charge of the final-state particles, and thus by the different detector components that will play a role in their detection.

%===========================================================

\subsubsection{$\Lambda \rightarrow p + \pi ^{-}$.}

For this process, there are only charged particles in the final state.  Therefore, one must rely on sub-components along the far-forward area, such as the B0 tracker, the Off-Momentum trackers, and Roman Pots for detection and reconstruction of the decay products.

\begin{figure}[!htbp]
\begin{tabular}{lll}
\parbox[c]{0.31\textwidth}{\includegraphics[width=0.31\textwidth,trim={3mm 0 5mm 10mm}, clip]{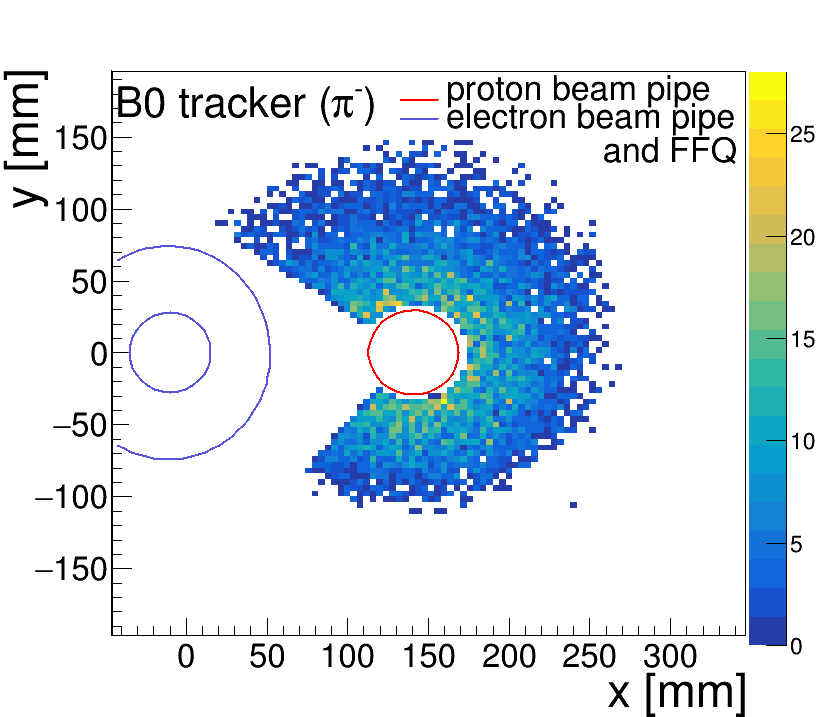}}
&
\parbox[c]{0.31\textwidth}{\includegraphics[width=0.31\textwidth,trim={3mm 0 5mm 10mm}, clip]{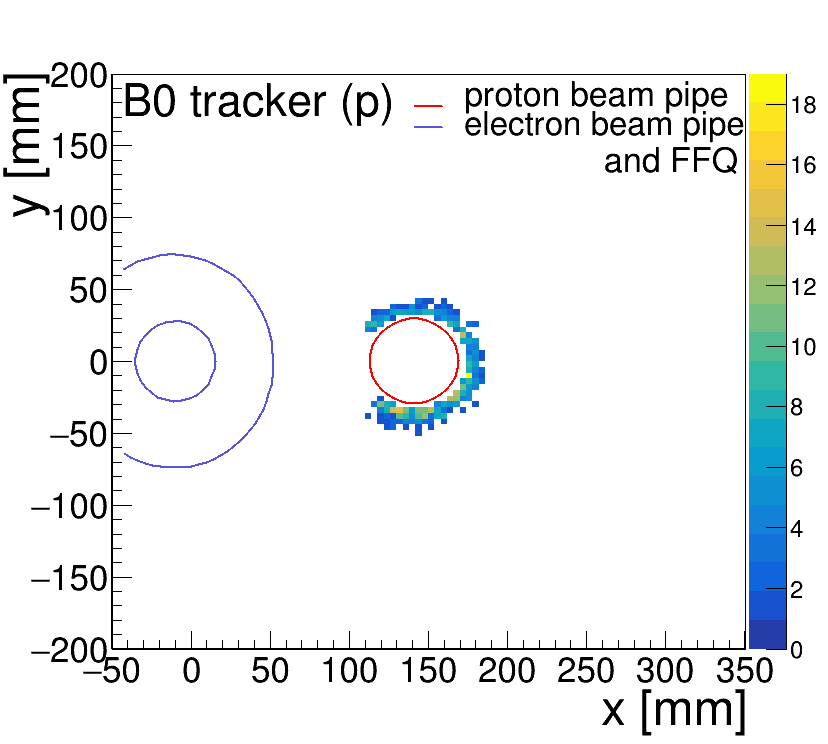}}
&
\parbox[c]{0.31\textwidth}{\includegraphics[width=0.31\textwidth,trim={3mm 0 5mm 10mm}, clip]{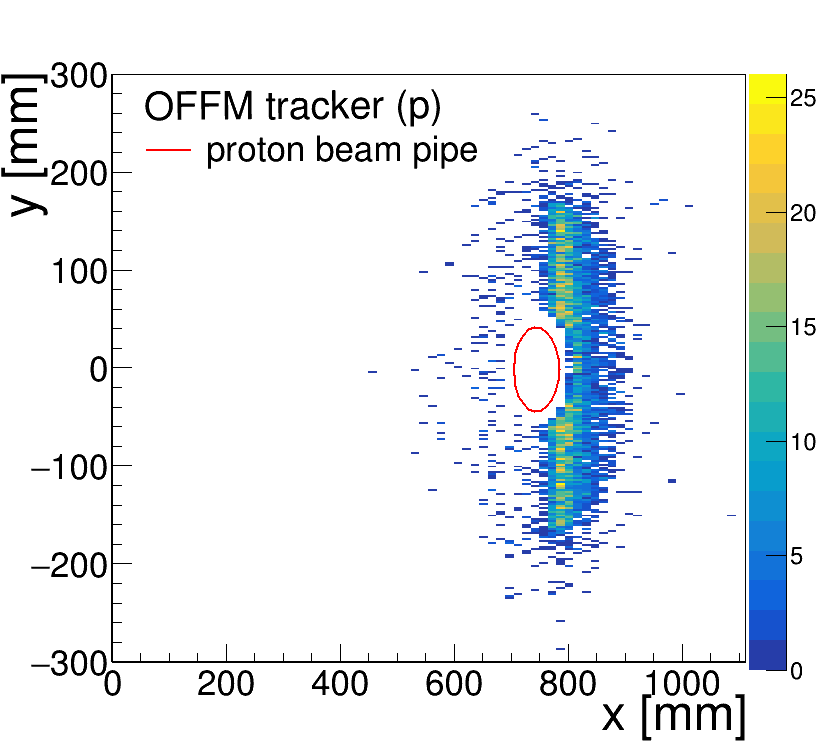}}
\\[-.2ex]
(a) & (b) & (c)
\end{tabular}
\caption[]{Occupancy plots for energy setting 5$\times$41 (a) for $\pi ^-$ in the B0 tracker, (b) for protons in the B0 tracker and (c) for protons in the Off-Momentum detectors. The red circle shows the beam pipe position and the blue circle shows the electron final-focus quadrupole (FFQ) aperture inside the B0 dipole.
\label{fig:occup_L_p_41}
}
\end{figure}

\begin{figure*}[htbp]
  \includegraphics[width=0.95\textwidth]{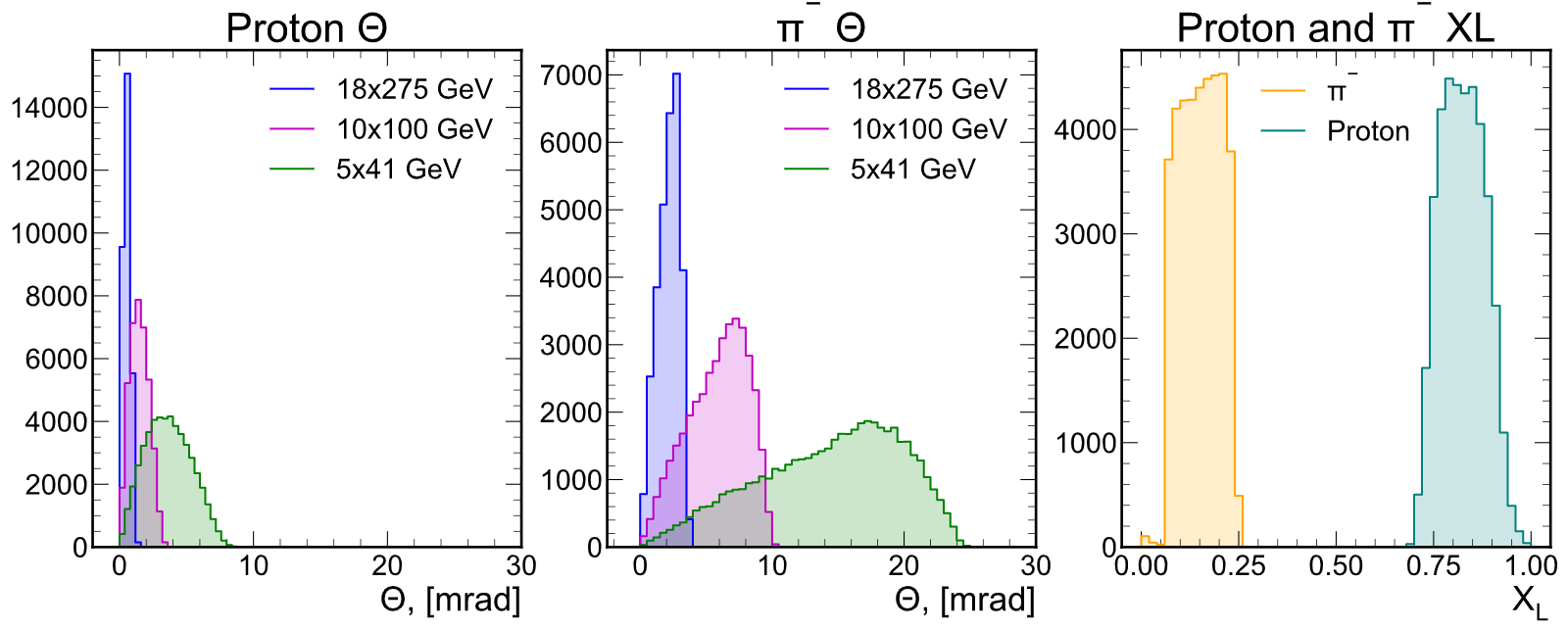}
\caption[]{Theta and $x_L$ distributions for detected decay products of $\Lambda$ particles for different beam energy combinations. Proton theta (left), $\pi^{-}$ theta (center),  proton and $\pi^-$ $X_{L}$ (right).

\label{fig:XL}
}
\end{figure*}

As an example, occupancy plots for the beam-energy setting of 5$\times$41 are shown in figure~\ref{fig:occup_L_p_41}. Since this is the lowest beam-energy setting, most of the $\Lambda$s would decay in the first metre (before the B0 magnet), and the $\Lambda$ decay products are expected to have low momenta. Therefore, as expected, protons coming from the $\Lambda$ decays will mostly be detected, owing to their lower rigidity, in the off-momentum detectors (c) and partially in a B0 tracker (b). While for pions, the tracker inside the B0 dipole will be the only detecting element (a). As one can also see from this figure, the proton-beam-pipe aperture inside the B0-dipole plays an important role and sets the detection efficiency for pions, as well as the azimuthal angle $\phi$-coverage of the detecting elements around the proton beam-pipe. Further information on the distributions for detected decay products at these lower beam energies of 5$\times$41 are given in figure~\ref{fig:XL}.

%=================

%===================

\begin{figure*}[htbp]
\begin{tabular}{lll}
\parbox[c]{0.31\textwidth}{\includegraphics[width=0.31\textwidth,trim={3mm 0mm 5mm 10mm}, clip]{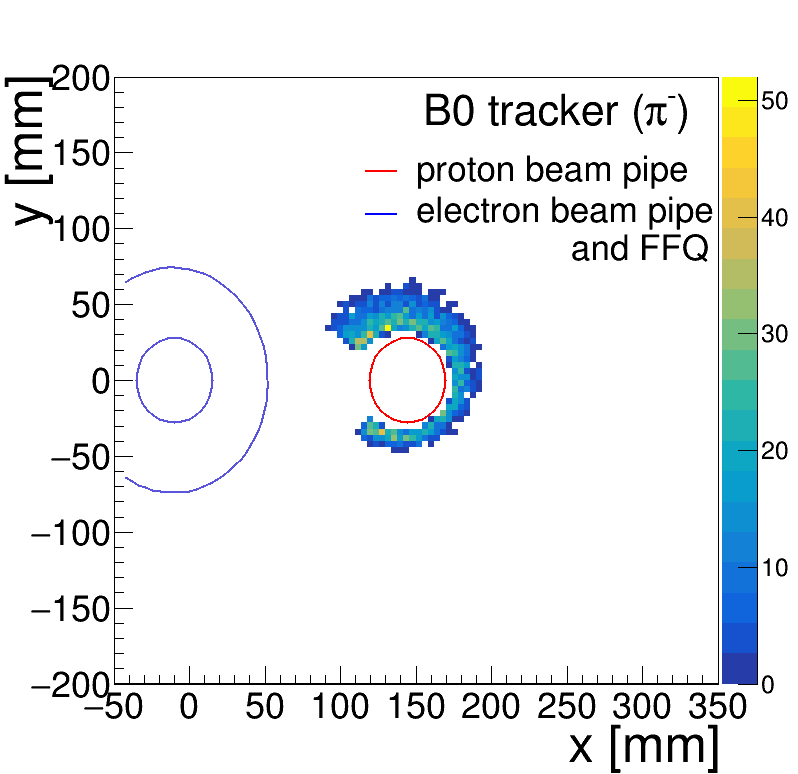}}
&
\parbox[c]{0.31\textwidth}{\includegraphics[width=0.31\textwidth,trim={3mm 0mm 5mm 10mm}, clip]{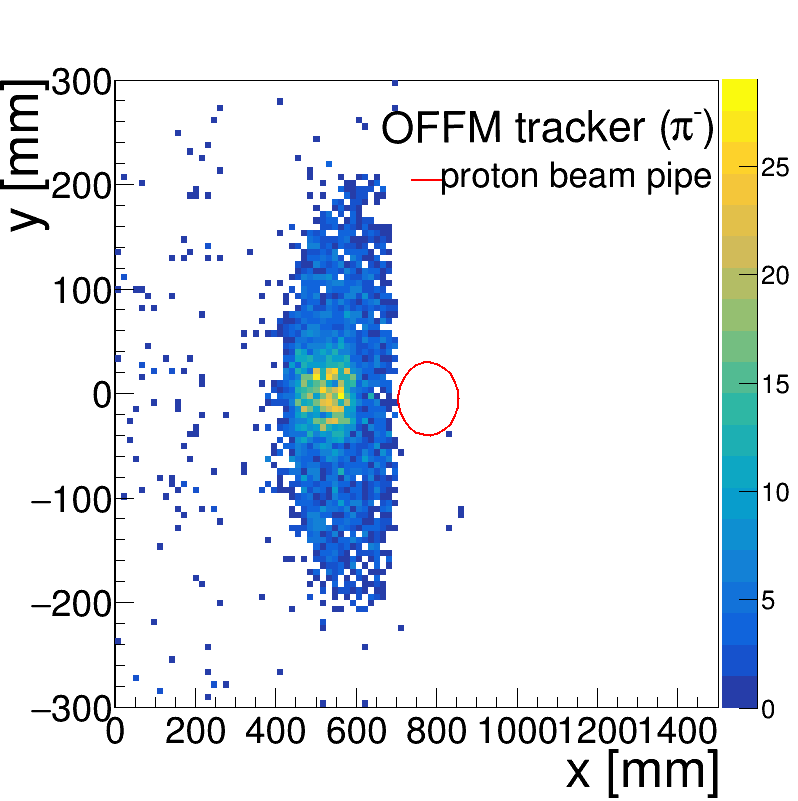}}
&
\parbox[c]{0.31\textwidth}{\includegraphics[width=0.31\textwidth,trim={3mm 0mm 5mm 10mm}, clip]{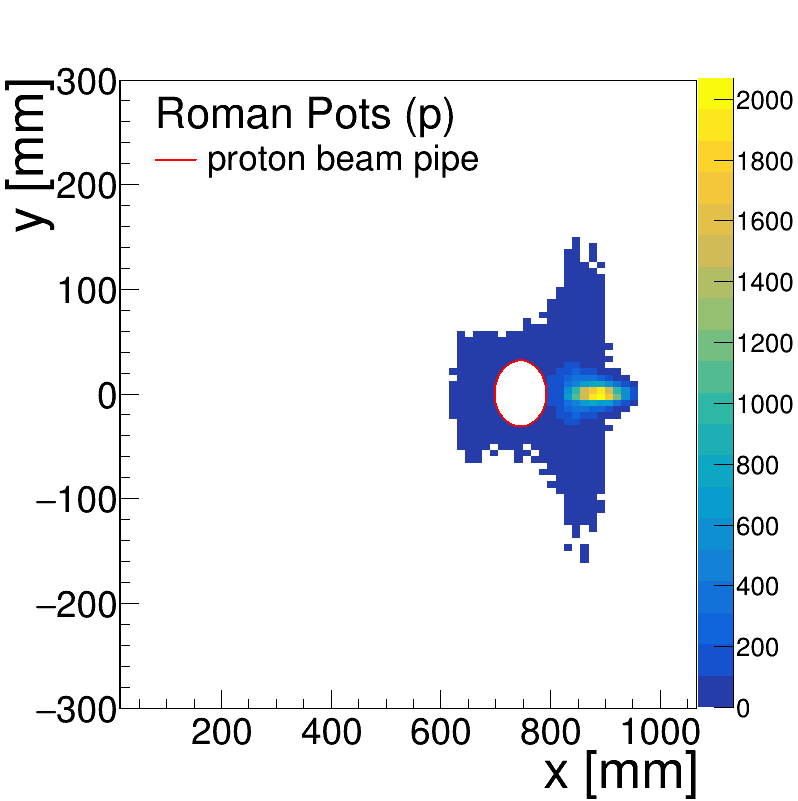}}
\\[-.2ex]
(a) & (b) & (c)
\end{tabular}
\caption[]{Occupancy plots for energy setting 10$\times$100 (a) for $\pi ^-$ in the B0 tracker, (b) for $\pi^-$ in the Off-Momentum tracker, and (c) for protons in the Roman Pots detectors. The red circle shows the beam pipe position and the blue circle shows the electron FFQ aperture inside the B0 dipole.
\label{fig:occup_L_p_100}
}
\end{figure*}

For the higher beam-energy settings, e.g.\ 10$\times$100, the protons are to be detected in the roman-pots (and partially in Off-Momentum detectors), see figure~\ref{fig:occup_L_p_100}. Pions originating from a $\Lambda$-decay with $Z_{vxt} < 4\,{\rm m}$ will only partially be detected in the B0-area, while most of them will go undetected  through the proton beam pipe. Pions with higher momentum and lower angles ($p_t$ or theta) can pass through the bores of the FFQs and be detected in the Off-Momentum detectors. Their detection represents the denser (light) area of detection in the Off-Momentum detectors (figure~\ref{fig:occup_L_p_100}(b)). Note that owing to the negative charge of the pions, they will experience an opposite bending in dipoles, as compared to protons (compare with the protons in the Off-Momentum detectors on figure~\ref{fig:occup_L_p_41}(b)). Therefore, in order to detect the $\Lambda$-decays in this channel, the Off-Momentum detectors need to provide a full azimuthal coverage, to establish a proper detection for the negatively-charged particles.

\begin{figure}[!ht]
 \centering
\includegraphics[width=0.49\textwidth, trim={0 0 0 10mm}, clip]{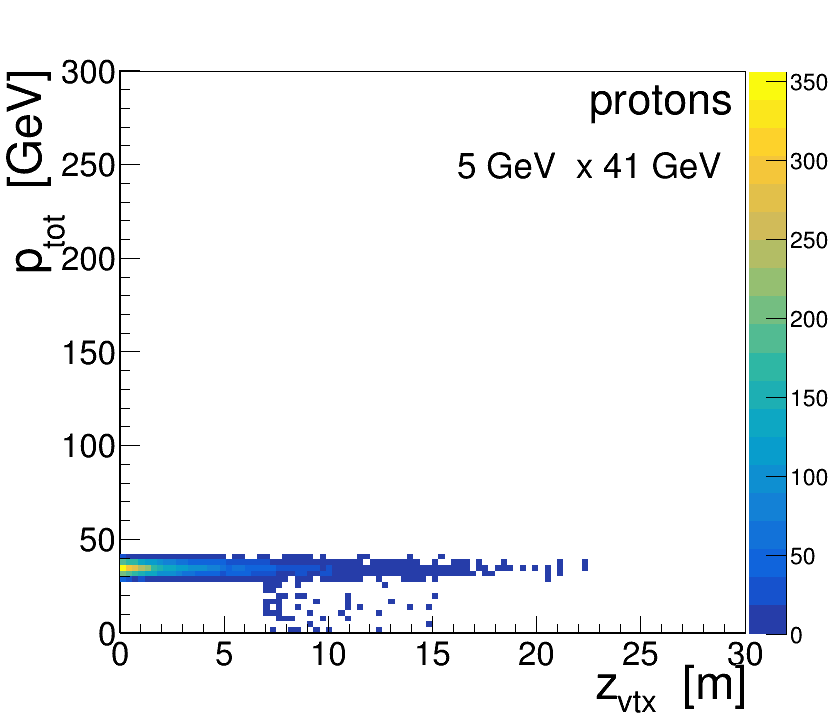}
\includegraphics[width=0.49\textwidth, trim={0 0 0 10mm}, clip]{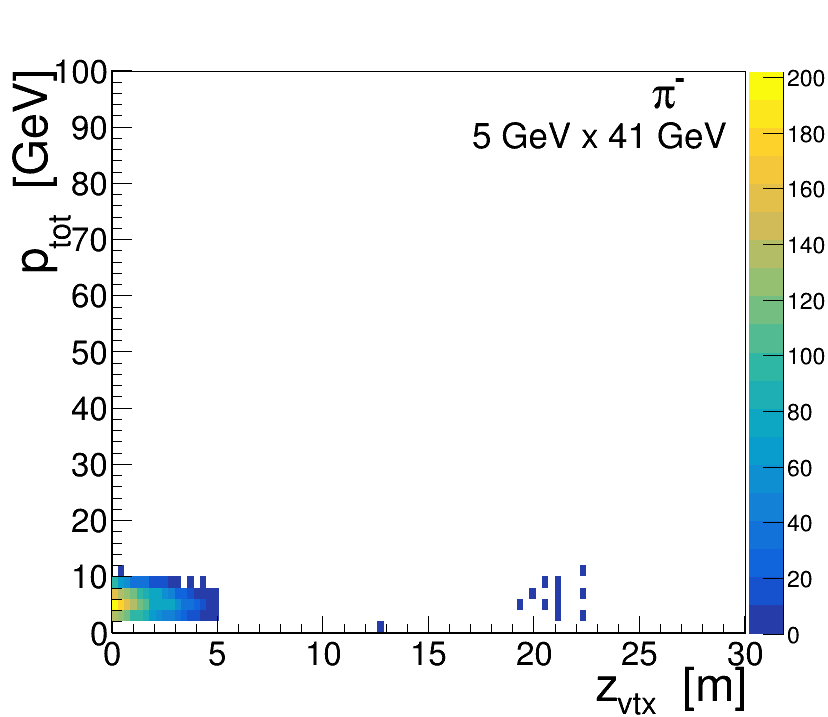}
\includegraphics[width=0.49\textwidth, trim={0 0 0 10mm}, clip]{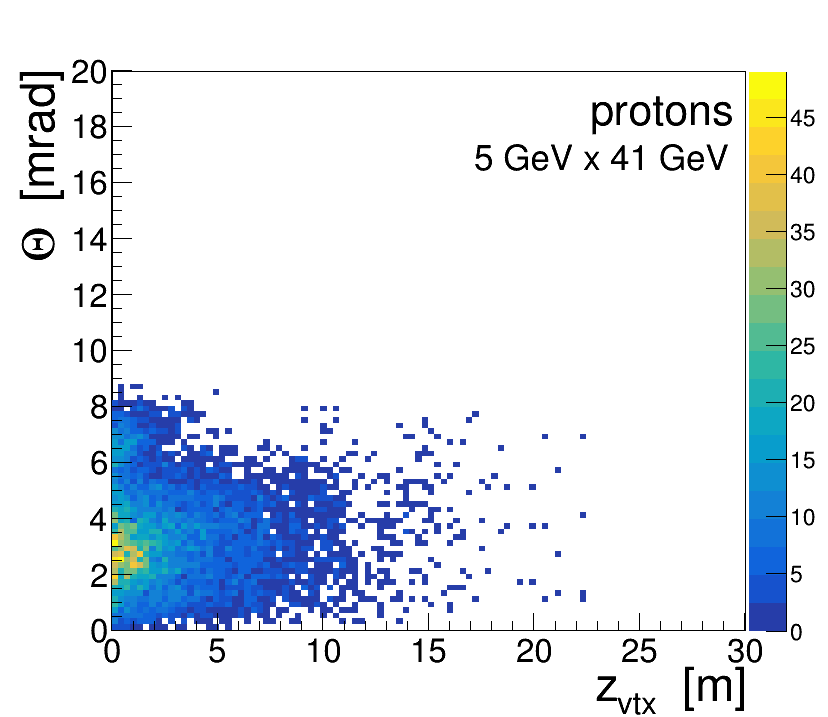}
\includegraphics[width=0.49\textwidth, trim={0 0 0 10mm}, clip]{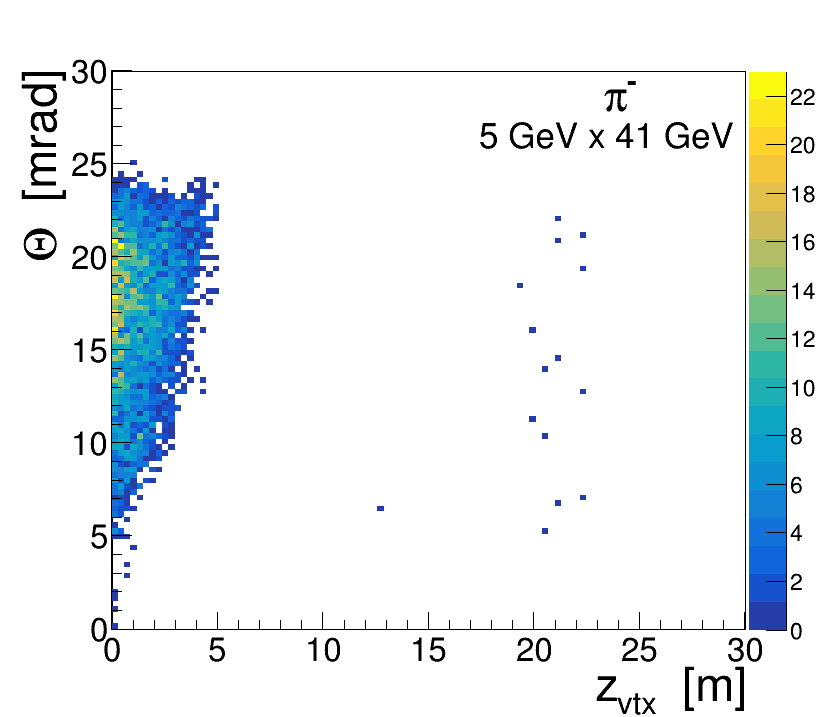}
% \begin{tabular}{lll}
%\parbox[c]{0.49\textwidth}{\includegraphics[width=0.49\textwidth]{pict/Zvtx_ptot_prot_41.png}}
%\parbox[c]{0.49\textwidth}{\includegraphics[width=0.49\textwidth]{pict/Zvtx_ptot_pi_41.png}}
%\\
%\parbox[c]{0.49\textwidth}{\includegraphics[width=0.49\textwidth]{pict/Zvtx_theta_prot_41.png}}
%\parbox[c]{0.49\textwidth}{\includegraphics[width=0.49\textwidth]{pict/Zvtx_theta_pi_41.png}}
%\end{tabular}
%\includegraphics[width=0.7\textwidth]{pict/reco-Lambda_theta_zvtx_5x41.png}
\caption{Momentum (top) and angular (bottom) distributions of protons (left) and $\pi^-$ (right) from the $\Lambda \rightarrow p + \pi^-$ decay at beam-energy setting 5$\times$41, as registered in the far-forward detectors as a function of their origination (the decay vertex).
 \label{fig:L-p_Zvtx_41}}
\end{figure}

For the 5$\times$41 beam-energy combination, figure~\ref{fig:L-p_Zvtx_41} shows the momentum (top panels) and angular (bottom panels) distributions of protons (left panels) and pions (right panels) from $\Lambda$-decay as a function of distance from the $\Lambda$-decay point, as detected in one of the beam line sub-detectors. This in turn illustrates which of the sub-detectors along the beam line detects the decay products. The protons carry most of the initial proton beam momentum and extend over the far-forward direction, with angles less than 8~mr. On the other hand, as one can clearly see from the high density of hits, the $\Lambda$-reconstruction efficiency will mainly depend on the efficiency for the detection of pions in the B0 area, with angles in the 5-25~mr range.

%\begin{figure}[!ht]
%  \centering
 % \includegraphics[width=0.8\textwidth]{pict/reco-L-p-Zvtx_41.png}
%  \includegraphics[width=0.7\textwidth]{pict/reco-Lambda_theta_zvtx_5x41.png}  \caption{Momentum (top) and angular (bottom) distributions of protons (left) and $\pi^-$ (right) from the $\Lambda \rightarrow p + \pi^-$ decay at beam energy setting 5$\times$41, as registered in the far-forward detectors as a function of their origination (the decay vertex).  }
%  \label{fig:L-p_Zvtx_41}
%\end{figure}

\begin{figure}[!ht]
\centering
\includegraphics[width=0.49\textwidth, trim={0 0 0 10mm}, clip]{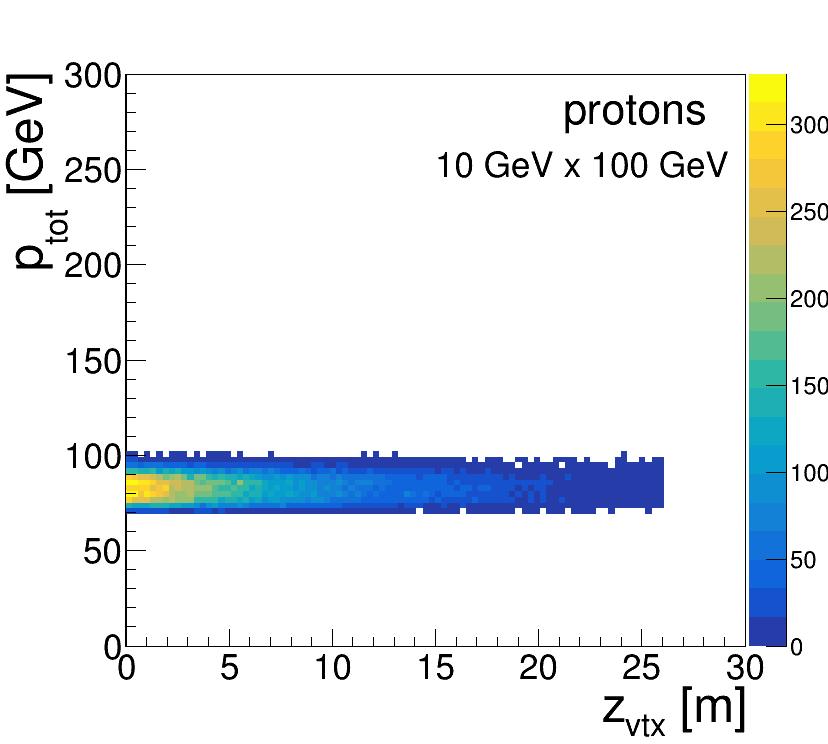}
\includegraphics[width=0.49\textwidth, trim={0 0 0 10mm}, clip]{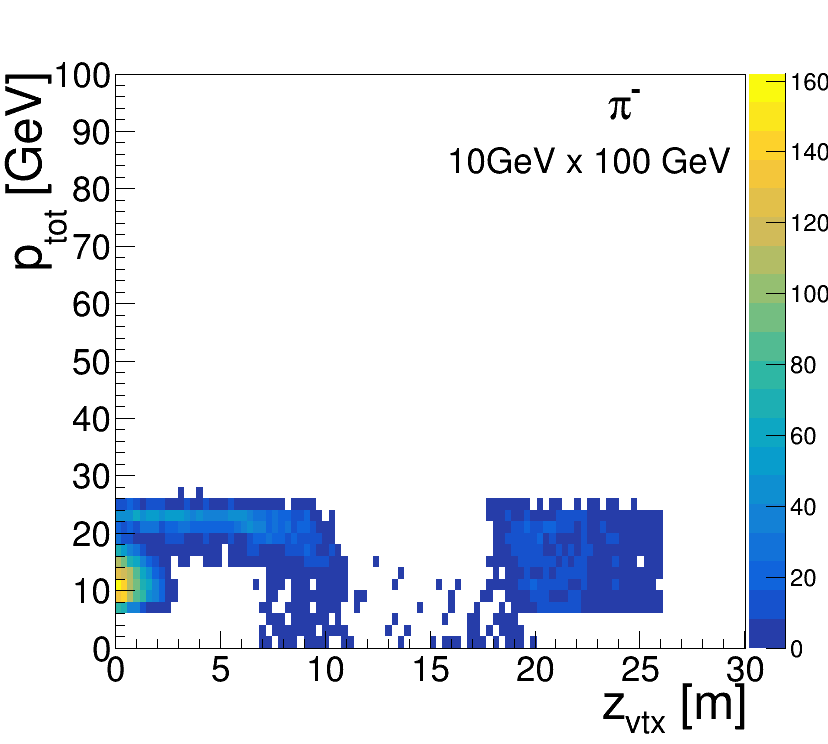}
\includegraphics[width=0.49\textwidth, trim={0 0 0 10mm}, clip]{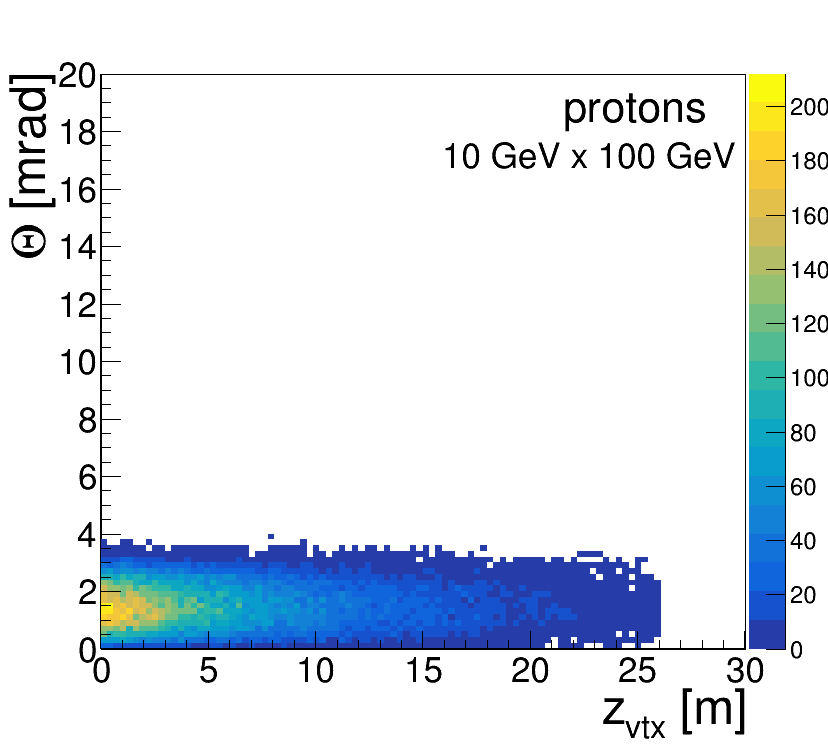}
\includegraphics[width=0.49\textwidth, trim={0 0 0 10mm}, clip]{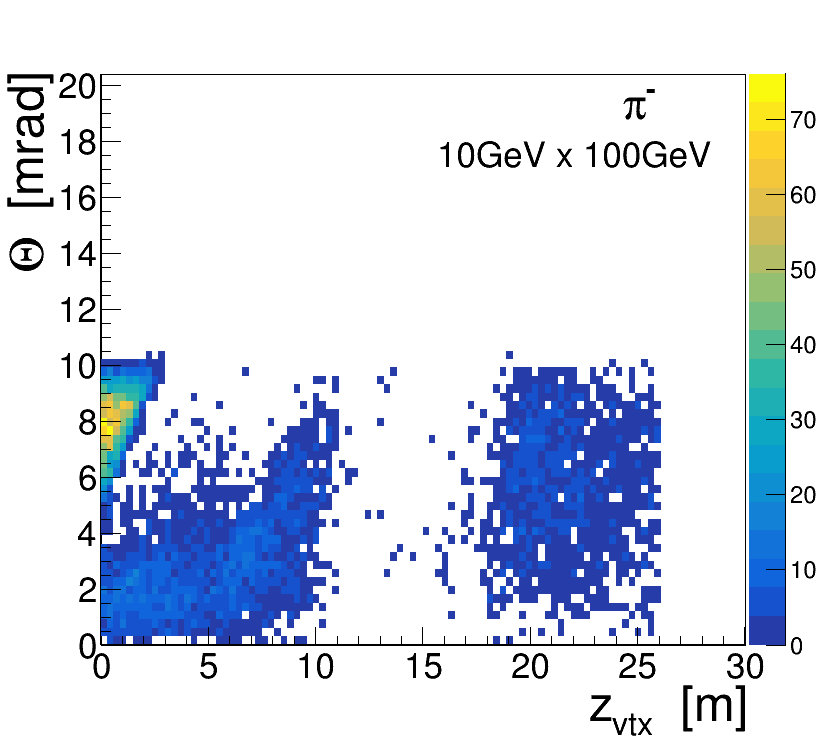}
%  \begin{tabular}{lll}
%\parbox[c]{0.45\textwidth}{\includegraphics[width=0.45\textwidth]{pict/Zvtx_ptot_prot_100.png}}
%\parbox[c]{0.45\textwidth}{\includegraphics[width=0.45\textwidth]{pict/Zvtx_ptot_pi_100.png}}
%\\
%\parbox[c]{0.45\textwidth}{\includegraphics[width=0.45\textwidth]{pict/Zvtx_theta_prot_100.png}}
%\parbox[c]{0.45\textwidth}{\includegraphics[width=0.45\textwidth]{pict/Zvtx_theta_pi_100.png}}
%\end{tabular}
%%  \includegraphics[width=0.7\textwidth]{pict/reco-L-p-Zvtx_10x100.png}
%%  \includegraphics[width=0.7\textwidth]{pict/reco-Lambda_theta_zvtx_10x100.png}
  \caption{Momentum (top) and angular (bottom) distributions of protons (left) and $\pi^-$ (right) from the $\Lambda \rightarrow p + \pi^-$ decay at beam-energy setting 10$\times$100, as registered in the far-forward detectors as a function of their origination (the decay vertex). For the $\pi^-$, one clearly sees the ``dead'' area in the FFQ magnet region, where placement of detectors is impossible.
  \label{fig:L-p_Zvtx_100}}
\end{figure}

\begin{figure}[!ht]
\centering
\includegraphics[width=0.46\textwidth]{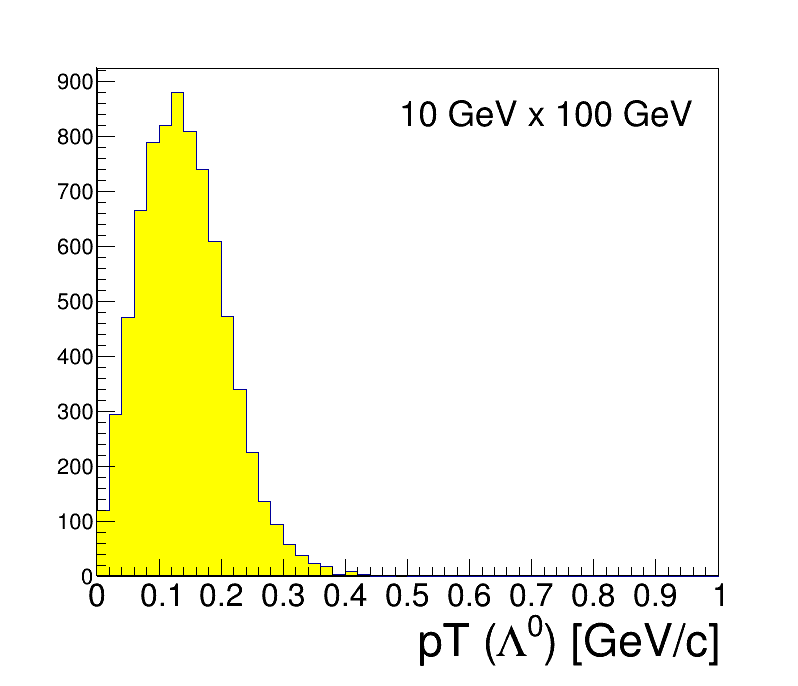}
\includegraphics[width=0.48\textwidth]{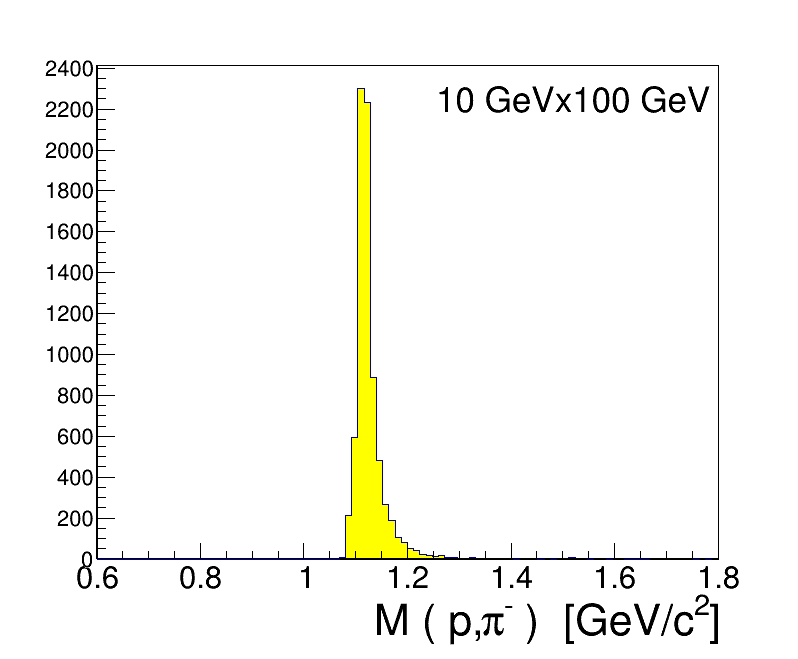}
\caption{$p_T$ (left) and invariant mass (right) of reconstructed $\Lambda$ particles for the 10$\times$100 beam energy setting.}
\label{fig:Mass}
\end{figure}

For the higher beam energy combination, for example 10$\times$100, the situation will be much different. Figure~\ref{fig:L-p_Zvtx_100}) shows the momentum and angular distributions for protons and $\pi^-$. For the latter, one can clearly see a ``dead" area appear along the beam line, where the FFQ beam elements are located, prohibiting placement of detectors and thus $\pi^-$ detection. This comes from the fact that these pions have significantly lower momentum and so are swept into the magnets and beam line. Those $\Lambda$s which decay after the set of FFQs will be tagged by the off-momentum detector, but since the $Z_{vtx}$ is unknown, it will be difficult to make a one-to-one correlation between the tagged position and the particle's momentum or angle. Therefore, for the final reconstruction of the $\Lambda$ invariant mass, one has to use only events with $Z_{vtx} < 3$-5 metres, to make this correlation possible. That this remains possible is revealed in figure~\ref{fig:Mass}\,--\,right, which shows the invariant mass spectra of the $\Lambda (p,\pi ^-) $ channel for this 10$\times$100 beam energy setting. The corresponding $p_T$ spectrum of the $\Lambda$ particles is shown on the left panel of figure~\ref{fig:Mass}.

We summarise this result in table~\ref{tab:lambda}, which shows the expected $\Lambda$ detection efficiency for the decay $\Lambda \rightarrow p + \pi^-$. A cut on decay within 4 meters, $Z_{vtx} < 4\,$m has been applied for this selection. The decrease in detection efficiency for the higher-energy settings comes mainly from this $Z_{vtx}$ cut, but is necessary to ensure $\Lambda$ mass reconstruction.

% CA: I CHANGED THE TABLE TO MAKE IT MORE PDG-ish
\begin{table}[hbt]
\centering
    \begin{tabular}  {   l c c c c  c   }
    \hline
%    \begin{tabular}  {  | c| c| c | c| c| c| c |   } %original
    \hline
%      & 5x41 GeV & 10x100 GeV & 18x275 GeV\\ %original
 Beam energies  & 5$\times$41 &  & 10$\times$100 &  & 18$\times$275\\
      \hline
 % \# protons &  2606   &  6261  &  2608\\
%   \hline
%   \# pions   &  1267   &  1045  &  103\\
 %  \hline
 % \# Lambdas &  1267    &  1045 &   103  \\
%     \hline
%Lambda Efficiency & 20\%  & 15\%  & 1\% \\  %original
Lambda Efficiency & 20\% & & 15\% & & 1\% \\
      \hline
      \hline
    \end{tabular}
    \caption{%The number of detected particles for 10k events, or
    $\Lambda$ detection efficiency as a function of energy setting, for $\Lambda$ detection with a cut on decay applied of $Z_{vtx} {\rm cut} < 4$\,m to ensure $\Lambda$-mass reconstruction.}
    \label{tab:lambda}
\end{table}

%===========================================================

\subsubsection{ $\Lambda \rightarrow n + \pi ^\circ$.}

\begin{figure*}[htbp]
 \includegraphics[width=0.98\textwidth]{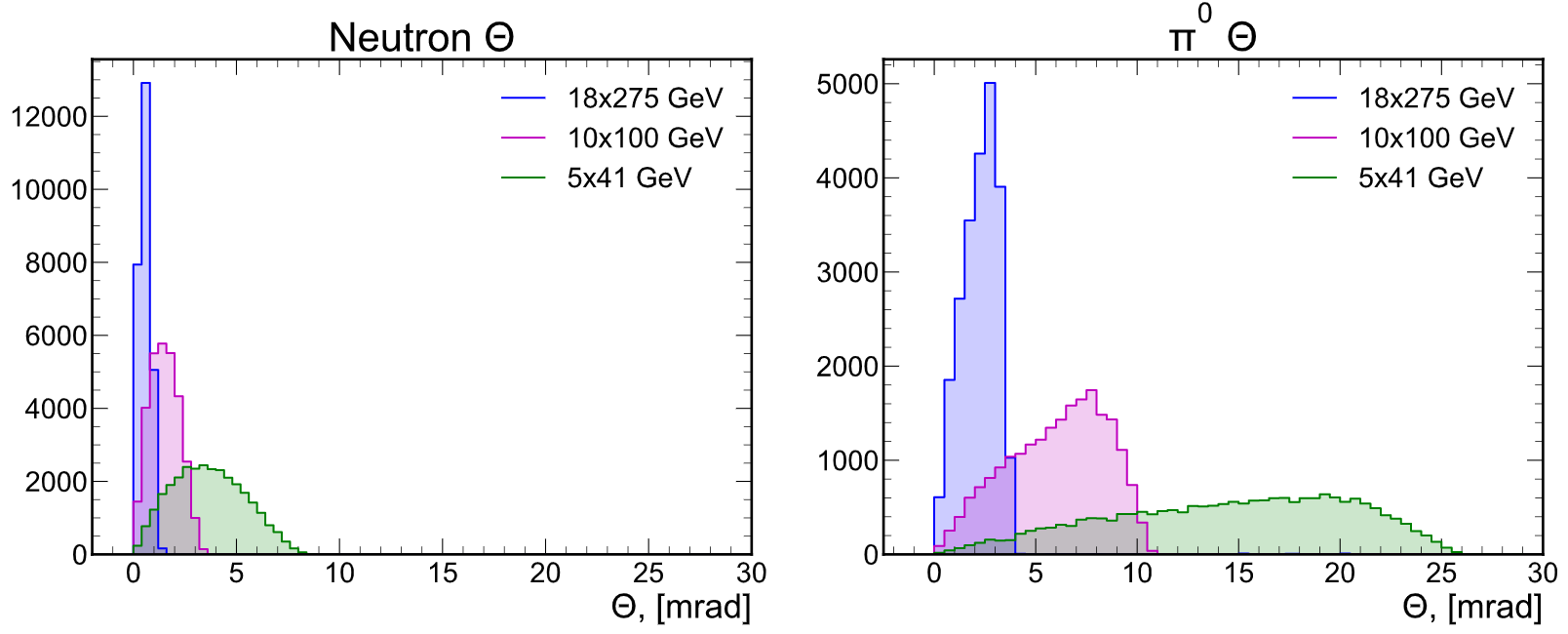}
     \caption[]{
        Angular distributions for detected decay products of $\Lambda \rightarrow n + \pi^{0}$: (a) neutrons;  and (b) $\pi^{0}$.  Beam energy settings: 18$\times$275, 10$\times$100, and 5$\times$41.
        \label{fig:n_pi0_theta}
    }
\end{figure*}

For this process, there are only neutral particles in the final state. The main scheme of detection for these particles will be the ZDC and/or some kind of electromagnetic calorimeter/photon detector in the B0 area. As with the $p + \pi^{-}$ decay mode, with lower beam energies, more particles can be detected in the central detector region. Figure~\ref{fig:n_pi0_theta} shows the angular ($\Theta$) distributions for $n$ and $\pi^{0}$ for different beam energies. It is furthermore assumed that the $\pi^{0}$ is reconstructed from $\pi^{0} \rightarrow \gamma\gamma$, where the photons are deposited in one of the corresponding detectors.

%\begin{figure*}[hb]
%    \includegraphics[width=\textwidth]{pict/reco-lambda_n_pi0_xl.png}
%    \caption[]{
%        $X_{L}$ distributions for detected decay $\Lambda$ products of : neutrons (a)  and  $\pi^{0}$ (b) for 10$\times$100 energy setting
%        \label{fig:n_pi0_xl}
%    }
%\end{figure*}

% Neutron and Pi0 Theta picture

\begin{figure*}[htbp]
    \includegraphics[width=0.98\textwidth]{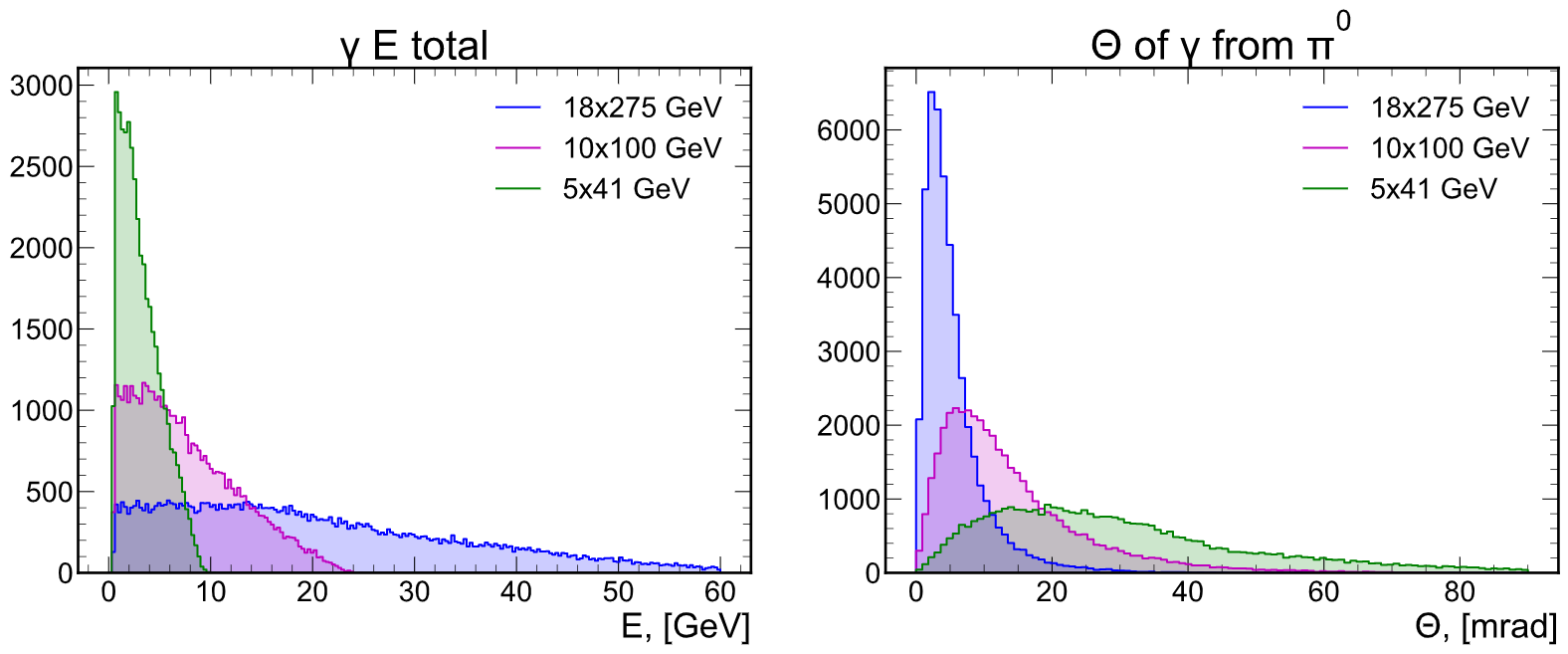}
    \caption[]{
        Energy and angular $\Theta$ distributions for the detected $\gamma\gamma$ to reconstruct the $\pi^{0}$ from a $\Lambda$ decay channel
        \label{fig:gam_etot_theta}
    }
\end{figure*}

\begin{figure*}[htbp]
    \includegraphics[width=0.98\textwidth]{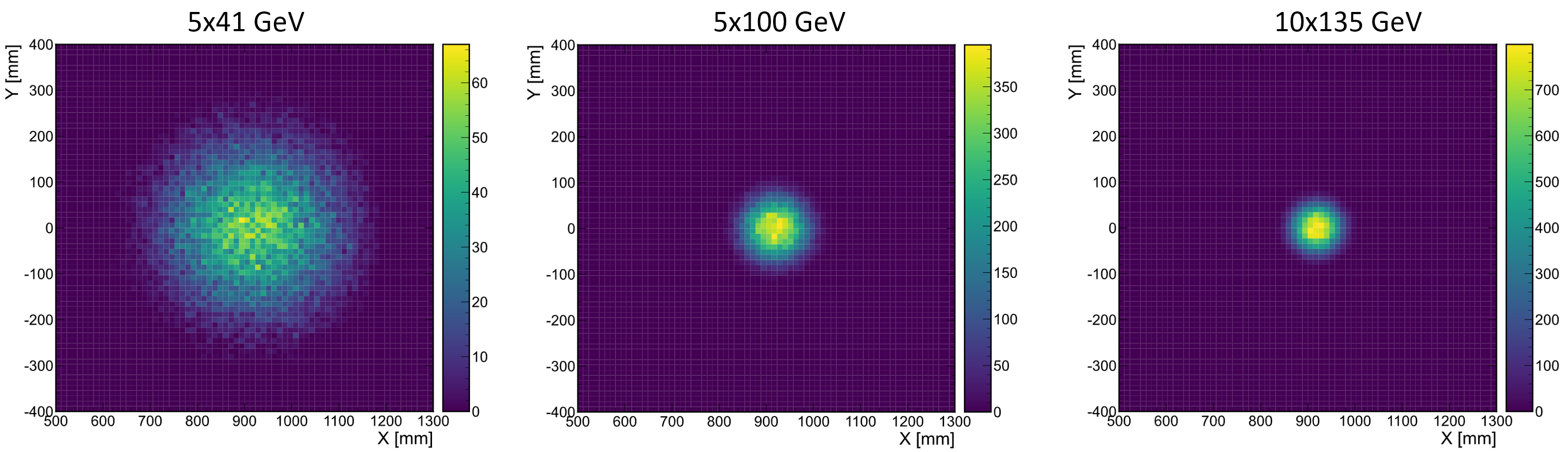}
%    \caption[]{
%        Occupancy distribution for detected neutrons from $\Lambda$ decay in ZDC for different beam energies
%        \label{fig:n_zdc_diff}
%    }
    \vspace*{\floatsep}% https://tex.stackexchange.com/q/26521/5764
    \includegraphics[width=0.98\textwidth]{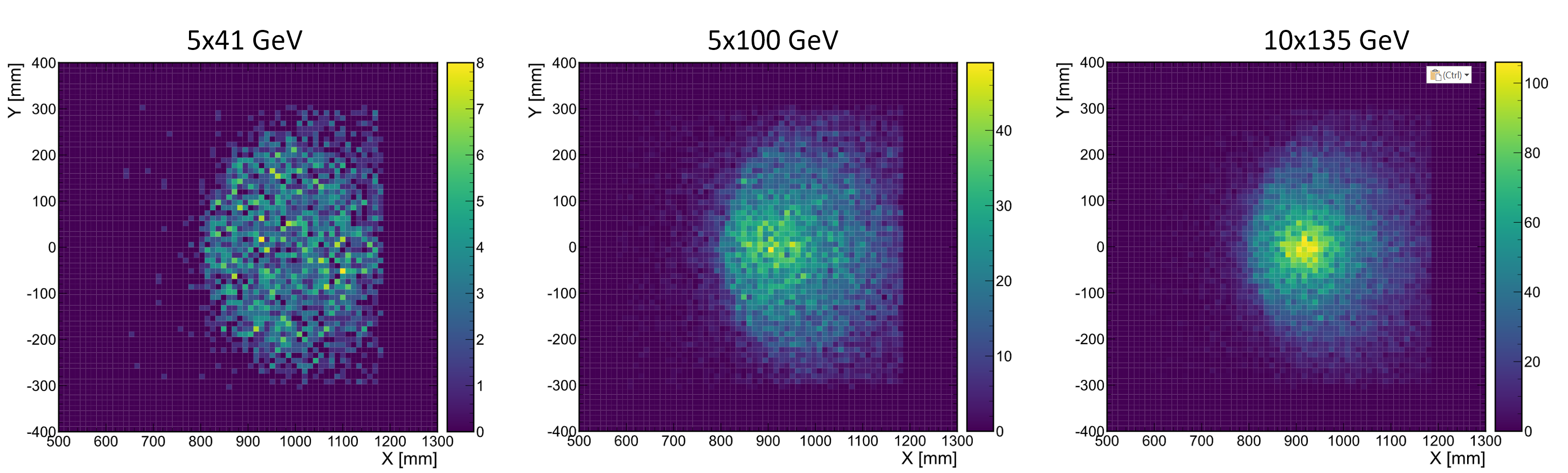}
    \caption[]{
        Occupancy distribution for neutrons (top panels) and $\gamma\gamma$ from $\pi^{0}$ decay (bottom panels) as detected in the ZDC for different beam energy settings.
        \label{fig:gam_zdc_diff}
    }
\end{figure*}

The energy and angular distributions of the two photons from the $\pi^{0}$ decay are shown in figure~\ref{fig:gam_etot_theta}, for various beam energy settings. At lower beam energy settings, some measurement to detect the larger-angle photons in the B0 area is required to recapture efficiency. As the beam energy increases, the ZDC starts playing the main role for detection of both neutrons and neutral-pions.
%Fig. ~\ref{fig:n_zdc_diff} and
This is illustrated further in figure~\ref{fig:gam_zdc_diff}, which shows the occupancy plots of the ZDC for both neutrons and the $\gamma\gamma$ from $\pi^0$ decay for different energy settings.
At the higher 10$\times$135 energies, the ZDC captures all photons from neutral-pion decay, while at the lower 5$\times$41 energies many photons are at larger angles reducing the detection fraction in the ZDC.

% Gammas Etotal Theta picture

% Occupancy plots in ZDC

%===========================================================
%\clearpage
%===========================================================

\begin{figure}[htbp]
    \centering
\includegraphics[width=1.00\linewidth]{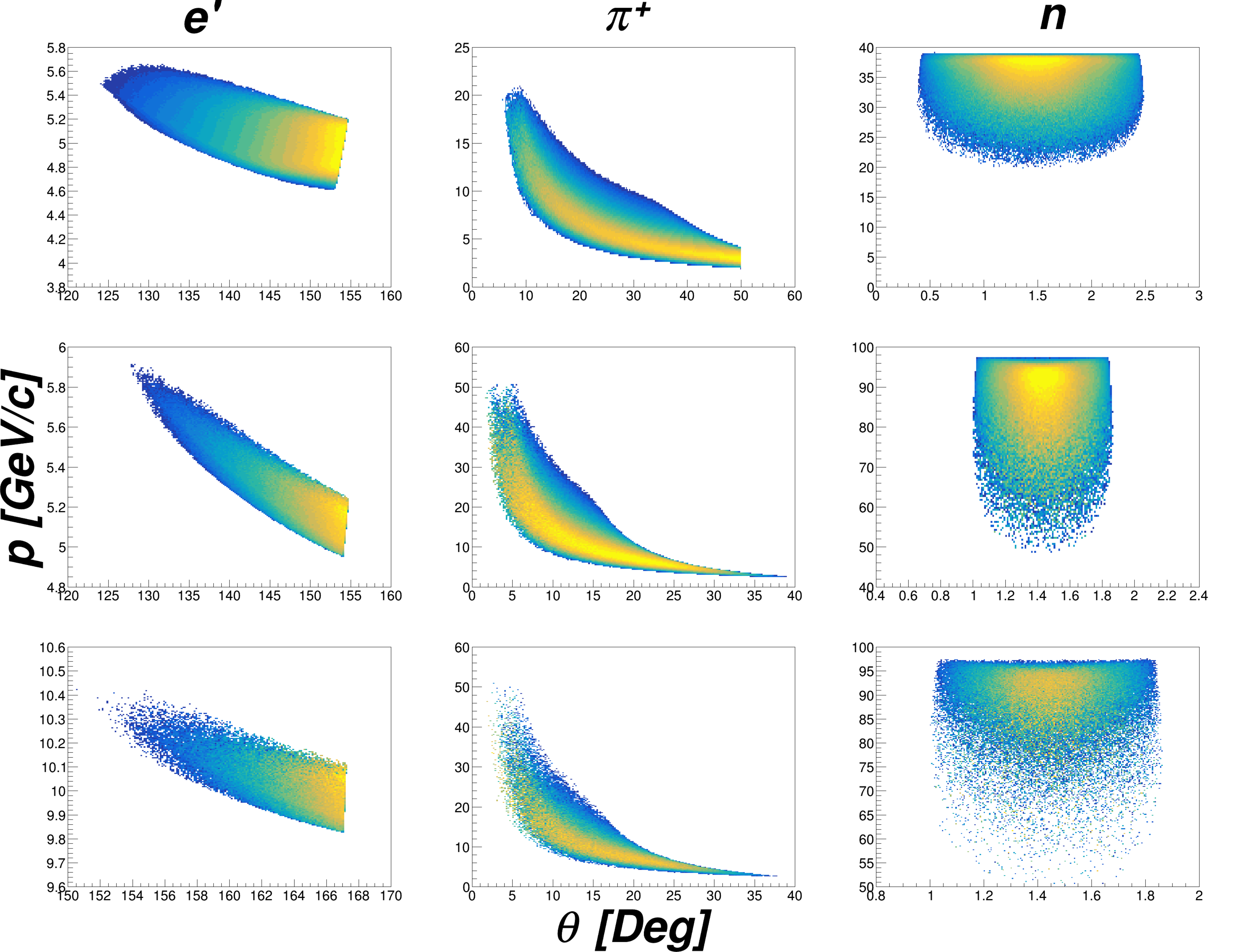}
% SK 18/01/21 - Old version with polar/individual images below

%\includegraphics[width=0.32\linewidth]{pict_alt/5_41_v1_electrons.pdf}
%\includegraphics[width=0.32\linewidth]{pict_alt/5_41_v1_pions.pdf}
%\includegraphics[width=0.32\linewidth]{pict_alt/5_41_v1_neutrons.pdf}

%\includegraphics[width=0.32\linewidth]{pict_alt/5_100_electrons.pdf}
%\includegraphics[width=0.32\linewidth]{pict_alt/5_100_pions.pdf}
%\includegraphics[width=0.32\linewidth]{pict_alt/5_100_neutrons.pdf}

%\includegraphics[width=0.32\linewidth]{pict_alt/10_100_electrons.pdf}
%\includegraphics[width=0.32\linewidth]{pict_alt/10_100_pions.pdf}
%\includegraphics[width=0.32\linewidth]{pict_alt/10_100_neutrons.pdf}

%\includegraphics[width=0.34\linewidth]{pict_alt/5_100_ePolar.png}
%\includegraphics[width=0.32\linewidth]{pict_alt/5_100_piPolar.png}
%\includegraphics[width=0.32\linewidth]{pict_alt/5_100_nPolar.png}

%\includegraphics[width=0.34\linewidth]{pict_alt/10_100_ePolar.png}
%\includegraphics[width=0.32\linewidth]{pict_alt/10_100_piPolar.png}
%\includegraphics[width=0.32\linewidth]{pict_alt/10_100_nPolar.png}
    \caption{Kinematic distributions for exclusive $p(e,e^\prime\pi^+n)$ events for $e^\prime$ (left), $\pi^+$ (center), and $n$ (right), at $5\times 41$ (top), $5\times 100$ (middle), and $10\times 100$ (bottom) beam energies. The neutron distribution is offset by 25\,mr owing to the beam crossing angle.}
    \label{fig:fpi_kin}
\end{figure}

\subsection{Exclusive $p(e,e^\prime\pi^+n)$ events}

The kinematic distributions for exclusive $p(e,e^\prime\pi^+n)$ events are shown in figure~\ref{fig:fpi_kin}. As for tagged DIS events, the neutrons assume nearly all of the proton  beam  momentum, and need to be detected at very forward angles in the ZDC. The scattered electrons and pions also have similar momenta as in the tagged DIS case, except that here the electrons are distributed over a wider range of angles. For instance, at the 5$\times$100 beam-energy setting, the $5-6\,$GeV electrons are primarily scattered 25$^\circ-45^\circ$ from the electron beam, while the $5-12\,$GeV $\pi^+$ are at 7$^\circ-30^\circ$ from the proton beam.
Further details of the exclusive events study, including the assumed requirements to separate exclusive events from non-exclusive background, and the pion form factor projections, are given in section~\ref{sec:fpi}.

\subsection{Accelerator and instrumentation requirements}
The physics simulation examples show that access to meson structure benefits greatly from EIC operations at the lower center-of-mass energies, with both $ep$ and $ed$ measurements at similar cm energies. Lower energies enhance the range of $Q^2$ at large $x_\pi$. Lower energies allow detection to uniquely tag kaon structure: this enhances $\Lambda$ decay probability at short distances and permits $\Lambda$-mass reconstruction to work from the detected decay products. To tag meson structure, proper instrumentation of B0 tracking detectors is needed, with full azimuthal coverage and perhaps a smaller proton-beam pipe diameter. Off-momentum detectors also have to provide full azimuthal coverage for detection of negatively-charged decay particles.

In terms of complimentary, an improved spectrometer along the beam line to enhance efficiency for detection of the low-momenta decay particles would be beneficial. The present beam line design leaves a large area with no possible detection, making $\Lambda$ tagging difficult for particles originating from larger $Z_{\rm vtx}$. This complicates access to meson structure at larger proton (ion) beam energies. Alternatively, a beam line design with an improved secondary focus could be beneficial for $\Lambda$ tagging.

\section{Physics projections}
% SJDK - Removed names from title, commented below as a reminder, delete once no longer needed
% ({\underline{Richard}}, Tim)
\subsection{Meson structure functions}
\label{sec:meson-structure-functions}

\subsubsection{Pion structure function projections}

A fast Monte Carlo was used for feasibility studies of $\pi$ and $K$ structure function measurements. The Monte Carlo is a C++ and ROOT based custom event generator~\cite{EIC_mesonMC} that uses the random number generator TRandom3 in ROOT. The inputs of the generator are minimum and maximum $Q^2$ and $x$ values, initial ion and electron beam energies, flags for initial beam smearing, and the number of events to simulate. The generator calls various quantities such as CTEQ6 PDF tables, nucleon structure functions, and the tagged $\pi$ and $K$ structure functions and splitting functions. The $\pi$ structure function, in particular, can be parametrised in many ways. Here, the $F_2^\pi$ structure function is calculated at NLO through the use of pion PDFs, which were determined in~\cite{Cao2020}.

\begin{figure}[!thbp]
\begin{center}
    \includegraphics[width=4.5in]{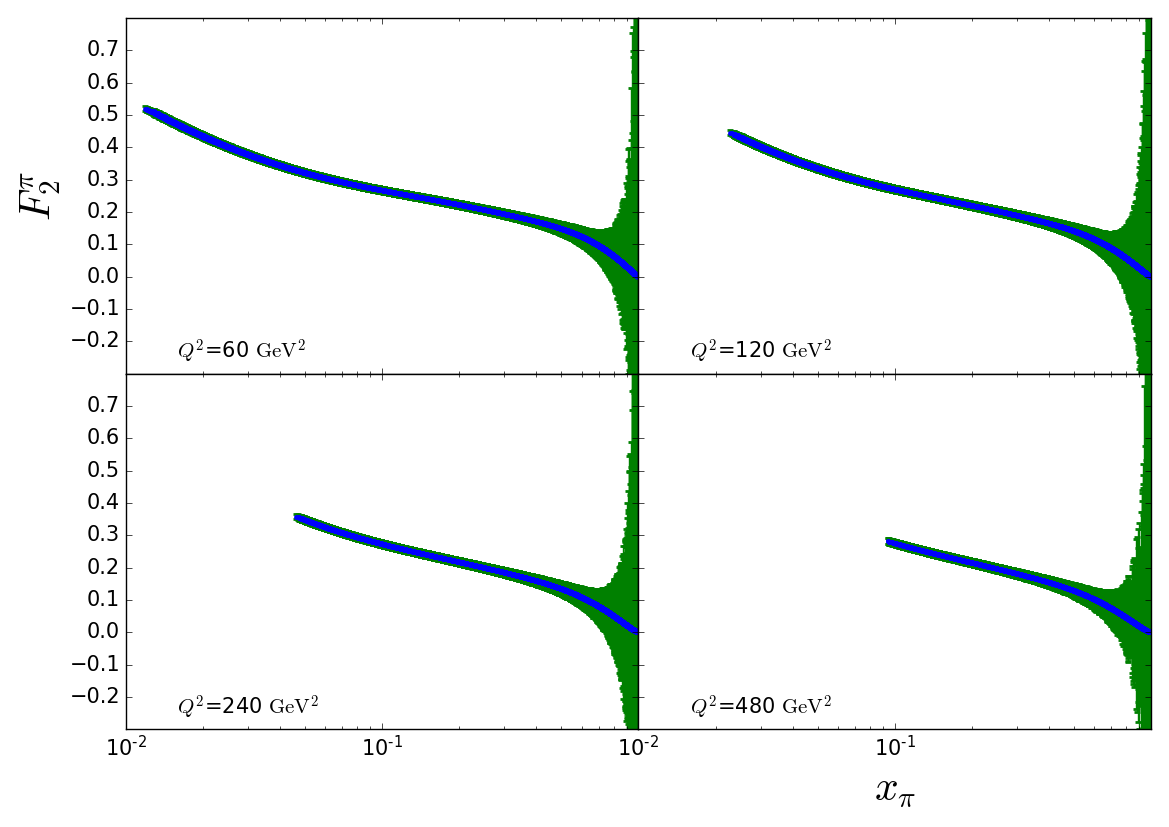}
    \caption{\label{fig:MC_fpi_logxpi_10on135}
    Monte Carlo projections of the pion structure function vs $x$ for a beam energy of 10$\times$135. The projected data is binned in $x$ and $Q^2$, with bin sizes of 0.001 and 10\,$\mathrm{GeV}^2$, respectively. The blue points are the Monte Carlo projections for $Q^2$ values of 60, 120, 240, 480\,GeV$^2/c$. The green bands are the statistical uncertainties for a luminosity of 100\,$\mathrm{fb}^{-1}$. 
    }
\end{center}
\end{figure}

The hadronic splitting function, ${\mathsf f}_i$, appearing in Eq.~(\ref{eq:TSF}) determines the meson flux essential to computing the leading-baryon production cross section of equation~(\ref{eq:Xsect}). For the sake of the simulations presented in this analysis, this flux was computed in the context of the single-meson exchange framework, which is valid for soft exchange momenta. The details of the hadronic splitting function were fixed to the relativistic vertex factor approach used in \cite{Hobbs:2014fya}, including a $s_{\pi N}$-dependent Gaussian interaction with ultraviolet regulator $\Lambda\! \sim\! 1\,$GeV. Although the details of the hadronic splitting were not varied in simulating EIC tagging measurements, it should be stressed that the EIC can be expected to be sensitive to the meson flux as well as the meson structure function.  A detailed examination of the sensitivity to the meson flux will be undertaken in the future.

\begin{figure}
\begin{center}
    \includegraphics[width=0.9\textwidth]{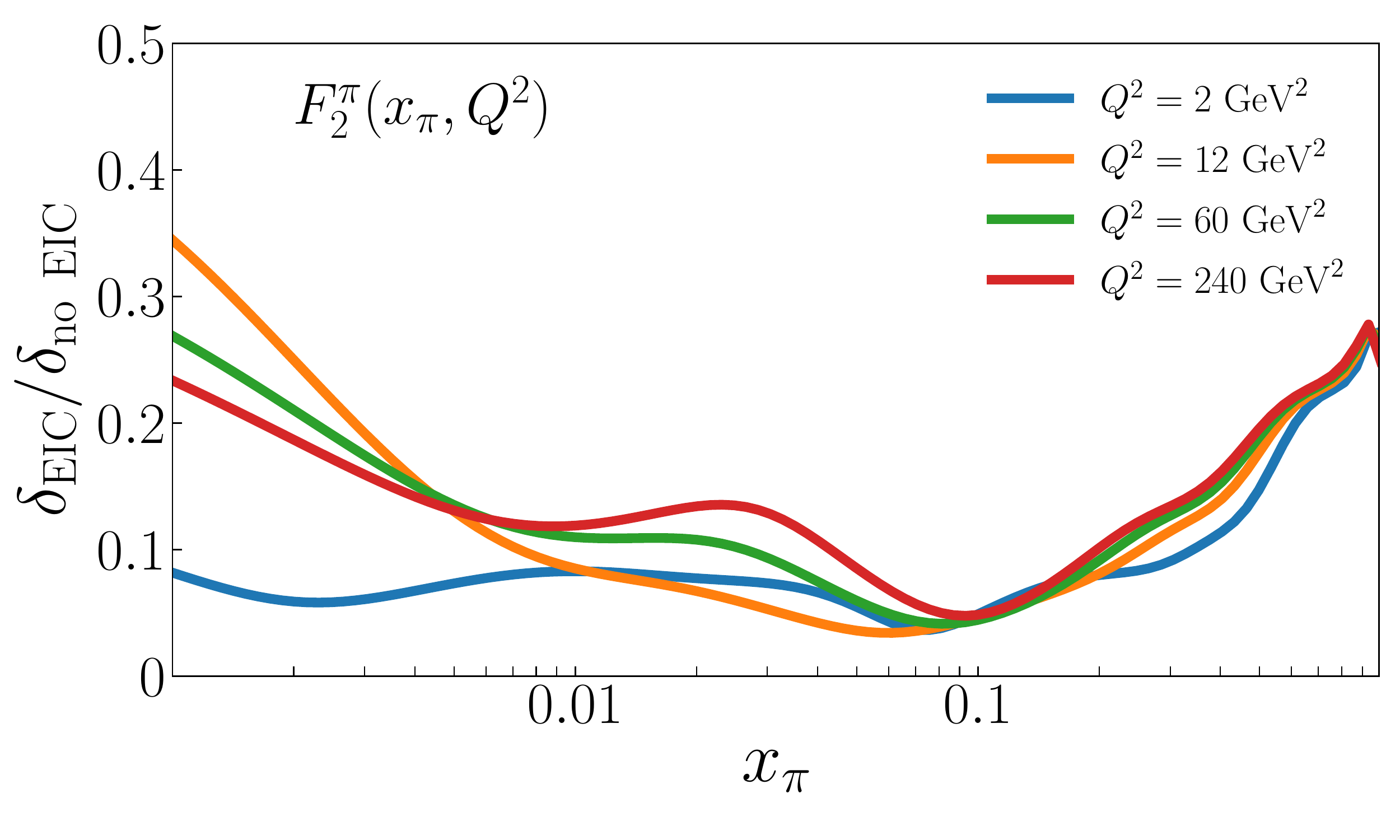}
    \caption{\label{fig:F2pi}
   Ratio of the uncertainty of the $F_2^\pi$ structure function from the global fit with and without including EIC projected data to the uncertainty of the $F_2^\pi$ as a function of $x_\pi$ for various $Q^2$ values.
    }
\end{center}
\end{figure}

%\begin{figure}[!thbp]
%\begin{center}
%    \includegraphics[width=4.5in]{pict_alt/MC_fpi_t_10on135.png}
%    \caption{\label{fig:MC_fpi_t_10on135}
%    Monte Carlo projections of the pion structure function vs t for a beam energy of 10$\times$135. The projected data is binned in x and $Q^2$ with bin sizes of 0.001 and 10 $\mathrm{GeV}^2$, respectively. The blue points are the Monte Carlo projections for $Q^2$ values of 60, 120, 240, 480~$GeV^2$/c. The green bands are the statistical uncertainties for a luminosity of 100~$fb^{-1}$.
%    }
%\end{center}
%\end{figure}

The plot in figure~\ref{fig:MC_fpi_logxpi_10on135} shows the reach in $x$ for four $Q^2$ bins at the 10$\times$135 energy setting.
The pion structure function simulations were validated by their agreement with the experimental HERA data~\cite{Chekanov:2002pf} in that regime, and with the GRV fit~\cite{Gluck:1991ey} at higher $x$.
Statistical uncertainties with the addition of the leading neutron detection fraction (discussed in the previous section) were incorporated to the overall uncertainty for a luminosity of $\mathcal{L}~=100\,\mathrm{fb}^{-1}$. For this energy, the coverage in $x$ extends down to 10$^{-2}$, with reasonable uncertainties in the mid-to-large $x$ region, increasing rapidly as $x\rightarrow1$. Even with these restrictions, the coverage in mid to high $x$ is unprecedented and should allow for detailed comparisons between pion and kaon structure.

%\textcolor{red}{In figure~\ref{fig:pion-pdf-impact},
%we showed the impact of EIC data on the pion PDFs themselves
%and their uncertainties.
%Figure~\ref{fig:F2pi} displays the ratio of the uncertainty
%of the $F_2^\pi(x_\pi,Q^2)$ structure function
%resulting from a global fit with EIC pseudodata to that without it.
%To investigate the $Q^2$ dependence on the impact, we show various $Q^2$ values of a wide range between a few ${\rm GeV}^2$ and a few hundred ${\rm GeV}^2$ over the range %$10^{-3}<x_\pi<1$.
%Strikingly, the $F_2^\pi$ structure function's uncertainties reduce
%by a dramatic 80-90\% in the range of $x_\pi$ between $3\times 10^{-3}$ and $0.4$ in the presence of EIC data, no matter the %values of $Q^2$.
%Within the whole range, the uncertainties reduce by 65\% or more.
%Below $x_\pi$ of 0.1, the $F_2^\pi$ structure function reduces by a factor of 10 for the case when $Q^2=2~{\rm GeV}^2$.
%As can be seen in figure~\ref{fig:cross_section_uncertainties},
%the data points at $x_\pi<0.1$ all had $Q^2<10~{\rm GeV}^2$,
%which indicates the constraining power at such low values of $Q^2$.
%The EIC provides an opportunity to improve our knowledge of the $F_2^\pi$.
%}
In Figure~\ref{fig:pion-pdf-impact} we showed the impact of EIC data on the pion PDFs themselves
and their uncertainties, folding in the estimated systematic uncertainty and the projected statistical uncertainties from the simulations (see Figure~\ref{fig:cross_section_uncertainties}). The resulting access to a significant range of $Q^2$ and $x$, for appropriately small $-t$, will allow for much-improved insights in the gluonic content of the pion.

Figure~\ref{fig:F2pi} displays the ratio of the uncertainty
of the $F_2^\pi(x_\pi,Q^2)$ structure function
resulting from a global fit with EIC projected data to that without it.
We show various $Q^2$ values of a wide range between a few ${\rm GeV}^2$ and a few hundred ${\rm GeV}^2$ over the range $10^{-3}<x_\pi<1$ to investigate the $Q^2$ dependence of the impact.
Strikingly, the $F_2^\pi$ structure function's uncertainties reduce
by 80-90\% in the range of $x_\pi$ between $3\times 10^{-3}$ and $0.4$ in the presence of EIC data, no matter the values of $Q^2$.
Within the whole range, the uncertainties reduce by 65\% or more.
Below $x_\pi$ of 0.1, the $F_2^\pi$ structure function reduces by a factor of 10 for the case when $Q^2=2~{\rm GeV}^2$. The constraining power at such low values of $Q^2$ is illustrated in figure~\ref{fig:cross_section_uncertainties}, where all data points shown are in the range $Q^2<10~{\rm GeV}^2$.
The EIC provides a unique opportunity to improve our knowledge of the $F_2^\pi$ over a large range in $Q^2$ and $x$.

As discussed in section~\ref{sec:sullivan}, theoretical calculations~\cite{Qin:2017lcd} predict that the Sullivan process should provide clean access to meson structure below a minimum value of $-t$. For the pion, this is $-t\leq0.6\,\mathrm{GeV}^2$. Similar, corrections for extracting the pion structure information from the theoretical backgrounds (absorptive corrections, higher meson-mass exchanges, etc., see section~\ref{sec:theory_background}) are minimized by measuring at the lowest $-t$ possible. Experimentally, this can be addressed by various tagged pion structure measurements as function of $-t$, for low $-t < \sim 0.6\mathrm{GeV}^2$, and verifying pion structure extraction.
%The full $-t$ range is shown in the plot of figure~\ref{fig:MC_fpi_t_10on135} for 10$\times$135\,GeV at the EIC. Specifically, this plot shows that in the accessible range of $-t$ there are reasonable uncertainties which would allow for an order-of-magnitude gain in statistics compared to HERA. The resulting access to a significant range of $Q^2$ and $-t$  values, including small $-t$, as well as access to a significant range of $x$, will allow for insights into the pion's gluon content.

\begin{figure}
\begin{center}
    \includegraphics[width=0.95\textwidth]{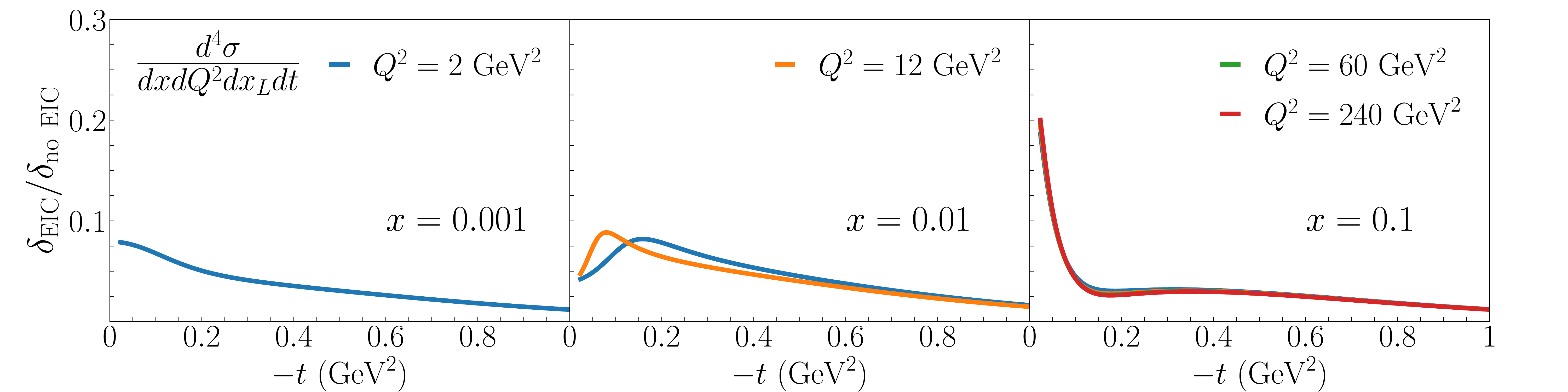}
    \caption{\label{fig:cross_section_ratio}
   Ratios of the uncertainty of the differential cross section $\frac{d^4\sigma}{dxdQ^2dx_Ldt}$ from the global fit including EIC projected data to the uncertainty of that without the EIC as a function of $-t \in [-t_{\rm min},1]$ for various $Q^2$ and for
   (\emph{left panel}) $x=0.001$, (\emph{middle panel}) $x=0.01$, and (\emph{right panel}) $x=0.1$.  For all calculations, the value $x_L=0.85$ was used.
    }
\end{center}
\end{figure}

%\textcolor{red}{In figure~\ref{fig:cross_section_ratio}, we show the reduction
%of the uncertainties of the measurable physical quantity
%$\frac{d^4\sigma}{dxdQ^2dx_Ldt}$ as a function of $-t$.
%While the one pion exchange will not be a feasible model for too large $|t|$,
%we show the ratio of cross section beyond that value up to $-t=1$.
%Following the impact of the EIC pseudodata on the PDFs and the $F_2^\pi$ structure functions,
%figure~\ref{fig:cross_section_ratio} provides insight to an actual physical observable.
%In the leading baryon electroproduction,
%neither the PDFs nor the structure functions are directly
%observable due to the pion flux as shown in equation~\ref{eq:TSF}.
%In the left and middle panels of figure~\ref{fig:cross_section_ratio},
%we show that as a function of $-t$, the uncertainties on the
%calculable differential cross section reduce by 90\%
%at $x=0.001$ and $x=0.01$.
%For the case when $x=0.1$, we see in the right panel that the values of $Q^2$ are insignificant in the ratio of uncertainties.
%However, at large $x$, the ratio of uncertainties rise when $-t$ is closer to $0$.
%Such an overall impact of 75\% improvement on the uncertainties
%indicates both that our knowledge currently is poor and that the EIC
%will provide good constraints for cross sections.
%}

Figure~\ref{fig:cross_section_ratio}, shows the reduction
of the uncertainties of the four-fold differential cross section,
$\frac{d^4\sigma}{dxdQ^2dx_Ldt}$. The impact is illustrated by means of the ratio of cross sections, including EIC projected data to the uncertainty of that without the EIC, as a function of $-t$ and 
up to $-t=1$. The left and middle panels of figure~\ref{fig:cross_section_ratio},
show that as a function of $-t$, the uncertainties on the differential cross section reduce by 90\%
at $x=0.001$ and $x=0.01$.
For the case when $x=0.1$ in the right panel the values of $Q^2$ are insignificant in the ratio of uncertainties.
At large $Q^2$, at $x$ $\sim$ 0.1, the ratio of uncertainties rises when $-t$ is closer to $0$ which is due to a reduced experimental phase space to the low$-t$ region for those kinematics. The overall impact of $>$75\% improvement on the uncertainties
indicates both that our knowledge currently is poor and that the EIC
will provide good constraints for cross sections. Furthermore, this underlines that the measured tagged cross sections as function of $-t$ can be used to confirm robustness of pion structure extraction.

\subsubsection{Kaon structure function and splitting function projections}
\label{sec:kaon}
The pion structure function analysis presented here can be extended to the kaon, as the single-meson exchange framework can be generalised to the flavour $\mathrm{SU}(3)$ sector with expected validity for soft exchange kaons.  Empirical knowledge of the kaon sector is even more sparse than the analogous information on the pion. As such, comprehensive data would be of great utility for unravelling the splitting function ratio, ${\mathsf f}_K/{\mathsf f}_\pi$ as well as the structure function of the kaon, $F_2^K$. For initial simulations, the splitting function ${\mathsf f}_K$ might be fixed at first-order to inclusive hadroproduction data, and $\Lambda$(uds) baryon production according to equation~(\ref{eq:Xsect}).  Ultimately, precise EIC data over a range of $x$, $Q^2$, $y$, and $t$ would be instrumental for the sake of unraveling and constraining the meson flux model from the structure function $F_2^K$.

%CA: what will be the standard for the Q2 units? GeV^2 or (GeV/c)^2?
%GH: I think Q2 and t should be in GeV2.  Momentum should be in GeV/c, to avoid confusion with energy.
% SJDK - Removed names from title, commented below as a reminder, delete once no longer needed
% (Garth, Stephen)
\subsection{Meson form factors
\label{sec:fpi}}

The experimental determination of the $\pi^+$ electric form factor ($F_{\pi}$) is challenging.
The best way to determine $F_{\pi}$ would be electron-pion elastic scattering.
However, the lifetime of the $\pi^+$ is only 26.0\,ns.
Since $\pi^+$ targets are impossible and $\pi^+$ beams with the required properties are not yet available for measurements at modest-to-large $Q^2$ values, one must employ exclusive electroproduction, $p(e,e^\prime \pi^+)n$.
This is best described as quasi-elastic ($t$-channel) scattering of the electron from the virtual $\pi^+$ cloud of the proton, where $t$ is the Mandelstam momentum transfer $t=(p_{p}-p_{n})^2$ to the target nucleon.
As discussed in section~\ref{sec:sullivan}, scattering from the $\pi^+$ cloud dominates the longitudinal photon cross section ($d\sigma_L/dt$) at sufficiently small $-t$.

To reduce background contributions, one normally separates the components of the cross section owing to longitudinal (L) and transverse (T) virtual photons (and the LT, TT interference contributions) via a Rosenbluth
separation.
However, L/T separations are impractical at the EIC, as one cannot reach sufficiently low $\epsilon$ data to provide a good lever arm.
Below, we propose an alternate technique to access $\sigma_L$ via a model, validated with exclusive $\pi^-/\pi^+$ ratios from deuterium.
Once $d\sigma_L/dt$ has been determined over a range of $-t$, from $-t_{\rm min}$ to $-t \approx 0.6$~GeV$^2$, the value of $F_{\pi}(Q^2)$ is determined by comparing the observed $d\sigma_L/dt$ values with the best available electroproduction model, incorporating off-shell pion and recoil nucleon effects.
In principle, the obtained $F_{\pi}$ values depend upon the model used, but one anticipates this dependence to
be reduced at sufficiently small $-t$.
Measurements over a range of $-t$ are essential as part of the model validation process.
The JLab 6\,GeV experiments were instrumental in establishing the reliability of this technique up to $Q^2=2.45$~GeV$^2$~\cite{Huber:2008id, Horn:2016rip, Horn:2007ug, Volmer:2000ek, Horn:2006tm, Tadevosyan:2007yd, Blok:2008jy, Huber:2014ius, Huber:2014kar}, and extensive further tests are planned as part of JLab experiment E12-19-006.

\subsubsection{Requirements for separating exclusive and SIDIS events.}

The exclusive $\pi^+$-channel cross section is several orders of magnitude smaller than the neighbouring SIDIS background; but it is distributed over a much narrower range of kinematics, and this is essential for the separation of the exclusive events from the background.  The exclusive $p(e,e^\prime\pi^+n)$ reaction is isolated by detecting the forward-going high-momentum neutron, i.e.\ $e-\pi^+-n$ triple coincidences.
Since the neutron energy resolution is not very good, the neutron hit is used as a tag for exclusive events.  The neutron momentum is otherwise not used in the event reconstruction.

Detector effects have been simulated via the following ad-hoc smearing functions.
The pion and electron angular resolutions were estimated by assuming a 10\,$\mu$m position resolution in a cylindrical silicon vertex tracker (comparable with ZEUS), and this Cartesian position uncertainty was propagated to polar coordinates $(\theta, \phi)$.
From this, $\delta p=250\mu$rad was conservatively assumed for all angles, for both the electron and the pion.
The pion and electron momentum resolution was estimated from track reconstruction in the magnetic field via the resolution in \cite{Gluckstern:1963ng}, assuming 5 position measurements in a 3T solenoidal field.
To simplify the MC study, $\delta p/p=2\%$ was conservatively assumed for all angles, for both the electron and the pion.
Since the neutron energy resolution in the ZDC is not very good, the neutron hit was used as a tag for deep exclusive meson production (DEMP) events.  The neutron momentum was not otherwise used in the event reconstruction.
Thus, the missing momentum is calculated as $p_{miss}=|\vec{p_e}+\vec{p_p}-\vec{p_{e'}}-\vec{p_{\pi}}|$.

The effectiveness of kinematic cuts to isolate the exclusive $\pi^+$ channel was evaluated by comparison with a simulation of $p(e,e^\prime \pi^+)X$ SIDIS events, including both detector acceptance and resolution smearing effects.
The most effective cuts are on the detected neutron angle ($\pm 0.7^o$ from the outgoing proton beam), the reconstructed $-t<0.5$~GeV$^2$,  
%GH: added text 21-feb-17
and the missing momentum defined above. The $p_{miss}$ cut is $Q^2$-bin dependent, where the value is chosen to optimize the signal/background ratio for each bin.  It ranges from $p_{miss}>95$~GeV/c at $Q^2=6$~GeV$^2$, to 77~GeV/c at $Q^2$=35~GeV$^2$, where all events are removed above the cut value. 
%GH: end new text
After application of these cuts, the exclusive $p(e,e'\pi^+n)$ events are cleanly separated from the simulated SIDIS events.

\paragraph{Determining the longitudinal cross section $d\sigma_L/dt$.}

After the exclusive $\pi^+ n$ event sample is identified, the next step is to separate the longitudinal cross section $d\sigma_L/dt$ from $d\sigma_T/dt$, needed for the exaction of the pion form factor.
However, a conventional Rosenbluth separation is impractical at the EIC owing to the very low proton beam energy required to access $\epsilon<0.8$.
Fortunately, at the high $Q^2$ and $W$ values accessible at the EIC, phenomenological models predict $\sigma_L\gg\sigma_T$ at small $-t$.
For example, the Regge-based model in \cite{Vrancx:2013fra} predicts $R=\sigma_L/\sigma_T >10$ for $Q^2>10\,$GeV$^2$ and $-t<0.06\,$GeV$^2$, and $R>25$ for $Q^2>25\,$GeV$^2$ and $-t<0.10\,$GeV$^2$.
Thus, transverse cross section contributions are expected to be 1.3--14\%, with the ratios becoming more favourable with increasing $Q^2$ and decreasing $-t$.
The most practical choice appears to be to use a model to isolate the dominant $d\sigma_L/dt$ from the unseparated cross section $d\sigma_{\rm uns}/dt$.

To control the systematic uncertainty associated with the theoretical correction to estimate $\sigma_L$ from $\sigma_{\rm uns}$, it is very important to confirm the validity of the model used.
This can also be done with EIC data, using exclusive $^2H(e,e^\prime\pi^+n)n$ and $^2H(e,e^\prime\pi^-p)p$ data in similar kinematics to the primary $p(e,e^\prime\pi^+n)$ measurement.
The ratio of these cross sections is
\begin{equation}
    R=\frac{\sigma[n(e,e^\prime\pi^-p)]}{\sigma[p(e,e^\prime\pi^+n)]} =\frac{|A_V-A_S|^2}{|A_V+A_S|^2},
\end{equation}
where $A_V$ and $A_S$ are the isovector and isoscalar amplitudes, respectively.
Since the pion pole $t$-channel process used for the determination of the pion form factor is purely isovector (owing to $G$-parity conservation), the above ratio will be diluted if $\sigma_T$ is not small or if there are significant non-pole contributions to $\sigma_L$.
The comparison of the measured $\pi^-/\pi^+$ ratio with model expectations, therefore, provides an effective means of validating the model used to determine $\sigma_L$~\cite{Huber:2014ius,Huber:2014kar}.
The same model, now validated, can likely be used to extract the pion form factor from the $\sigma_{\rm uns}$ data.

\subsubsection{$\pi^+$ form factor projections.}

\begin{figure}[!thbp]
\begin{center}
%\vspace{-0.5cm}
\includegraphics[width=4.0in]{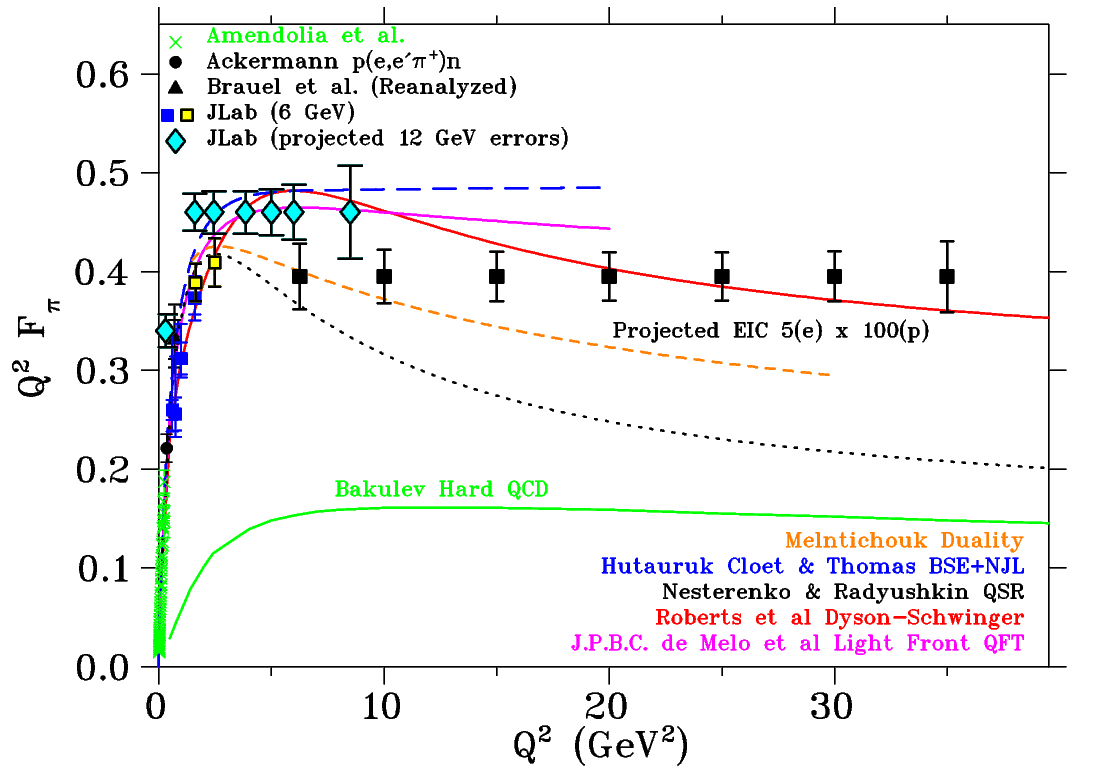}
%\vspace{-0.2cm}
\caption{\label{fig:eic_fpi}
%GH: caption info added 21-feb-17
Existing data (green \cite{Amendolia:1984nz,Amendolia:1986wj}; 
black \cite{Ackermann:1977rp,Brauel:1979zk,Huber:2008id};
blue and yellow \cite{Horn:2006tm,Tadevosyan:2007yd,Huber:2008id})
and projected uncertainties for
future data on the pion form factor from JLab (cyan~\cite{E12-19-006}) and EIC (black),
in comparison to a variety of hadronic structure calculations
(green solid \cite{Bakulev:2004cu};
orange dash \cite{Melnitchouk:2002gh};
blue long-dash \cite{Hutauruk:2016sug};
black dot \cite{Nesterenko:1982gc};
red solid \cite{Chang:2013nia};
violet solid \cite{Mello:2017mor}).  
%GH: end added info
The EIC projections
clearly cover a much larger $Q^2$ range than the JLab measurements.}
\end{center}
\end{figure}

As already discussed above, the value of $F_{\pi}(Q^2)$ can be determined by comparing the measured $\sigma_{uns}$ at small $-t$ to the best available electroproduction model, incorporating pion pole and non-pole contributions and validated with $\pi^-/\pi^+$ data.
The model should have the pion form factor as an adjustable parameter, so that the best fit value and its uncertainty at fixed $(Q^2, W)$ are obtained by comparison of the magnitude and $t$-dependences of model and data.
If several models are available, the form factor values obtained with each one can be compared to better understand the model-dependence.
The importance of additional $p(e,e'\pi^+n)$ model development to improve knowledge of pion form factors cannot be overestimated, and additional activity in this area should be encouraged.

Using this technique, the EIC can enable a pion form factor measurement up to $Q^2=35\,$GeV$^2$, as shown in figure~\ref{fig:eic_fpi}.
The errors in the yields are based on the following assumptions:
cross sections parametrised from the Regge model in \cite{Choi_2015},
integrated luminosity of 20\,fb$^{-1}$ for 5$\times$100 GeV measurement,
clean identification of exclusive $p(e,e^\prime\pi^+n)$ events by tagging the forward neutron,
and a cross section systematic uncertainty of 2.5\% point-to-point and 12\% scale.
One should then apply the following additional uncertainty, since the form factor will be determined from unseparated, rather than L/T-separated data:
$\delta R=R$ systematic uncertainty in the model subtraction to isolate $\sigma_L$, where $R=\sigma_L/\sigma_T=0.013-0.14$ at $-t_{\rm min}$.
The model fitting procedure is finally used to extract $F_{\pi}(Q^2)$ from the $\sigma_{uns}$ data, where one assumes the applied model is validated at small $-t$ by comparison to data.
Additional model uncertainties in the form factor extraction are not estimated here, but the EIC should provide data over a sufficiently large kinematic range to allow the model-dependence to be quantified in a detailed analysis.

\subsubsection{$K^+$ form factor.}

The reliability of the electroproduction method to determine the $K^+$ form factor is not yet fully established.
A recent extraction of the kaon form factor from electroproduction data at $Q^2=1.00, 1.36, 1.90, 2.07, 2.35\,$GeV$^2$ is discussed in \cite{Carmignotto:2018uqj}.
The L/T separated kaon electroproduction cross sections were extracted at different values of $-t$ using data from JLab~\cite{Mohring:2002tr,Coman:2009jk,Horn:2006tm} and the successful method from \cite{Horn:2006tm, Blok:2008jy} was applied to determine the kaon form factor.
JLab E12-09-011~\cite{E12-09-011} acquired data for the $p(e,e^\prime K^+)\Lambda$, $p(e,e^\prime K^+)\Sigma^0$ reactions at hadronic invariant mass $W=\sqrt{(p_{K}+p_{\Lambda,\Sigma})^2}>2.5$~GeV, to search for evidence of scattering from the proton's ``kaon cloud''.
The data are still being analysed, with L/T-separated cross sections expected in the next $\sim 2$ years.

If the anticipated data confirm that the scattering from the virtual $K^+$ in the nucleon dominates at low  four-momentum transfer to the target  $|t|\ll m_p^2$, the experiment will yield the world's first quality data for $F_K$ above $Q^2>0.2$~GeV$^2$.
This would then open up the possibility of using exclusive reactions to determine the $K^+$ form factor over a wide range of $Q^2$ at higher energies.
While the general technique will remain the same, the $\pi^-/\pi^+$ validation technique to confirm the $\sigma_L$ extraction cannot be used for the $K^+$.
One possibility could be for $\Lambda/\Sigma^0$ ratios to play a similar role.
However, conditions under which the clean separation of these two channels may be possible at the EIC requires further study.
These studies are planned for the near future.

\section{Summary and prospects}

%Light pseudoscalar meson structure relevance
%
%Rich science questions
%
%Synergy experimental data, QCD phenomenology, Lattice QCD, Phenomenology
%
%Sullivan process, "off-shell", tagged data observables is relevant either way
%
%Drives Far-Forward beam line detection scheme at EIC
%
%Excellent prospects due to synergy 12-GeV, COMPASS++/AMBER, EicC, EIC
%
%Must-know information on meson structure and emergent mass

After more than seventy years, there is now a growing realisation that the first-ever discovered mesons hold the keys to our further understanding of the vast bulk of visible mass in the Universe.  The pion was the first discovered meson, in 1947 \cite{Lattes:1947mw}, soon followed in the same year by the kaon \cite{Rochester:1947mi}, the first strange particle \cite{Christy:1957lsa, Yamanaka:2019}. These Nambu-Goldstone bosons would be massless if Nature expressed chiral symmetry simply; and would remain massless in the absence of quark couplings to a Higgs boson. Yet, these light pseudoscalar mesons are intimately linked to confinement; their structure is complicated; and their masses, although uncommonly light, are not zero, being generated by constructive interference between an emergent mass mechanism, expressed in dynamical chiral symmetry breaking, and the Higgs mechanism.  Some 100 MeV of the 494 MeV kaon mass, with its heavier strange quark, or 20\%, may be attributed directly to the Higgs mechanism.

The emergence of the bulk of visible mass and its manifestations in the existence and properties of hadrons and nuclei are profound questions that probe into the heart of strongly interacting matter. What Nature provided as properties of pions and kaons, the Standard Model's would-be Nambu-Goldstone modes, are tell-tales of the emergent hadron mass and structure mechanisms, and the required interplay with the Higgs mechanism. For example, the quark and gluon energy contributions to pion and kaon masses give information on the balance of these mechanisms, the magnitude and scale-dependence of pion and kaon form factors inform about the size and range of the interference between emergent mass and the Higgs-mass mechanisms, the pressure distribution and transverse momentum distributions in pions as compared to protons inform about universality of the attractive and repulsive forces inside hadrons.

Understanding fundamentally requires a synergistic effort that combines experiment, theory, computing, and phenomenology, with the aims being to reveal how the roughly 1 GeV mass-scale that characterises atomic nuclei appears and why it has the observed value; why ground-state pseudoscalar mesons are unnaturally light in comparison; and to elucidate the role of the Higgs boson in forming hadron properties. Pions, kaons, protons and their counterpart neutrons provide the building blocks of the visible universe. Their exact QCD substructures have now readily become available by marked progress in theory and computing, and their further understanding in turn will shed light on how Nature created mass and visible structure.

%Answering these questionsThis fundamentally requires a synergistic effort that combines experiment, theory, computing, and phenomenology in order to reveal how the roughly 1 GeV mass-scale that characterises atomic nuclei appears; why it has the observed value; why ground-state pseudoscalar mesons are unnaturally light in comparison; and the role of the Higgs boson in forming hadron properties. Pions, kaons, protons and their counterpart neutrons provide the building blocks of the visible universe. Their exact QCD sub- structures have now readily become available by marked progress in theory and computing, and their further understanding in turn will shed light on how Nature created mass and visible structure.
%ADD SOME SENTENCES FROM 2.4 - (RE) used phrasing from that in abstract in order not to change flow here

The foreseen Electron-Ion Collider (EIC) will be a real game changer for experimental data on pion and kaon structure. In specific kinematic regions, an electron scattering process coupled with the observation of recoil nucleons (N) or hyperons (Y) receiving sufficiently low four-momentum transfer, $-t$, can reveal features associated with the ``meson cloud''of the nucleon. With proper theoretical understanding on the interpretation of the off-shell pion (kaon) target and possible theoretical backgrounds as function of $-t$, experimental access to a physical pion (kaon) target is enabled. This can be further experimentally validated by constraining theoretical backgrounds by using both $ep$ and $ed$ scattering data, and by ensuring interpretation independent of $-t$. Based on present theoretical and experimental guidance, this may be possible for $-t <$ 0.6 (0.9) GeV$^2$ for pion (kaon) targets. For elastic scattering, this Sullivan process carries information on the pion or kaon form factor. For deep inelastic scattering, the process informs about the mesonic parton content. Regardless of interpretation, the various tagged pion (kaon) cross section data as a function of $-t$ are valuable in their own right.

Since only a small fraction of the nucleons emit a virtual meson and only small momentum transfers from the nucleon to the resulting baryon allow the interpretation in terms of a real pion (kaon), the highest luminosities of $\sim$10$^{34}$ electron-nucleons cm$^{-2}$ s$^{-1}$ are necessary. Owing to the long lifetime of the $\Lambda$, lower collision (or rather proton/ion beam) energies are slightly favored at the EIC to tag kaon structure. The need to efficiently tag pion and kaon structure is further fundamentally intertwined with the integration of the EIC detector in the interaction region, especially related to any far-forward (in the direction of the proton/ion beam) detection scheme of recoil nucleons and hyperons. All sub-components of the far-forward area play an important role to detect forward-going protons (in Roman pots) and neutrons (in Zero-Degree Calorimeters), and to detect hyperon decay products: protons and negatively charged pions at opposite sides of the beam line, and neutrons and photons originating from neutral pion decays in zero-degree and electromagnetic calorimeters. It is shown that appreciable detection efficiencies are thus achieved, with exception of kaon tagging at the highest EIC proton beam energies (275 GeV). The kaon tagging scheme could be improved by an alternate magneto-optics design, removing a large area with nearby and therefore integrated magnets in the present beam line, and hence no possible detection of decay products, or alternately an improved secondary focus.

This detection scheme allows, through the Sullivan process, excellent prospects for pion (and kaon) structure function measurements over a large range of $x$ and $Q^2$, approaching the vast ($x,Q^2$) landscape of the HERA proton structure function measurements. The tagged pion (kaon) cross section data remain precise over a large range of $-t$, up to $-t \sim$ 0.6 GeV$^2$ for the pion. This could allow further study of more exclusive semi-inclusive and deeply-virtual Compton scattering data, towards transverse momentum and pressure distributions in the pion. To access the pion form factor, an alternate technique to access the longitudinal cross section is used, via a model, validated with exclusive $\pi^+/\pi^-$ ratios from deuterium. Scattering from the pion cloud dominates the longitudinal cross section at low $-t$, and, if dominant, this ratio would approach unity. This could allow precise pion form factor determination at EIC up to $Q^2 \sim$ 35 GeV$^2$. The reliability of a similar $\Lambda/\Sigma$ ratio method to extract the kaon form factor has not yet been established, but it may be studied from a Jefferson Lab 12-GeV kaon electroproduction experiment that ran in 2018/2019.

The EIC will play a key role to access pion and kaon structure over a wide range of centre-of-mass (CM) energies, $\sim 2-140\,$GeV, to chart in-pion and in-kaon distributions of mass, charge, magnetisation, and perhaps angular momentum. Nonetheless, to provide experimental measurements guiding theoretical understanding requires a coherent, worldwide effort. Jefferson Lab will provide, at its CM energy $\sim 5\,$GeV, data for the pion (kaon) form factor up to $Q^2 \sim 10 (5)\,$GeV$^2$ and for insights into mechanisms competing with the Sullivan process, and early measurements of the pion (kaon) structure functions at large-$x$ ($>$ 0.5).  AMBER can provide pion, and especially much-needed kaon, Drell-Yan measurements in the CM energy region $\sim 10-20\,$GeV. These measurements are critical elements in a global effort on pion and kaon structure function measurements. They also allow an independent determination of the ``pion flux'' for EIC Sullivan process measurements. An Electron-Ion Collider in China (EicC) is under consideration, with CM energy $\sim 10-20\,$GeV perfectly attuned to AMBER and forming a bridge from Jefferson Lab to EIC.

Successful completion of the programme sketched herein will deliver deep, far-reaching insights into the distributions and apportionment of charge, mass, and spin within the pion and kaon; the similarities and differences between such distributions in these (almost) Nambu-Goldstone modes and the benchmark proton; the symbiotic relationship between the emergence of hadron mass and confinement; and the character and consequences of constructive interference between the Standard Model's two mass-generating mechanisms.  It has the potential to finally complete a chapter in science whose first lines were written more than eighty years ago and contained the prediction of the pion's existence \cite{Yukawa:1935xg}.

% SJDK 18/01/21 - Info on acknowledgements from the IOP guidelines is included below.
%\subsection{Acknowledgments}
%Authors wishing to acknowledge assistance or encouragement from
%colleagues, special work by technical staff or financial support from
%organizations should do so in an unnumbered `Acknowledgments' section
%immediately following the last numbered section of the paper. In \verb"iopart.cls" the
%command \verb"\ack" sets the acknowledgments heading as an unnumbered
%section.

%Please ensure that you include all of the sources of funding and the funding contract reference numbers that you are contractually obliged to acknowledge. We %often receive requests to add such information very late in the production process, or even after the article is published, and we cannot always do this. Please %collect all of the necessary information from your co-authors and sponsors as early as possible.

% SJDK 18/01/21 Acknowledgements section below. Fill in as your funding agency requires and add any further thanks you want.
\ack
% TH 01-24-2021
This work was supported in part by
%% List of DOE grants
the U.S. Department of Energy, Office of Science, Office of Nuclear Physics, under contracts DE-AC05-06OR23177, DE-FG02-03ER41260;
 %% List of NSF grants,
the US National Science Foundation under grants PHY-1653405, PHY-1714133, PHY-2012430,;
%% List of all other grants
the Natural Sciences and Engineering Research Council of Canada (NSERC), FRN: SAPIN-2016-00031;
%% ... CDR
the Jiangsu Province \emph{Hundred Talents Plan for Professionals};
% 25/01/21 added by J. Segovia
Ministerio Espa\~nol de Ciencia e Innovaci\'on, grant no. PID2019-107844GB-C22;
Junta de Andaluc\'ia, contract nos.\ P18-FRJ-1132;
Operativo FEDER Andaluc\'ia 2014-2020 UHU-1264517;
% 25/01/21 added by T. Frederico
 Conselho Nacional de Desenvolvimento Cient\'ifico e Tecnol\'ogico (CNPq) under grant no. 308486/2015-3;
 Funda\c c\~ao de Amparo \`a Pesquisa do Estado de S\~ao Paulo (FAPESP) under the thematic project  grant 2017/05660-0;
 INCT-FNA project 464898/2014-5;
 Coordena\c c\~ao de Aperfei\c coamento de Pessoal de N\'ivel Superior (CAPES—Finance Code 001);
 Research  Corporation  for  Science  Advancement through the Cottrell Scholar Award. %Lin
The work of NS was supported by the DOE, Office of Science, Office of Nuclear Physics in the Early Career Program.
% Wally special mention
% RE 19/01/21 added Wim and Doug
Contributions of and discussions with Wally Melnitchouk are gratefully acknowledged.
The authors would like to thank Wim Cosyn and Doug Higinbotham for helpful comments during this work and preparation of this manuscript.
The authors would also like to thank Zafar Ahmed, Rory Evans and Wenliang (Bill) Li for their work and assistance on the DEMP event generator.

%%%%%%%%%%%%%%%%%%%%%%%%%%%%%%%%%%%%%%%%%%%%%%%%%%%%%%%%%%%%%%%%%%%%%%%%%%%%%%%%%%%%%%%%%%%%%%%%%%%%%
%\section*{References}
%CA there was a conflict due to the use of natbib package defined in the header
% SJDK - The only conflicts/warnings I get are due to undefined citations, looks like some are missing entirely and others are missing some info

\bibliographystyle{unsrt}
\bibliography{bibliography}

\newpage

\setcounter{tocdepth}{2}
\tableofcontents

\end{document}